\documentclass {article}
\usepackage{amsfonts,amsmath,latexsym}
\usepackage{amstext,amssymb}
\usepackage{amsthm}
\usepackage {epsf}
\usepackage {epsfig}
\usepackage[scanall]{psfrag}
\usepackage {graphicx}

\newcommand {\len}{\operatorname{length}}
\newcommand {\Area}{\operatorname{Area}}

\newcommand {\dist}{\operatorname {dist}}

\newcommand{\diver}{\mathop\text{div}\nolimits}

\newcommand{\CCM}{\mathcal C_M}

\newcommand{\bnu}{{\text{\boldmath$\nu$}}}

\renewcommand {\line}{\hbox to\linewidth}
\newtheorem {thm} {Theorem}

\newtheorem {lemma} [thm] {Lemma}
\newtheorem {cor} [thm] {Corollary}
\newtheorem {defn} {Definition}

\begin{document}

\title{{\normalsize Submitted to {\it International Journal of Solids and Structures.}}\\ \ \\ \ 
Compatibility Conditions for  Discrete Planar Structures}
\author{Andrejs Treibergs\thanks{\tt treiberg@math.utah.edu} \ \ \  Andrej Cherkaev\thanks{Supported by  grant DMS NSF, Award 1515 125.\ \ \ {\tt cherk@math.utah.edu}}\ \ \  Predrag Krtolica
\thanks{\tt krtolica@math.utah.edu}
\\ {\small\it Department of Mathematics, University of Utah, 155 South 1400 East, Salt Lake City, Utah 84112}
\\ \vtop{{\small{\bf Key words:} Compatibility conditions, discrete structures, Cauchy Green strain,\newline\phantom{nuts!} damage, continuum limit, compatibility integral, Maxwell number}}
}
\maketitle

\begin{abstract}
Compatibility conditions are investigated for planar network structures consisting of nodes and connecting bars. These conditions restrict the elongations of bars and are analogous to the compatibility conditions of deformation in continuum mechanics. Two problems are considered: the discrete  problem for structures with prescribed lengths and its linearization, the discrete problem of prescribed elongations. These problems approximate two continuum problems,
the nonlinear continuum problem of given Cauchy Green tensor and its linearization, the continuum problem of prescribed strain. 
 The requirement that the deformations remain planar imposes  solvability conditions of all four problems.   For triangulated structures, compatibility conditions for the nonlinear problem are expressed as a polynomial equation in the lengths of edges of the star domain surrounding each interior node. In the linearized discrete problem, the compatibility conditions become linear relations for the elongations of edges on the same domains. The continuum limits of the compatibility conditions for both discrete problems are proved to be the compatibility conditions for the continuum problems.
 
The compatibility equations may be summed along a closed curve to give conditions supported on a strip along the curve. Similarly, for continuous materials, the compatibility equation for the prescribed strain problem may be integrated along a closed curve to provide an integral condition, 
analogous to how the prescribed Green tensor problem may be integrated to give the Gauss-Bonnet integral formula.

Compatibility conditions are investigated for general trusses such as
plates connected by girders or triangulated domains with holes or missing edges. Compatibility conditions on general trusses may be non-local. There may be rigid trusses without compatibility conditions in contrast to continuous materials. The number of compatibility conditions is the number of bars that may be removed from a structure and still keep it rigid. This number measures the resilience of the structure. The compatibility equations around a hole involve the edges in the neighborhood surrounding the hole. An asymptotic density of compatibility conditions for periodically damaged triangular structures is found to be sensitive to the number and location of the removed edges.
\end{abstract}

\tableofcontents 

\section{Introduction}
Two overdetermined problems for the deformation of material are considered,  the discrete problem for structures with prescribed lengths and its linearization, the discrete problem of prescribed elongations. These problems approximate two continuum problems used for reference,
the nonlinear continuum problem of given Cauchy Green tensor and its linearization, the continuum problem of prescribed strain. Their compatibility conditions are investigated,  mainly in the discrete linear situation, where they are conditions on the inhomogeneous term of an overdetermined matrix equation.

The compatibility equations of a general discrete structure may involve data from widely separated points and are not usually local.
However, for a truss to approximate a material,  the compatibility conditions must have a local nature and limit to the continuum compatibility equations at all points as the discretization is refined. In a continuous material, the compatibility equations express how nearby deformations influence deformations at a point. For  discrete structures that approximate a continuous material, the compatibility equations must be supported on diminishing neighborhoods of most points, which we call \emph{material points.}
 A class of structures which approximate the material in planar domains are the  \emph{triangulated structures.} Compatibility conditions of a triangulated structure are localized to triangles neighboring a vertex, thus as the triangulation is refined, all interior vertices are material points approximate all points in the domain.

\subsection{Motivation--compatibility equations in a truss}
\begin{figure}[h]
      \begin{center}
          \scalebox{0.5}{\includegraphics{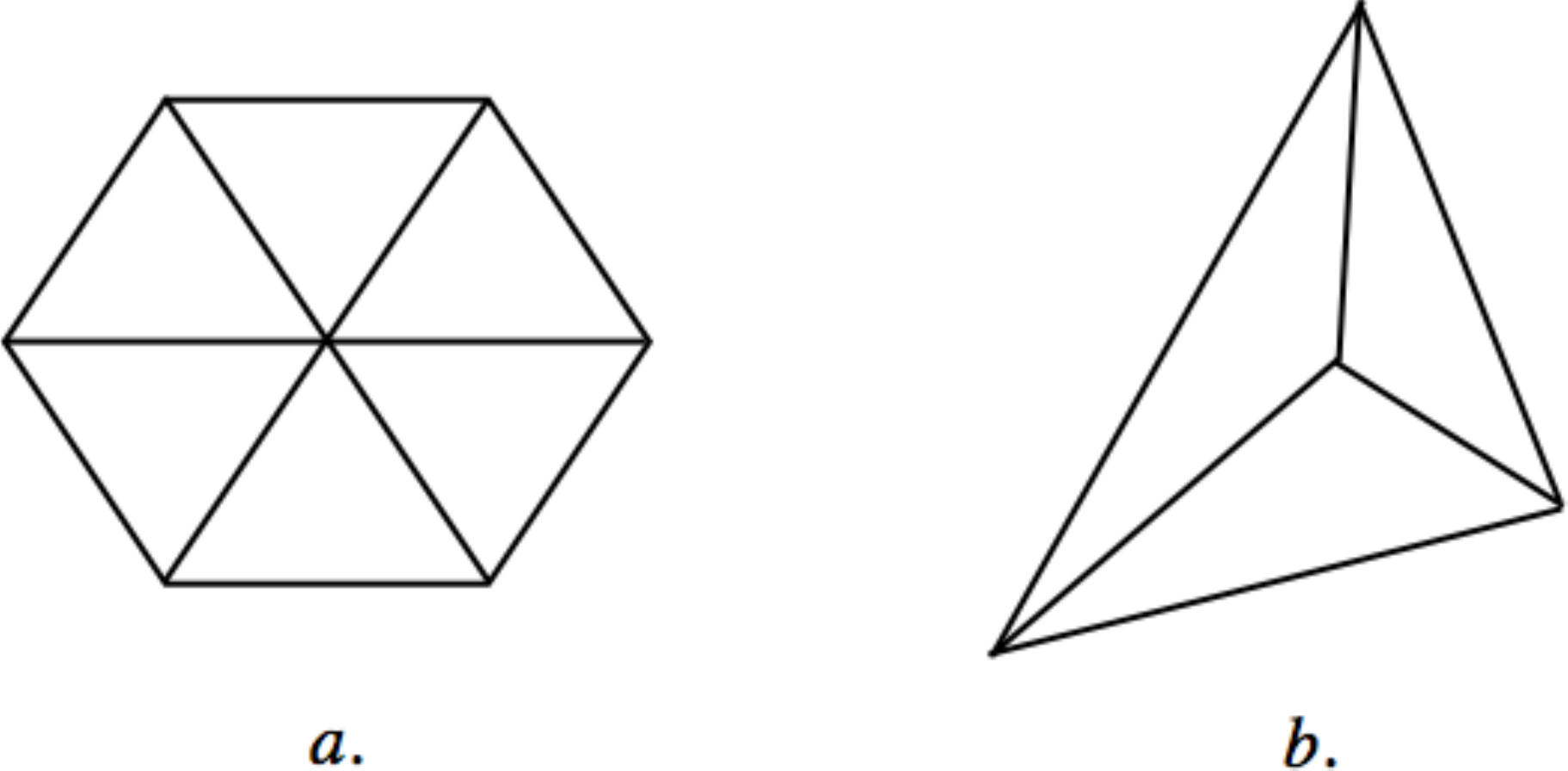}}
 \caption{Interior node compatibility condition:   neighborhood stays rigid after removing an edge. \label{fig_innerpoint}}
 \end{center}
       \end{figure}

Suppose that one is to arrange given rigid bars (edges) into a triangulated structure in the plane. The lengths of edges cannot be arbitrary. The relations between the bar lengths that allow them to fit as edges in a structure are called discrete compatibility conditions for this structure.  As a simple but essential example, consider a neighborhood of an interior node $V$ of a triangulated surface, that is, a node surrounded by a closed chain of nodes and edges joined by radial edges to the center node; all edges are rigid as in Figure~\ref{fig_innerpoint}. 
The union of triangles that meet the center node is called a \emph{star neighborhood}.
If one edge is removed, the distance between its ends is still fixed by the remaining edges; therefore its length is related to the lengths of other edges. This dependence is a compatibility condition. It is defined at each interior node of the triangulated surface structure, depends only on the triangles neighboring the interior node and follows from neighborhood remaining planar; namely, the sum of angles between edges going around the vertex is $360^{\circ}$. Hence the curvature atom $K(V)=0$ vanishes at the center (see equation \eqref{eq_Katom}).  We call the linearized compatibility condition at an interior node a \emph{wagon wheel condition.}

\subsection{Cauchy-Green Deformation  Tensor} 
Let us recall the basic definitions of the kinematics of a deformable body~\cite{MH:1983}.
  Let $\mathcal B \subset \mathbf E^2$ be a Euclidean material disk domain with piecewise smooth boundary and coordinates $(X^1, X^2)$, and $\mathcal S\subset\mathbf E^2$ the target domain with coordinates $(x^1,x^2)$. An in-plane displacement
$$
\phi : \mathcal B \to\mathcal S
$$
has a prescribed Green Deformation tensor (Right Cauchy-Green Deformation Tensor)  $. \zeta$
$$
\text{(NC)}\qquad\qquad\qquad\qquad
C_{AB}=\sum_iF^i{}_A\, F^i{}_B=\zeta_{AB},\qquad \text{where $F^i{}_A=\dfrac{\partial \phi^i}{\partial X^A}$}.\qquad\qquad\qquad\qquad
$$
The problem (NC) is to find the deformation $\phi$ given the positive definite  Green tensor $\zeta$. Hence $C$ is a Riemannian metric on $\mathcal B$ pulled back from the Euclidean metric at $\phi(X)$. At each point of $\mathcal B$, $\zeta$ is a symmetric $2\times 2$ matrix whose three entries depend on the two components of the vector $\phi$, therefore are not independent but are constrained by a differential compatibility condition. Because the deformation remains planar,  the compatibility condition is equivalent to that the  Gaussian curvature $\mathcal K(\zeta)$ of the metric being zero.

\subsection{General, triangulated and triangular structures}\label{ss_gentristr}

\begin{defn}\label{def_str}  There are three types of discrete structures.
\begin{itemize}
\item A \emph{general discrete structure} or \emph{abstract truss} $\mathcal S$\rm is 
a connected finite graph:   the vertices $V_i\in \mathcal V$ form a finite set   of points in the plane;  and the edges $\mathcal E$ form a finite collection  of pairs of distinct vertices $E_{ij}$  linking  $V_i$ to $V_j$. 
\end{itemize}\rm
Connected means there is an edge path between each pair of vertices in the graph connecting one vertex to the other.
A  general discrete structure need not be a planar graph.  The nodes and links may coincide, may overlap and there may be more than one edge linking a pair of vertices. 
\begin{itemize}
\item A \emph{triangulated structure} is obtained by the vertices and edges of piecewise linear triangulation of a planar domain. 
\end{itemize}
 A triangulation of a domain is a tiling by non-overlapping closed triangles whose union is the closure of the domain. The triangles are allowed to intersect only in a common vertex or along a whole common edge. 
 A triangulated structure is made up of the nodes and edges of a triangulated domain. It is assumed that edges of the triangles are straight line segments.  The union of vertices and edges of triangulation is sometimes called the \emph{one-skeleton} of the triangulation~(e.g.,~\cite[p. 5]{Ha:2002}). The trusses made of non-intersecting links may be viewed as triangulated structures with missing links.
\begin{itemize}
\item A \emph{triangular structure} is a triangulated structure whose links are unit edges of the triangular lattice.
\end{itemize}
\end{defn}

The triangular lattice is the set of points in the plane of the form $k\mathbf e_1+\ell \mathbf e_2$ where $k$ and $\ell$ are integers and  vectors $\mathbf e_1=(1,0)$ and $\mathbf e_2=\left(1/2,{\sqrt 3}/2\right)$. All links have unit length and the angles between neighboring links are all $60^{\circ}$.

Some of our theorems apply for general structures. The majority of theorems apply to triangulated structures since the formulation of compatibility depends on triangles around interior nodes forming a planar neighborhood of the node. Others are proved only for triangular trusses. Because the triangular trusses are the simplest to understand, many notions will be explained for triangular trusses first.

\subsection{Energy of approximating discrete structures}
If the deformation is elastic, one can associate with it an energy density $e(\zeta)$ which depends on the deformation and may be expressed as the infimum over displacements $\phi$
$$
W(\mathcal B)= \inf_{\phi}\int_{\mathcal B} e(\zeta) \, dx,\qquad \text{where $\zeta=(\nabla \phi)^T\, \nabla \phi$ and subject to boundary conditions.}
$$
This is equivalent to the constrained infimum over Cauchy Deformation tensors
$$
W(\mathcal B)= \inf_{\zeta}\int_{\mathcal B} e(\zeta) \, dx,\qquad\text{$\zeta$ is subject to $\mathcal K(\zeta)=0$ and boundary conditions,}
$$
 but the latter is frame-independent which may be preferable for some problems.

Suppose that the links of a discrete structure have
 a prescribed length $\ell_{ij}=\len(E_{ij})$. The nonlinear existence problem (ND) is whether this structure may be realized as points in the Euclidean plane such that distances between endpoints equal the prescribed lengths of the edges~\eqref{eq_prescell}. 
When the structure is the triangulation of a planar domain, the compatibility condition states that the curvature atom (angle excess which depends on the $\ell_{jk}$) vanishes at each interior vertex $K(V_i)=0$, (see section \eqref{s_ND}). 
In section~\ref{ss_Kpoly}, it is shown that for triangulated structures compatibility conditions for (ND) may be expressible as a polynomial equation in $\ell_{ij}$.

Let $e(\ell_{ij})$ denote the energy associated to the length of a link, which is a convex function of the length such that $e(\ell)>0$ if $\ell>0$ and $e(0)=\infty$. The elastic energy of the whole structure is 
$$
W(\mathcal S)=\sum_{i<j} e(\ell_{ij}),\qquad\text{subject to $K(V_i)=0$ at interior vertices.}
$$
Let a sequence of triangulated structures approximate a material domain such that  the diameter of the triangles tends uniformly to zero.
In the continuum limit, the compatibility conditions for (ND) tend to Christoffel-Riemann compatibility conditions of (NC), namely, that the Gauss curvature of $\zeta$ vanishes (see Section~\ref{ss_NDapproxNC}). 

Being a function of lengths, the expression for energy is frame independent, that is invariant under translation and rotation of coordinates. By a
minimization of $W(\mathcal S)$ plus the work of external forces with respect to $\ell_{ij}$ that satisfy the compatibility equations, one finds the equilibrium configuration expressed in terms of optimal lengths of edges.  The energy and the equilibrium equations can be represented in a coordinate-free form which can be convenient for calculating the state of structures undergoing large deformation.

In the linearized theory (see Section~\ref{ss_LinDesc}), the energy is quadratic in the elongations $\lambda_{ij}$ of the edges and the compatibility conditions (wagon wheel conditions) become linear. The elastic energy is
$$
W_{LD}(\lambda)=\frac 12 \sum_{i<j} c_{ij} \, \lambda_{ij}{}^2, \qquad\text{subject to $ B\lambda=0$,}
$$
where $B$ is the matrix of compatibility coefficients.

\subsection{Small Deformations and Strain}\label{ss_sdas}

If the strain is small we have $\phi(x)=x+u(x)$ so that $\nabla\phi=I+\nabla u$. The $2\times 2$ symmetric matrix $\epsilon$ of strain is defined to be the linearization of the symmetric square root  $\sqrt{\zeta}-I$ near the identity
$$
\sqrt \zeta= [(I+\nabla u)^T(I+\nabla u)]^{\frac 12} = I + \frac 12[\nabla u + (\nabla u)^T] + \mathbf o(|\nabla u|)
$$
or by the strain tensor
$$
\epsilon = \binom {\epsilon_{11}\quad\epsilon_{12}}{ \epsilon_{21}\quad\epsilon_{22}} = \frac 12[\nabla u + (\nabla u)^T] 
$$
where
$$
\epsilon_{11}=\frac{\partial u_1}{\partial x_1},\qquad
\epsilon_{12}=\epsilon_{21}=\frac 12\left[\frac{\partial u_1}{\partial x_2}+\frac{\partial u_2}{\partial x_1}\right],\qquad
\epsilon_{22}=\frac{\partial u_2}{\partial x_2}.
$$
Again, since the three entries $\epsilon_{ij}$ are determined by two $u_1$, $u_2$, they satisfy a compatibility condition~\eqref{eq_deriveCC}
$$
\operatorname{Ink}(\epsilon)=\frac{\partial^2 \epsilon_{11}}{\partial x_2{}^2}-2\frac{\partial^2 \epsilon_{12}}{\partial x_1\, \partial x_2}+\frac{\partial^2 \epsilon_{22}}{\partial x_1{}^2}=\nabla\cdot \eta^T\epsilon \eta=0
$$
which holds at all points of $\mathcal B$, where $\eta$ is the rotation given by~\eqref{eq_RotM}.

\subsection{Existence of deformations} 
The deformation of a continuous material is a vector field of deflections that is described by a symmetric right Cauchy Green tensor. Being planar means that the Green tensor is subject to pointwise differential constraints, the compatibility conditions.
The linearization of the problem to find configurations with prescribed Right Green-Cauchy tensor  (NC) is the problem of prescribed strains (LC).
 For the discrete approximating structures (trusses) the nonlinear problem is to find a configuration in the plane whose links have prescribed length (ND). Its linearization is the prescribed elongations problem (LD). Both are overdetermined and also require compatibility conditions for their solution. 
 The compatibility conditions restrict how the edges can be deformed. The same compatibility conditions are valid for the deformed and undeformed configurations; therefore they represent constraints for any deformed structure. When the deformations are small, we arrive at linearized compatibility constraints.

If the corresponding compatibility conditions hold, then all four problems may be solved for the deformation, at least locally. The solvability of (ND) under the condition that curvature atoms vanish at interior vertices is a simple case of Alexandrov's theory of polyhedral approximation and is discussed in Section~\ref{s_ND}, \ref{ss_NCapprox}. That the vanishing of the Riemannian curvature is sufficient that (NC) can be solved locally is classical~\cite{L:1926}. The local solutions are piecewise Euclidean. Thus the global solubility of (NC) follows from the global solubility of (ND) in Section~\ref{ss_solnNC}. The global solubility of (LC) may be obtained similarly to (NC)~\cite{So:1956}.  (LD) problem is an overdetermined matrix equation which is soluble if the compatibility equations hold in Section~\ref{ss_LinDesc}.

\subsection{Compatibility conditions with local support and material points} 
\begin{figure}[h]
      \begin{center}
          \scalebox{0.5}{\includegraphics{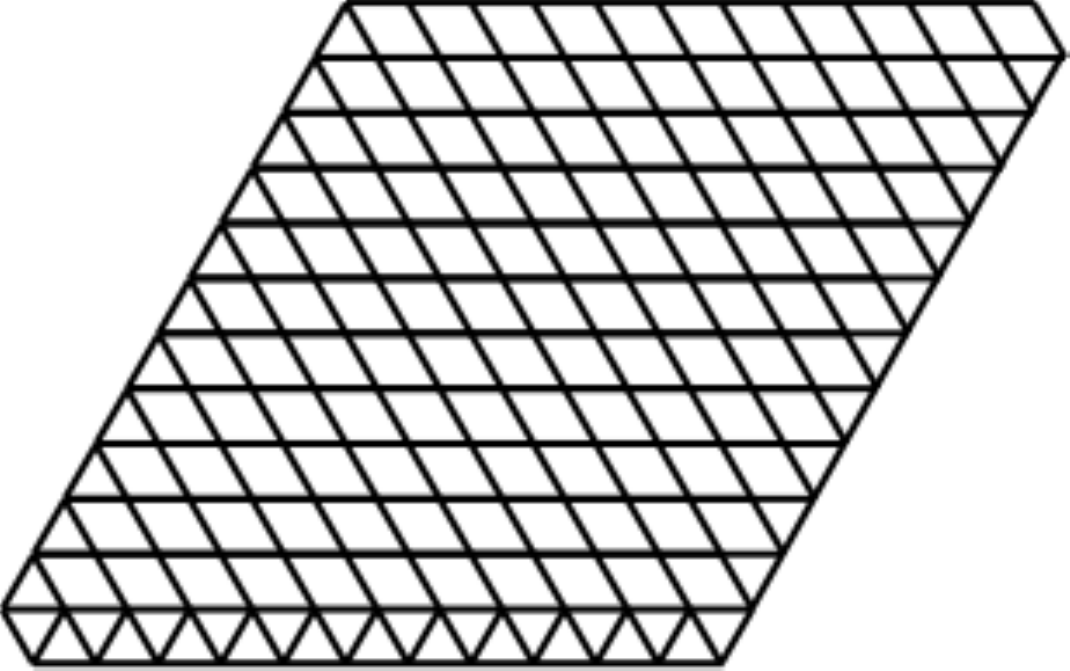}}
 \caption{Rhombus $\mathcal P_n$ with maximally many links removed maintaining infinitesimal rigidity.}  \label{fig_maximalremoved}
 \end{center}
       \end{figure}

  The compatibility conditions of a general structure may be nonlocal, as in Figure~(\ref{fig_BTP}c), where $Q$ may be far separated from rest of the truss. However, only specific structures approximate a material.  For a material, compatibility conditions express how the deformations of a material in the neighborhood of a point influence the deformation at the point; thus they are local in nature.  The support of compatibility conditions in a sequence of approximating structures must tend to material points of the material.
  For example, the sequence of trusses as in Figure~\ref{fig_maximalremoved} with just a single northeast link added in the northeast corner will create a single non-local compatibility condition whose support does not converge to a material point.


(LD) is obtained by linearizing (ND) rather than discretizing (LC); nevertheless,
 discrete and continuous compatibility conditions are consistent. The solution of the discrete problem (ND) approximates the solution of (NC) (Section~\ref{ss_NCapprox}). We show that the continuum limit of vanishing of curvature compatibility constraints in a triangulated structure gives the continuum vanishing of curvature compatibility condition of (NC) (Section~\ref{ss_NDapproxNC}). 
  The continuum limit of linearized compatibility constraints,  the wagon wheel condition at interior vertices (section~\ref{ss_CCforLD}), is the compatibility condition of strains, $\operatorname{Ink}(\epsilon)=0$.

\subsection{Compatibility conditions  (Wagon wheel conditions)  in the discrete linear problem (LD)}
The number of compatibility conditions is the codimension of the range of the prescribed elongations problem (LD). We call a structure {\it generic} if this dimension is given by the Maxwell Dimension, $\CCM$ \eqref{eq_CCm}, which is merely the excess in the number of equations over the number of variables (Section~\ref{ss_maxDimDef}).  For a triangulated structure without holes, $\CCM$ is given by the number of interior nodes~\eqref{eq_CCmholes}. Equivalently, a structure is generic if it is infinitesimally rigid. For triangulated structures which are generic, there is an independent compatibility equation supported in the star-neighborhood of every interior vertex, the wagon wheel condition, which is explained in Section~\ref{ss_CCforLD}. In this paper, we explore the meaning of $\CCM$ for generic triangulated structures.

\begin{itemize}
\item  
 In a triangulated structure, the number of compatibility conditions is equal to the number of edges that can be removed keeping the structure rigid. The presence of additional edges increases the resilience of the structure and helps to maintain its structural integrity when damaged. $\CCM$ provides a quantitive measure for resilience and use it to analyze the damaged by a crack, multiple faults, etc. (Section~\ref{ss_AC}). 
 
\item $\CCM$ is the sharp upper bound on how many links may be removed from a truss before it loses rigidity.
\item The minimal number of links that can be removed from a structure to make it flexible may be considerably smaller than the Maxwell Dimension $\CCM$. For example, in a triangular truss, this number is two. At any convex corner which is connected to the rest of the truss by two or three links, removing one or two will free up the remaining attached link.

\item  In a triangulated structure, $\CCM$ for (LD) may be easily computed from the topology of a structure which facilitates analysis of damaged structures (as in  Section~\ref{ss_AC}.)

\end{itemize}

\subsection{Compatibility as a measure of resilience  of a periodic structure} 
In the periodic triangular structures in which we consider repeated fixed period cells with holes, we define the asymptotic compatibility condition as the homogenized limit of the number of compatibility conditions divided by the area as the number of period cells tends to infinity~\eqref{eq_defAC}. This number depends on both the size of holes in the period cell as well as the geometry of the hole.  A periodic material with a single long hole in the period cell is weaker than if the same sized hole were round, and this is weaker than many small holes of the same area.

\subsection{Integral form of compatibility conditions} 
There is an integral compatibility condition for all simple closed curves in the domain.
In the linearized problem (LD) for triangulated structures, when we sum up the  compatibility condition elements localized at nodes interior to a closed curve, we obtain a relation between the elongations supported on a strip along the curve. Lemma~\ref{lem_vanish} shows that the sum of all localized compatibility conditions of (LD) cancel on vertices interior to a closed loop. Theorem~\ref{th_bounlay} shows that the sum of the interior compatibility conditions gives a boundary compatibility condition that is supported on the double layer, the edges entirely within one link of the loop. The similar boundary compatibility condition for (ND) is just that for a connected domain, the total turning angle going around a loop, which can be computed from edge lengths of the double layer, is just $2\pi$.

The compatibility condition for (LC) is expressible as the vanishing of an exact two-form and may be integrated analogously to integrating a gradient field over the boundary curve of a subdomain. In Section~\ref{ss_doublay}, applying Stokes's Theorem shows that there is a boundary compatibility condition that is also a double layer: it depends on both the boundary curvature and the normal derivative of the prescribed strain. The similar condition for (NC) is the Gauss-Bonnet Theorem for connected domains: the integral of the curvature of the outer boundary as a plane curve plus the angle excesses at the vertices equals $2\pi$~\eqref{eq_GaussBonnet}. The expressions for these quantities in the same local curvilinear coordinates is presented in Section~\ref{ss_NCdoublay}.

\subsection{The genericity of structures}
A structure is generic if it is infinitesimally rigid. 
The computation of the number of compatibility conditions is simple for generic triangulated structures, but which structures are generic?
For triangular structures, genericity is proved by showing that the compatibility conditions given by the wagon wheel conditions supported in the neighborhoods of all interior points plus the conditions coming from ring-girders around the holes form a basis for all compatibility conditions (Theorem~\ref{th_multiple}). 
The genericity of Bigon-Triangle-Prism Structures (BTP Structures) of Section~\ref{ss_defBTP}, a large family of structures including triangulated structures,   is proved in Section~\ref{ss_BTP}. Such structures are built out of smaller infinitesimally rigid structures, and their $\CCM$ can be computed from its pieces.  

A structure whose geometry is regular may be far weaker than one whose edges take lots of directions. For example, if the structure consists of two rigid pieces that are connected by $n$ links from one piece to the other and these connectors are parallel then the structure is not rigid at all. However, roughly speaking,  if all the connecting links have pairwise independent directions, then one must remove $n-2$ of them before the structure loses its rigidity along the seam.  Thus it has $n-2$ additional compatibility conditions. This is the prism construction, proved in Sections~\ref{ss_defBTP} and~\ref{ss_BTP}.

\subsection{Previous Results}

Compatibility conditions are routinely used in calculation of the stresses of loaded frames. The elongations of several rods that end at a node are determined by the deflection of that node and compatibility conditions hold if several nodes are interconnected.  These conditions are commonly expressed through the deflections where they play an auxiliary role in the determination of the stress state of the structure. Here we study geometrical aspects of compatibility conditions in complex networks.

Network structures have been studied by mechanical, material, and physical scientists as well as by mathematicians. In the nineteenth century, Maxwell found conditions under which mechanical structures made out of bars joined together at their ends would be stable~\cite{M:1864}. He used the method of constraint counting to estimate the dimension (Maxwell dimension) of infinitesimal deformations for generic structures. Recently, Maxwell's ideas were revived by Thorpe and collaborators in studies of network glasses. Jacobs and Hendrickson~\cite{JH:1997} developed an algorithm, the pebble game, to compute this dimension exactly. They applied this algorithm to study percolation of rigidity, the transition of a floppy structure to a rigid one (see the overview in \cite{TJCR:1999}). These studies count the nullity in the case when the system~\eqref{eq_preslam2} for (LD) is underdetermined. In the present paper, we are concerned with the opposite problem for rigid structures and study the degree to which the structure is overdetermined.

Using network models to approximate materials is fairly common~\cite{BKN:2013} although their use to
approximate Green Tensor equations is rare. The spring network model of Hrennikoff~\cite{Hr:1941} is pioneering both approximating elasticity as well as using finite elements.
The modern finite element formulation is based on lattice models, see for example~\cite{Bd:2007},~\cite{BS:2010}.
However, the problem of compatibility of the links is usually avoided by the description the deformation through the positions of nodes, so the compatibility is automatically satisfied.

Network models were used to study the resilience of lattices. 
These models describe the transition of the network when damaged elements are replaced by initially inactive ``waiting links'' resulting in waves of damage, see Cherkaev and Zhornitskaya~\cite{CZ:2003},  Cherkaev, Vinogradov and Leelavanichkul~\cite{CVL:2006} and Cherkaev and Leelavanichkul~\cite{CL:2012}. During the transition, compatibility varies. The compatibility and self-deformation of various triangular lattices with two kinds of fixed length links were studied by
Cherkaev, Kouznetsov, and Panchenko where the local compatibility conditions \cite{CKP:2010} were derived. The study of compatibility conditions for node and bar structures was initiated by Krtolica~\cite{K:2016} who proved that the continuum limit of the discrete compatibility condition is the continuous one (Theorem~\ref{th_pedja}). 

The continuum limit of a triangulated structure is a planar solid; its deformation obeys the (NC) continuum compatibility condition $\mathcal K=0$ (see Section~\ref{ss_NPGDCTC}). We show that the discrete compatibility condition of (ND) tends to this condition. Similarly, we show the (LD) compatibility condition tends to the one for (LD).
Our interpretation of the network approximation (ND) of the prescribed Green's tensor problem (NC) as a polyhedral metric approximating a Riemannian surface metric is based on A.~D. Alexandrov, {\it e.g.,}~\cite{AZ:1962}.
An alternative homogenization procedure  of discrete models uses $\Gamma$-convergence (see books by Braides~\cite{Br:2006},~\cite{Br:2014}). For example, it is applied to spring networks~\cite{AS:2017} and beam networks~\cite{SAdI:2011} by Seppecher and co-authors and to triangulated planar bar lattices by Raoult~\cite{R:2010}.

\section{Nonlinear Discrete Prescribed Length Problem (ND).}\label{s_ND}

Let $v$ be the number of vertices and $e$ the number of edges in a general structure.
In addition to this combinatorial data, we associate a length $\ell_{ij}$ to each edge. If the structure is concretely realized as points and segments of the plane, each edge has a positive length $\ell_{ij}$ induced from the Euclidean metric. 
\begin{equation}\label{eq_prescell}
\text{(ND)}\qquad\qquad\qquad\qquad\qquad
|V_i-V_j|=\ell_{ij}\qquad \text{for all $E_{ij}\in\mathcal E$.}\qquad\qquad\qquad\qquad\qquad
\end{equation}
The realization problem asks to find the positions of the vertices in the Euclidean plane if the lengths are prescribed for given combinatorial data. The lengths are assumed to satisfy the {\it triangle inequality}: whenever $V_i$, $V_j$ and $V_k$ are the  vertices of a triangle, then 
\begin{equation}\label{eq_triineq}
\ell_{ik}\le\ell_{ij}+\ell_{jk}.
\end{equation}
Equality in the triangle inequality corresponds to a degenerate triangle with three collinear points.  The realization of a truss may have multiple vertices at the same position in the plane, multiple edges connecting a pair of vertices and degenerate triangles. We consider flexing a truss in the plane in such a way as to preserve the lengths. The abstract truss may not be really constructible as a linkage that flexes because the links may have to pass through themselves in order to do so.

To an abstract truss which is combinatorially a triangulated surface $\mathcal B$, we can associate a piecewise linear surface. Triangles with prescribed edge lengths $\ell_{ij}$  may be filled in by triangular pieces of the Euclidean plane with the given edge lengths. The metrics from the triangular pieces glue together to form a piecewise-linear (PL) metric on the filled in truss $g_{\mathcal B}$.  The existence problem asks whether an abstract triangulation of a disk with a given metric comes from a truss in the Euclidean plane. We say that a PL immersion $x:\mathcal B\to\mathbf E^2$ is a {\it realizable configuration} if the metric induced on $\mathcal B$ by the Euclidean structure $g_{\mathcal B}$ is the pull-back of the  Euclidean metric
\begin{equation}\label{eq_pullba}
g_{\mathcal B}(x) = x^*(ds^2_{\mathbf E^2}(X(x)).
\end{equation}
In other words, it is possible to map the filled in triangulated $\mathcal B$ to the plane in such a way that the lengths measured in the Euclidean plane correspond to filled in lengths so that the image of edges has the Euclidean length given by $\ell_{ij}$?

\subsection{Solving (ND) for Triangulated Structures}
Suppose that the truss is an abstract triangulated planar domain, namely, it is the one skeleton (vertices and edges) of a triangulated domain, embedded in the plane and bounded by $g+1$  closed curves ($g$ is the {\it genus}). For simplicity, the curves are required to be disjoint and simple. Simple means that the curves have no self intersections, so the domain has no pinch points. Let $f$ be the number of triangular faces.  The Euler Characteristic (e.g., \cite[p. 378]{O:1966}) for a triangulated disk  is given by the formula
$$
\chi =1-g= f-e+v=1,
$$
where $v$ is the number of vertices, $e$ the number of edges and $f$ the number of faces (triangles).
If $v_b$ and $v_i$ denote the number of interior and boundary nodes, and $e_b$ and $e_i$ the number of boundary and interior edges, we have for  
\begin{equation}\label{eq_simpcond}
e=e_b+e_i,\qquad v=v_b+v_i,\qquad e_b=v_b,\qquad 3f=e_b+2e_i
\end{equation}
Substituting Euler's formula it follows that
\begin{equation}\label{eq_chi}
3\chi = e_b-e_i+3v_i.
\end{equation}

 The {\it Curvature atom} $K(V_i)$ at a node, which is the discrete analog of Gaussian Curvature, is defined to be the angle excess~\cite[p. 8]{AZ:1962}. If $V_i$ is an interior vertex and $V_{i_1},\ldots,V_{i_k}$ are its adjacent vertices taken cyclicly, then
\begin{equation}\label{eq_Katom}
K(V_i)=\sum_{j=1}^k \angle(V_{i_j},V_i,V_{i_{j+1}}) - 2\pi
\end{equation}
where $V_{i_{k+1}}=V_{i_1}$,  $k$ is the valence (number of links) at the vertex $V_i$, and $\angle(A,B,C)$ is the Euclidean angle included between the vectors $BA$ and $BC$. In the piecewise Euclidean metric $g_{\mathcal B}$, this is determined from the side lengths by the cosine law
\begin{equation}\label{eq_coslaw}
\angle(ABC)=\cos^{-1}\left( \frac{\ell_{AB}^2+\ell_{BC}^2-\ell_{AC}^2}{2\ell_{AB}\ell_{BC}}\right).
\end{equation}
A necessary and sufficient condition that the nonlinear realization problem is solvable is that the curvature atom vanishes at each interior vertex.

\medskip
\begin{thm} \label{th_realize} A necessary and sufficient condition that the edge lengths of a combinatorial triangulated disk may be nondegeneratevely developed (mapped jigsaw puzzle wise triangle by triangle) in the Euclidean plane is that the strict triangle inequality
$$
\ell_{ik}<\ell_{ij}+\ell_{jk}
$$
 hold (compare \eqref{eq_triineq}) for all triangles and that the curvature atoms vanish
\begin{equation}\label{eq_0KAtom}
K(V_i)=0,\qquad\text{for all interior vertices $V_i$}
\end{equation}
The development is unique up to reflection or rigid motion.
\end{thm}
This theorem is proved in the Appendix, Section~\ref{ss_threal}.

\subsection{Polynomial Expression for the Compatibility Conditions for (ND)}\label{ss_Kpoly}

In this section, we show that the compatibility conditions for the prescribed lengths problem (ND) of a triangulated truss may be expressed as polynomial equations in the lengths of edges. 
We use the standard procedure for reducing systems of equations involving radicals to polynomial systems.
\begin{thm} Let $V_i$ be an interior node of a triangulated truss. For the prescribed lengths problem~(ND), the compatibility condition that the vanishing of curvature~\eqref{eq_0KAtom}  may be expressed as  a polynomial equation in the lengths of the edges of the triangles that  meet $V_i$.
\end{thm}
\begin{proof} 
 If $V_i$ is an interior vertex and $V_{i_1},\ldots,V_{i_k}$ are its adjacent vertices taken cyclicly, let $\phi_i= \angle(V_{i_j},V_i,V_{i_{j+1}})$. Since $V_i$ is interior  point of a planar triangulation, $0\le \phi_i\le \pi$. Then by \eqref{eq_Katom}, the condition $K(V_i)=0$ is equivalent to
\begin{equation}\label{eq_1iscos}
1=\cos\left(\sum_{i=1}^k\phi_k\right).
\end{equation}
Expanding using the standard addition formulas, this gives a homogeneous trigonometric polynomial in sines and cosines of order $k$. The cosines $\cos(\phi_i)$ are  rational, homogeneous of degree zero functions of the lengths by the cosine law~\eqref{eq_coslaw}.  The sines satisfy $\sin\phi_i=\sqrt{1-\cos^2\phi_i}$. Multiplying through by the denominators, the condition may be written
\begin{equation}\label{eq_Qrad}
Z= P_{10} + \sum_{j=1}^N P_{1j}\sqrt{Q_{1j}}=0
\end{equation}
where  $P_{1j}$ and $Q_{1i}$ are homogeneous polynomials in the lengths  and where  there are  $N=2^{k-1}$ terms. To eliminate the irrational terms, raise $Z$ to powers $2,3,\ldots 2^{k-1}$.  Note that even powers of $\sqrt{Q_{1j}}$ are polynomial and odd powers are a product of a polynomial and 
$\sqrt{Q_{1j}}$. Counting the number of types of products of radicals in the multinomial expansions, for each type there are
$$
\sqrt{Q_i}\quad\text{$k$ terms},\qquad
\sqrt{Q_iQ_j}, i\ne j,\quad\text{$\binom k2$ terms},\qquad
\cdots\qquad, \sqrt{\prod_{i=1}^kQ_k},\quad\text{$1$ term.}
$$
All together, there are $2^k$ different types. Writing these types $\hat Q_j$ we get
\begin{equation}\label{eq_sysQtypes}
Z^n= P_{n0} + \sum_{j=1}^{2^k-1} P_{nj}\sqrt{\hat Q_{j}}=0,\qquad n=1,\ldots,2^k-1.
\end{equation}
which is a linear system for the $\sqrt{\hat Q_{j}}$'s. Solving the system gives rational expressions for $\sqrt{ Q_{j}}$ which may be substituted into \eqref{eq_Qrad}. Multiplying through by the denominator we arrive at the desired polynomial equation.
\end{proof}

\paragraph{Triangle example of polynomial equivalent of $K(V_i)=0$.} There are short cuts for the three sided star about the interior vertex $V$ surrounded by $V_1$, $V_2$ and $V_3$ (Figure~\ref{fig_innerpoint}b.). The sum of the angles $\phi_1+\phi_2+\phi_3=2\pi$ where $\phi_i=\angle V_iVV_{i+1}$ so that
$$
\cos(2\pi-\phi_3)=\cos(\phi_1+\phi_2).
$$
We obtain
$$
\cos\phi_3=\cos\phi_1\cos\phi_2-\sin\phi_1\sin\phi_2.
$$
Squaring and excluding sine terms we come to the relation
$$
\cos^2\phi_1+\cos^2\phi_2+\cos^2\phi_3-2\cos\phi_1\cos\phi_2\cos\phi_3=1.  
$$
By substituting squares of radial lengths $p_i=|V_i-V|^2$ and circumferential lengths $q_{ij}=q_{ji}=|V_{i}-V_j|^2$, this transforms to the sixth order polynomial equation in the lengths
\begin{align*}
0&=\sum_i\Bigl\{p_i^2q_{jk}+p_iq_{jk}^2 +p_jp_kp_{jk}^2\Bigr\}
-\sum_i\sum_{j\ne i}\Bigl\{ p_ip_jq_{jk}+p_iq_{ij}q_{jk}\Bigr\}
+q_{12}q_{23}q_{31}
\end{align*}
where $i\ne j\ne k\ne i$ so in the first sum $j=i+1\!\!\mod 3$, $k=i+2\!\!\mod 3$  and in the second sum $k(i,j)=6-i-j\!\!\mod 3$ is the   index other than $i$ and $j$.
This expression is invariant under the permutation of $V_i$'s.

\paragraph{Hexagon polynomial version of $K(V_i)=0$.}
For the central point of a hexagonal star, the condition~\eqref{eq_1iscos} becomes
the sum of products of six sines or cosines. One can check that the sines occur in even powers. Fifteen terms have two sines, and four cosines, fifteen terms with four sines and two cosines, one term has all cosines and one term all sines; $32$ terms in all. 
The expression \eqref{eq_Qrad} is the homogeneous function of degree six in the squares of the lengths of the edges. Since there are only $32$ different types of terms, the elimination procedure above requires raising this expression to powers $2,3,\ldots 31$.  The solution of the resulting system~\eqref{eq_sysQtypes}   is substituted into~\eqref{eq_1iscos} to get a rational expression of the compatibility condition which is rational in the lengths of the twelve sides of the hexagon. 

For a rough estimate of the degree of the polynomial, note that the highest degree of the coefficients of the equation~\eqref{eq_sysQtypes} is $6n$ in the squares of the lengths. Thus the degree in the squares of the lengths of the polynomial compatibility condition is bounded by the degree of the coefficient times the degree of the determinant of system
$$
\text{degree in squares of the lengths}\le 6\times\left(6+6\cdot 2+\cdots + 6\cdot{31} \right)=17,856.
$$


\section{The Linearized Discrete Problem of Prescribed Elongations (LD).}\label{ss_LinDesc}

The linearization of the discrete nonlinear realization problem (ND) is to prescribe the infinitesimal elongations along edges and solve for the infinitesimal displacements.


\subsection{(LD) equations and the number of compatibility conditions for a generic truss.}
The infinitesimal deformations may be regarded as velocities of the nodes.
 To derive the equations, suppose that we have a time-dependent immersion of an abstract truss $x(X,t):\mathcal B\times(-\delta,\delta) \to\mathbf E^2$ where $x(X,0)$ is the reference configuration. Differentiating \eqref{eq_prescell} with respect to time, for all edges $V_iV_j$
\begin{equation}\label{eq_preslam}
(V_i-V_j)\bullet (U_i-U_j)=\ell_{ij}(0) L_{ij}=\lambda_{ij}
\end{equation}
where
$$
\left.\frac d{dt}\right|_{t=0}V_i(t)=U_i;\qquad
\left.\frac d{dt}\right|_{t=0}\ell_{ij}(t)=L_{ij}.
$$
Stacking the displacements we get the unknown vector $U$ of dimension $2v$. The linear system \eqref{eq_preslam} may be written 
\begin{equation}\label{eq_preslam2}
\text{(ND)}\qquad\qquad \qquad\qquad\qquad\qquad\qquad AU=\Lambda \qquad\qquad\qquad\qquad\qquad \qquad\qquad\qquad
\end{equation}
where $A$ is a $v \times 2e$ dimensional matrix and $\Lambda$ is $e\times 1$ column matrix.

The nullspace of $A$, the velocities that don't change lengths up to first order are called the  \emph{infinitesimal deformations,}  which have dimension $n$, the nullity of $A$.  In the Euclidean plane, velocities of rigid motions are three dimensional, consisting of two translation dimensions and one rotation dimension. Velocity fields of rigid motions are always infinitesimal deformations and are, therefore, in the null space. Hence $n\ge 3$.  $n$ degrees of freedom of motions must be pinned down by boundary conditions to determine the displacements uniquely.

\begin{figure}[h]
      \begin{center}
          \scalebox{0.5}{\includegraphics{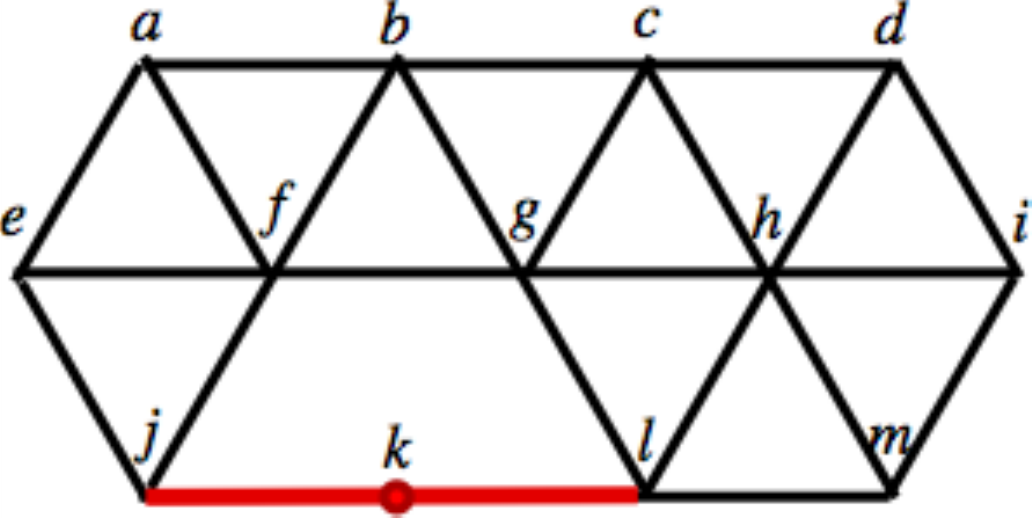}}
 \caption{A vertical flex occurs at $V_k$ since  edges $E_{jk}$ and $E_{kl}$ are parallel.}\label{fig_nongeneric}
 \end{center}
       \end{figure}

A truss is called {\it infinitesimally rigid} if the infinitesimal deformations are three dimensional: they consist exactly of velocity fields of rigid motions. An infinitesimal deformation which does not come from rigid motion is called a {\it nontrivial flex.} 
Usually the system \eqref{eq_preslam2} is over-determined. If the truss is infinitesimally rigid with the fewest possible edges, that is if $A$ has full rank and $2v-3=e$, then we say the structure is {\it statically determined.} 
 An example of a rigid truss that is not infinitesimally rigid is given in Figure~\ref{fig_nongeneric}. In such infinitesimally flexible trusses, there is equality in the triangle inequality for the nodal distances at the flexing vertex.

Not all $\Lambda\in\mathbf R^e$ may be realized as elongations of edges when a truss undergoes deformation. The codimension of the range of $A$ is called the number of compatibility conditions $c$. In fact, we may find a $c\times e$ matrix $B$ of independent rows, called the {\it compatibility conditions}, such that 
$$
\operatorname{range} A = \ker B.
$$

Since all the geometry is encoded in $A$, it is natural that the matrix $B$ may be computed from $A$. 

\begin{lemma} For any (not necessarily triangulated) truss in the plane, the compatibility matrix may be computed as the first $e$ columns of
$$
\tilde B = \tilde I -\tilde A(\tilde A^T\tilde A)^{-1}\tilde A^T,    
$$
where $\tilde A$ is the matrix $A$ augmented by a rank three $3\times 2v$ matrix annihilating the rigid motions~\eqref{eq_augA}.
\end{lemma}
\begin{proof}
Let $A_0$ be any rank three $3\times 2v$  matrix that annihilates  the translations and rotation and $B_0$ be a $CC \times 3$ matrix. If we augment the matrix, the right side and compatibility,
\begin{equation}\label{eq_augA}
\tilde A=\binom{A}{A_0}, \qquad
\tilde \Lambda=\binom{\Lambda}{0},\qquad \tilde B=(B,B_0)
\end{equation}
then the kernel $\ker\tilde A$ vanishes if and only if  the truss is infinitesimally rigid. In that case, 
\begin{equation}\label{eq_tildeAU}
\tilde A U =\tilde\Lambda
\end{equation}
 so
\begin{equation}\label{eq_solLD}
U=(\tilde A^T\tilde A)^{-1}\tilde A^T\tilde\Lambda.
\end{equation}
Excluding it from \eqref{eq_tildeAU} we formulate the compatibility condition
$$
\tilde B \tilde \Lambda=0,\qquad \tilde B = \tilde I -\tilde A(\tilde A^T\tilde A)^{-1}\tilde A^T.
$$
$B$ is then just the first $e$ columns of $\tilde B$.\end{proof}

\subsection{The Maxwell Number of compatibility conditions in a generic truss.}\label{ss_maxDimDef}

In this section, we assume that the trusses come from triangulated planar domains. 
Using $e_b=v_b$ and \eqref{eq_chi}, the quantity 
\begin{equation}\label{eq_Eu}
3+e-2v=3+e_i-2v_i-v_b=3-3\chi +v_i=3g+v_i\ge 0.
\end{equation}
Since it is nonnegative,   we can estimate of the number of compatibility conditions as the  number of variables minus the number of equations plus the dimension of rigid motions
\begin{equation}\label{eq_CCm}
\CCM=e-2v+3,
\end{equation}
the {\it Maxwell Dimension}  observed by James Clerk Maxwell~\cite{M:1864}. Thus, a simply connected truss which is statically determined has $\CCM=0$. The number of compatibility conditions is the maximal number of edges that may be removed from the truss while keeping it infinitesimally rigid.
In terms of the dimension of infinitesimal deformations (nullity of $A$) we have  $\operatorname{rank}(A)=2v-n$ so
$$
c=e-2v+n\ge \CCM.
$$
We shall call the truss {\it generic} if its number of compatibility conditions is the Maxwell number. A truss is generic if and only if it is infinitesimally rigid. We will characterize the number of compatibility conditions for a large family of trusses in Theorem~\ref{th_multiple} geometrically.

For a triangulated disk, the genus $g=0$. By \eqref{eq_Eu},
\begin{equation}\label{eq_CCmholes}
c\ge \CCM=e-2v+3=v_i,
\end{equation}
thus the Maxwell Dimension equals the number of interior vertices.

\paragraph{An example of a non-generic truss.} In the infinitesimally flexible
 unit-length triangular truss illustrated in Fig.~\ref{fig_nongeneric}, there are $v=13$ vertices, $e=24$ edges so Maxwell dimension $\CCM=e-2v+3=24-26+3=1$. However, The kernel admits an nontrivial infinitesimal flex at vertex $k$ in the vertical direction so $n=4$ and $c=2$. Adding a link $E_{fk}$ or $E_{gk}$ would make the truss infinitesimally rigid.

\subsection{Recoverable and Unrecoverable damage.}

Given a triangulated truss, we may suppose that some subset of the links, that we designate as
 \begin{figure}[h]
      \begin{center}
          \scalebox{0.625}{\includegraphics{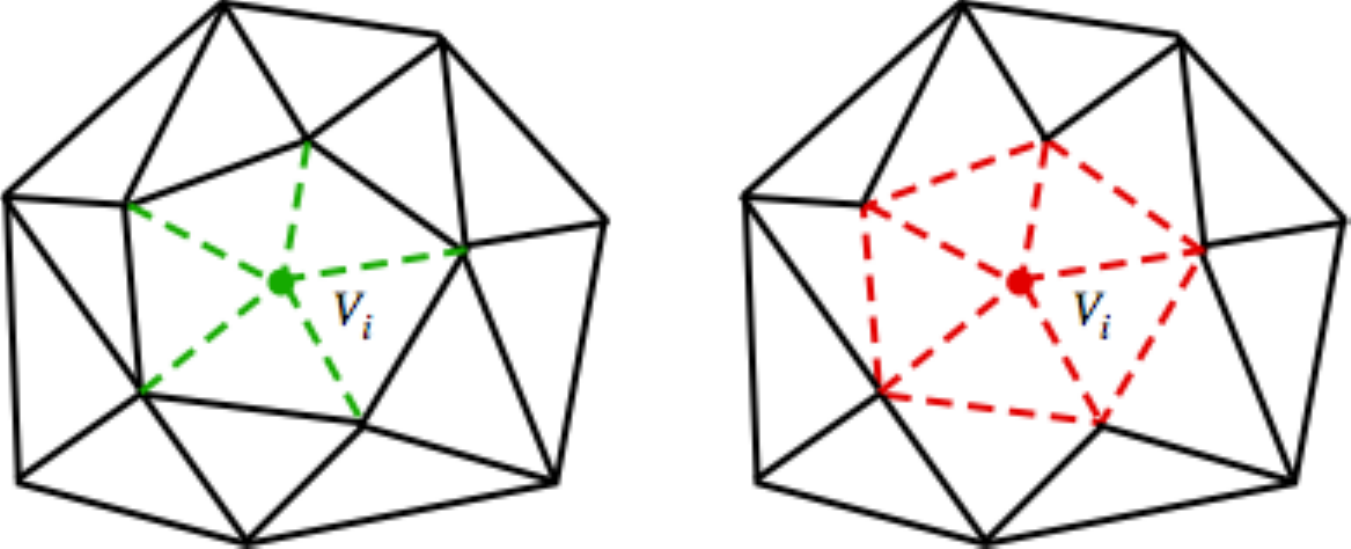}}
 \caption{Recoverable and Unrecoverable Damage}\label{fig_damned}
 \end{center}
       \end{figure}
 being damaged, are removed from the truss. If the resulting remaining subtruss is still rigid, we call this {\it recoverable damage.} In other words, the deformations at the undamaged nodes are still determined. If on the other hand, if this subtruss is flexible, then the damage is {\it not recoverable.}  Thus any rigid truss may be regarded as a triangulated disk with recoverable damage.

 In Figure~\ref{fig_damned}, removing more of the middle of the truss (dashed lines) results in a flexible truss. As another simple example, suppose one of the boundary vertices is a convex corner connected to the rest of the truss by two edges.  If one of the edges is damaged, then this corner is free to flex. The remaining free edge cannot be stretched, it carries no energy and thus is invisible from an elastic model.
 If the damage is interior to a truss (Figure~\ref{fig_damned}b), it may only mean that some inner vertices are not determined from the others, and from the outside, the subtruss looks rigid.  The deformation at the inner vertex $V_i$ cannot be determined. Still, the damaged triangulated truss may be viewed as a general truss. Moreover, any rigid triangulated truss may be viewed as a triangulated disk with recoverable damage.

In the linear theory, we may suppose that we remove $k$ links from the truss. This is equivalent to removing $k$ rows from $A$. Let $A_0$ be a $3\times 2v$ dimensional matrix that annihilates the rigid motions. 
$$
\tilde A =\begin{pmatrix} A'\\ A''\\A_0\end{pmatrix}
$$
 where $A''$ is a $k\times 2v$ dimensional matrix corresponding to the removed edges and $A'$ is the $(e-k)\times 2v$ dimensional matrix corresponding to the remaining links. If the damage is recoverable, the null space of $A'$ consists only of rigid motions. Put
 $$
 \hat A = \binom{A'}{A_0},\qquad \hat \Lambda = \binom{\Lambda'}{0}
 $$
 If the damage is recoverable, $\hat A$ has full rank and $\hat A U = \hat\Lambda$ may be solved as in~\eqref{eq_solLD}. Hence the elongations of the removed links may be expressed as
 $$
 \Lambda''=A''(\hat A^T\hat A)^{-1}\hat A^T\hat\Lambda
 $$
 and the compatibility condition for the reduced system are
 $$
\hat B \hat \Lambda=0,\qquad \hat B = I- \hat A(\hat A^T\hat A)^{-1}\hat A^T.
 $$
 $B'$, the compatibility matrix is just the first $e-k$ columns of $\hat B$.  Thus any truss mat be viewed as a damaged triangulated truss.

 For a simply connected triangular truss, it is shown in Theorem~\ref{th_simple} that a basis for the compatibility conditions is given by the wagon wheels centered on interior vertices. The support of a wagon wheel condition is localized to the triangles immediately neighboring the vertex. In case the truss suffers recoverable damage,  the compatibility conditions are different.   For example, if several links are removed from the middle of the truss making a hole, then several compatibility conditions have to be combined to account for the hole.

\subsection{Relation to Elasticity Theory}

 We focus on the geometry of the strain of a material. However, in the constitutive theory,  deformations are useful because the elastic energy in a link is often modeled as a function of the deformation. 
 The infinitesimal displacements are related to elongations via $AU=\Lambda$.
In the linear case, let
$$
C=\operatorname{diag}(c_{11},c_{22},\ldots)
$$
 be the $e\times e$ elasticity matrix of positive spring constants for the edges. By Hooke's Law, the forces along the edges $C\Lambda$ are proportional to the elongations. Then $A^TC\Lambda$ are forces at the vertices, and $K=A^TCA$ is the stiffness matrix which is nonnegative definite with rank $2v-n$.
If $F$ is the $2v\times 1$ vector of forces applied at the vertices, then force balance is $A^TC\Lambda=F$.

 The equation for balanced forces may be solved for elongations or displacements.
 \begin{align*}
A^TC\Lambda&=F;&A^TC\Lambda&=F\\
B\Lambda&=0.& \Lambda&=Au
\end{align*}

The analogous equations for linearized elastostatics are
in terms of strains $\epsilon$ or infinitesimal displacements $u$, defined in Section~\ref{ss_sdas} or~\eqref{eq_deriveCC}, are

\begin{align*}
\diver\cdot \,\mathbf c \cdot \epsilon &=\rho f;&\diver\cdot\, \mathbf c \cdot \epsilon &=\rho f\\
\operatorname{Ink}(\epsilon)=\epsilon_{11,22}-2\epsilon_{12,12}+\epsilon_{22,11}&=0.& \epsilon &= \frac 12(\nabla^Tu + \nabla u)
\end{align*}
where $\rho$ is mass density, $f$ is an external body force and $\mathbf c(x)$ is the elasticity tensor. $A^T$ is the analog of $\diver$ and $B$ is the analog of $\operatorname{Ink}(\epsilon)$.

\subsection{Compatibility conditions for the triangulated truss.}\label{ss_CCforLD}

In this section, we consider triangulated trusses only.  We analyze compatibility conditions on the elongations $\lambda_{ij}$, that is, necessary conditions that are satisfied when the linearized discrete problem is soluble. 

\medskip
\line{(LD)\hfill$
(V_i-V_j)\bullet (U_i-U_j)=\ell_{ij}(0) L_{ij}=\lambda_{ij}$\hfill}
\medskip

\noindent The conditions are derived from the fact that the neighboring triangles around a vertex are embedded in the plane.

\paragraph{The wagon wheel condition for a triangular truss.}  Let $V_0$ be an interior vertex of a triangular truss. The union of triangles containing $V_0$ is a regular unit hexagon. 
Let  $\lambda_{0,i}$ (called $\lambda_i$ for short) denote the radial elongations and  $\lambda_{i,i+1}$ for $i=1,\ldots,6$ taken modulo~$n=6$ denote the concentric elongations.
The condition may be deduced as the linearization of the compatibility condition $K(V_0)=0$ for the discrete nonlinear problem. The cosine law gives
$$
\ell_{i,i+1}^2=\ell_i^2+\ell_{i+1}^2-2\ell_i\ell_{i+1}\cos\alpha_i.
$$
where $\alpha_i=\angle V_iV_0V_{i+1}$.
Differentiating, 
$$
\ell_{i,i+1}L_{i,i+1}=\ell_iL_i+\ell_{i+1}L_{i+1}-(\ell_iL_{i+1}+L_i\ell_{i+1})\cos\alpha_i+\ell_i\ell_{i+1}\sin(\alpha_i)\left.\frac d{dt}\right|_{t=0}\alpha_i.
$$
Since the sum of the angles is constant $2\pi$, the compatibility condition is
the vanishing of the angle sum, 
\begin{equation}\label{eq_anglesum}
0=\sum_{i=1}^n\frac{\ell_{i,i+1}L_{i,i+1}-\ell_iL_i-\ell_{i+1}L_{i+1}+(\ell_iL_{i+1}+L_i\ell_{i+1})\cos\alpha_i}{\ell_i\ell_{i+1}\sin\alpha_i}.
\end{equation}
Substituting $\ell_i=\ell_{i,i+1}=1$ for the lengths and $\alpha_i=60^{\circ}$ for the angles, this equation becomes
\begin{equation}\label{eq_WW}
0=\sum_{i=1}^6 L_{i,i+1}-\sum_{i=1}^6 L_{i},
\end{equation}
It relates the elongations along the  spokes  to the elongations of the rim, so we call it the {\it wagon wheel condition}  at the vertex $V_0$.

\paragraph{Generalized wagon wheel condition for a triangulated truss.}
Suppose that $V_0$ is an interior vertex of valence $n$ in a triangulated truss, and that $V_1,\ldots,V_n$ are the adjacent vertices going around in order. It turns out that the compatibility equation is again that a weighted sum of the radial elongations $L_{i}$  equals a weighted sum of the concentric elongations, $L_{i,i+1}$ for $i=1,\ldots,n$ taken modulo~$n$. 
Since the sum of the angles is constant $2\pi$, we again obtain~\eqref{eq_anglesum}. By regrouping the sum, this becomes
\begin{equation}\label{eq_genWW}
0=\sum_{i=1}^n\frac{\ell_{i,i+1}}{\ell_i\ell_{i+1}\sin\alpha_i}L_{i,i+1} - \sum_{i=1}^n\left\{\frac{\ell_i-\ell_{i+1}\cos\alpha_i}{\ell_i\ell_{i+1}\sin\alpha_i}+\frac{\ell_{i}-\ell_{i-1}\cos\alpha_{i-1}}{\ell_{i-1}\ell_{i}\sin\alpha_{i-1}}\right\}L_{i}
\end{equation}
The wagon wheel condition may be rewritten in a simpler form. If we denote $\beta_i=\angle V_0V_iV_{i+1}$ and $\gamma_i=\angle V_0V_{i}V_{i-1}$, then the area of the triangle 
$$
2\Area(\bigtriangleup V_0V_iV_{i+1})=\ell_i\ell_{i+1}\sin\alpha_i=\ell_i\ell_{i,i+1}\sin\beta_i=\ell_{i+1}\ell_{i,i+1}\sin\gamma_{i+1}.
$$
It follows that 
\begin{equation}\label{eq_spth}
\frac{\ell_{i,i+1}}{\ell_i\ell_{i+1}\sin\alpha_i}=
\frac{\ell_{i,i+1}}{\ell_i\ell_{i,i+1}\sin\beta_i}=\frac 1{\ell_i\sin\beta_i}=\frac 1{h_i}
\end{equation}
where 
$$
h_i=\ell_i\sin\beta_i=\ell_{i+1}\sin\gamma_{i+1}
$$
 is the support distance,  the distance of the of line through the $\ell_{i,i+1}$ side to $V_0$. Also, subtracting the projection of $\ell_{i+1}$ on $\ell_i$ we obtain
 $$
 \ell_i-\ell_{i+1}\cos\alpha_i=\ell_{i,i+1}\cos\beta_i.
 $$
Then
$$
\frac{\ell_i-\ell_{i+1}\cos\alpha_i}{\ell_i\ell_{i+1}\sin\alpha_i}=
\frac{\ell_{i,i+1}\cos\beta_i}{\ell_i\ell_{i,i+1}\sin\beta_i}=\frac{\cos\beta_i}{h_i}
$$
and
$$
\frac{\ell_{i}-\ell_{i-1}\cos\alpha_{i-1}}{\ell_{i-1}\ell_{i}\sin\alpha_{i-1}}=
\frac{\ell_{i-1,i}\cos\gamma_{i}}{\ell_{i}\ell_{i-1,i}\sin\gamma_i}=\frac{\cos\gamma_i}{h_{i-1}}.
$$
This gives  the  final wagon wheel condition.
\begin{thm} In a triangulated truss, the compatibility condition at an interior vertex has the form
\begin{equation}\label{eq_genWW2} 
0=\sum_{i=1}^n\frac{ L_{i,i+1}}{h_i}-\sum_{i=1}^n\left\{\frac{\cos(\beta_i)}{h_i}+\frac{\cos(\gamma_{i})}{h_{i-1}}\right\}L_i,
\end{equation}
where $h_i$ is defined by~\eqref{eq_spth}.
\end{thm}
Rearranging the wagon wheel condition~\eqref{eq_genWW} into a triangle-wise sum
$$
0=\sum_{i=1}^n\frac 1{h_i}\biggl\{ L_{i,i+1}-\cos(\beta_i)L_i-\cos(\gamma_{i+1})L_{i+1}\biggr\}.
$$
In other words, on average, the projected elongations of the radial components onto the circumferential line cancels the circumferential component of that line.

\paragraph{Necessity of the wagon wheel condition for the triangular truss.}
On the triangular truss   the necessity computation is facilitated by the fact that the spokes are equal and evenly spaced so $W_{i-1}+W_{i+1}=W_i$. Thus using~\eqref{eq_WW} and $\ell_i=\ell_{i,i+1}=1$,
\begin{equation}\label{eq_affinehex}
\begin{aligned}
\sum_{i=1}^n{\ell_{i,i+1} L_{i,i+1}}&=
\sum_{i=1}^n{W_{i+1}\bullet Z_{i+1}-W_i\bullet Z_{i+1}-W_{i+1}\bullet Z_i-W_i\bullet Z_i}\\ &=
\sum_{i=1}^n\left\{ W_i-W_{i-1}-W_{i+1}+W_i\right\}\bullet Z_i
=
\sum_{i=1}^nW_i\bullet Z_i
=
\sum_{i=1}^n\ell_iL_i.
\end{aligned}
\end{equation}
Thus the wagon wheel condition is a consequence of satisfying the equations~(LD).
The same compatibility condition applies to affinely distorted hexagons. Under the change of variables $Y_i=MW_i$ where $M$ is an invertible matrix, we still have $Y_{i-1}+Y_{i+1}=Y_i$ so that the same computation also works.

\paragraph{Necessity of the wagon wheel condition for the triangulated truss.}
The compatibility conditions were derived by differentiating $K(V_i)=0$. Here we show they are necessary consequences of equations~(LD).
Rewrite the equation using $W_i=V_i-V_0$ and $Z_i=U_i-U_0$, and recover~\eqref{eq_genWW} from the prescribed elongations equations~\eqref{eq_wagonparts}.
\begin{equation}\label{eq_wagonparts}
\begin{aligned}
W_i\bullet Z_i&=\ell_{i} L_{i}\\
(W_{i+1}-W_i)\bullet (Z_{i+1}-Z_i)&=\ell_{i,i+1} L_{i,i+1}
\end{aligned}
\end{equation}
Using the coefficients suggested by the calculation above, \eqref{eq_wagonparts} and   $W_{i+1}\bullet Z_i=\ell_{i+1}L_i\cos\alpha_i$,
\begin{align*}
\lefteqn{
\sum_{i=1}^n\frac{\ell_{i,i+1} L_{i,i+1}}{\ell_i\ell_{i+1}\sin\alpha_i}=
\sum_{i=1}^n\frac{W_{i+1}\bullet Z_{i+1}-W_i\bullet Z_{i+1}-W_{i+1}\bullet Z_i+W_i\bullet Z_i}{\ell_i\ell_{i+1}\sin\alpha_i}}\\
&=\sum_{i=1}^n\left\{
\frac 1{\ell_{i-1}\ell_{i}\sin\alpha_{i-1}}+\frac 1{\ell_i\ell_{i+1}\sin\alpha_i}\right\}W_i\bullet Z_i
- \left\{\frac{W_{i-1}}{\ell_{i-1}\ell_{i}\sin\alpha_{I-1}}+\frac{W_{i+1}}{\ell_i\ell_{i+1}\sin\alpha_i}
\right\}\bullet Z_i\\
&=\sum_{i=1}^n\left\{
\frac {\ell_{i} }{\ell_{i-1}\ell_{i}\sin\alpha_{I-1}}+\frac{\ell_{i} }{\ell_i\ell_{i+1}\sin\alpha_i}\right\}L_{i}
- \left\{\frac{\ell_{i-1}\cos\alpha_{i-1}}{\ell_{i-1}\ell_{i}\sin\alpha_{I-1}}+\frac{\ell_{i+1}\cos\alpha_i}{\ell_i\ell_{i+1}\sin\alpha_i}
\right\}L_i
\end{align*}
Thus the wagon wheel condition in  is a consequence of satisfying the (LD) equations~\eqref{eq_preslam2} in a triangulated truss as well.


\section{Integral and sum of compatibility around a curve} \label{ss_boundarycomp}

The compatibility condition may be integrated over any simple closed curve of the structure.
Consider a simply connected subdomain bounded     
using~\eqref{eq_RotM}. Thus it can be integrated to give a  compatibility condition on the curve in Theorem~\ref{th_boundaryintCL}.
 This is reminiscent of applying the divergence theorem to a solution of the second order divergence equation $-\diver A(x)\nabla u = 0$ on a simply connected subdomain $\Omega$ to get a boundary integral involving the solution
$$
\int _{\partial\Omega}A(x)\nabla u  \,\bullet\, {\text{\boldmath$\nu$}}(x)\, ds=0
$$ 
where {\boldmath$\nu$} is the inner unit normal. For the problem (NC), the integral compatibility condition is  the Gauss Bonnet formula~\cite[p. 375]{O:1966} 
\begin{equation}\label{eq_GaussBonnet}
\iint_{\Omega} \mathcal K\, dM+\int_{\partial \Omega} \kappa_g\, ds +\sum_i\alpha_i= 2\pi,
\end{equation}
where $\mathcal K$, $\kappa_g$, $dM$, $ds$ and $\alpha_i$ are the Gauss curvature, the geodesic curvature of the boundary curve, the area form, the arclength and  angle changes at the corners  expressed {\it in terms of the} $\zeta$ {\it metric}. 
The compatibility condition for prescribed deformation tensor $\zeta$, the vanishing of the Gauss curvature $\mathcal K=0$, gives integral compatibility around a curve.
 An analogous equation holds for the discrete problem (ND) (see Section~\ref{ss_NDGB}).

In a triangulated structure, the compatibility conditions are localized at interior points. 
 For the linearized problem (LD), the wagon wheel condition is supported on the closed star neighborhood of an interior point. We will show that for interior edges that are contained in four stars, the sum of their compatibility conditions is zero on that edge. Thus the sum over all interior stars results in a compatibility condition whose support consists of a double layer, the edges on the boundary or one link away from the boundary of the subdomain. Moreover, in Theorem~\ref{th_Dbouncc} it is shown that the continuum limit of the boundary compatibility condition of (LD) is the compatibility condition of (LC).

\subsection{Sum of (LD) compatibility along a curve for a triangular truss.}

Consider the union of hexagons $\mathcal P$ in a triangular truss whose boundary curve is a single simple closed curve. Let $\sigma(L)$ denote the functional on elongations given by the sum of the wagon wheels conditions for the hexagons of $\mathcal P$. For those edges $E_{ij}$ that are included in four hexagons, as a radial edge for the hexagons centered at the endpoints and as a circumferential edge for those hexagons centered on the opposite vertices of triangles containing the edge, the sum cancels and the coefficient of $L_{ij}$ is zero in $\sigma(L)$. Thus, only the edges whose endpoints are in a double layer,  at most one unit from $\partial P$,  contribute to $\sigma(L)$.

\begin{figure}[h]
      \begin{center}\label{fig_hexwt}
          \scalebox{0.5}{\includegraphics{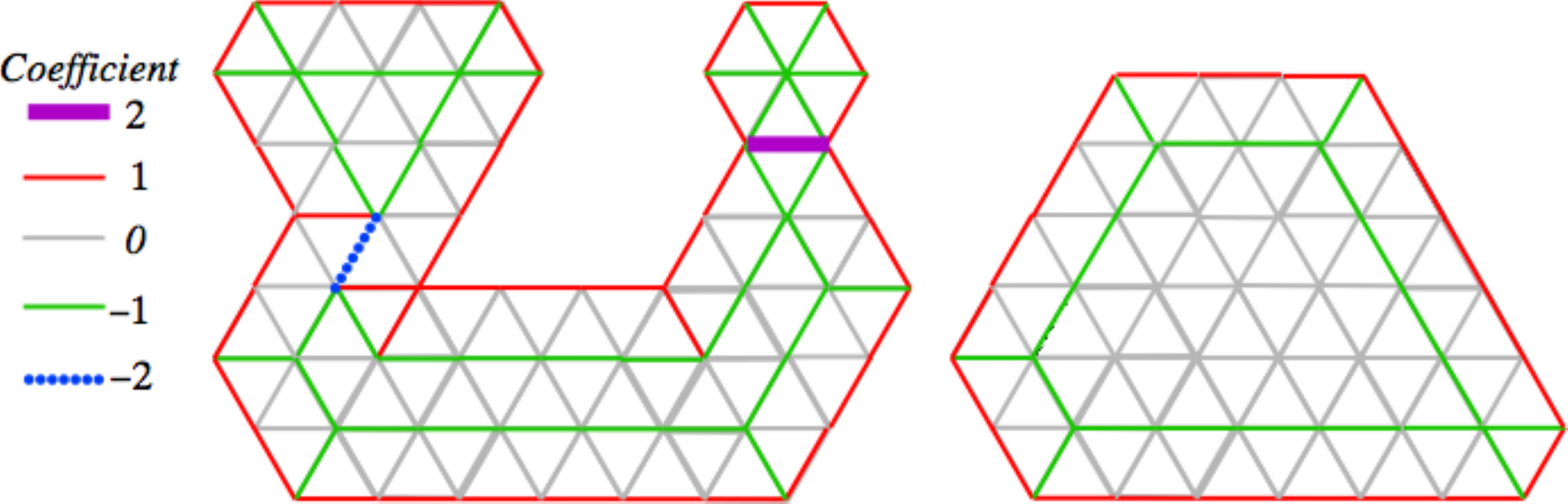}}
 \caption{Double layer in triangular domains.}
 \end{center}
       \end{figure}

The weights depend on which hexagons contain the edge. For example, if an edge is circumferential to two hexagons and radial for none, the weight of $L_{ij}$ would be $2$. The boundary edges receive a coefficient $+1$. The parallel curve, consisting of edges one link inside the boundary receives the coefficient $-1$. At convex corners, the unique incoming curve receives $-1$. At boundary points with zero curvature, the incoming edges receive zero. At concave corners, the incoming middle curves receive $+1$.

 We summarize in the table for various kinds of edges.

\begin{center}
\begin{tabular}{lc}
Edge Type of $E_{ij}$ & Coefficient of $L_{ij}$\\
\hline
Boundary& $1$\\
Isthmus & $2$\\
Unique incoming &$-1$ \\
Extreme of multiple incoming & $0$\\
Middle of multiple incoming & $1$ \\
Interior spine &  $-2$\\
Interior & $-1$
\end{tabular}
\end{center}

  In a simply connected domain, the difference of the lengths is accounted for by the curvature of the boundary. Denote the parallel curve $\tau$. Denote
and the curvature atom at a boundary vertex $V_j$ by $\kappa_j$, which for our hexagon domains takes values in $\{ \frac{\pi}2,0,-\frac{\pi}3, -\frac{2\pi}3\}$.  Then by Steiner's formula~\cite[pp.~7--8]{Sa:1976}
$$
\len(\partial \mathcal P) =\len(\tau) + \frac 3{\pi}\sum_{V_j\in\partial\mathcal P}
\kappa_j.
$$
Equivalently, 
$$
\sum_{\text{$E_{ij}$ is boundary edge}}\sigma(E_{ij})=-\sum_{\text{$E_{ij}$ is interior edge}}\sigma(E_{ij}) - \sum_{\text{$E_{ij}$ is incoming edge}}\sigma(E_{ij}).
$$

In the case that the $\mathcal P$ is a convex union of regular hexagons, as in Figure~\ref{fig_hexwt}b,   $\sigma(L)=0$ and~\eqref{eq_dbCC}  simplify because there are only boundary edges, a single incoming edge at  corners and boundary parallel interior edges.
\begin{equation}\label{eq_sdbCC} 
\sum_{\text{$E_{ij}$ is boundary edge}}L_{ij}=\sum_{\text{$E_{ij}$ is boundary parallel edge}}L_{ij} + \sum_{\text{$E_{ij}$ is incoming edge}}L_{ij}.
\end{equation}
As a simple application of Theorem~\ref{th_bounlay}, we can conclude that if there are no elongations on the boundary of a domain, then there cannot be only positive elongations in the neighboring edges of the boundary layer.
\begin{cor} Let $\mathcal P$ be a convex union of hexagons  structure with interior points. Suppose that the elongations $L_{ij}$ are zero on the boundary $\partial \mathcal P$ and positive on edges within one link of the boundary. Then $L$ cannot satisfy the compatibility conditions at all interior points of $\mathcal P$.
\end{cor}
\begin{proof}The weights $\sigma(E_{ij})$ of~\eqref{eq_dbCC} are positive on boundary edges and nonpositive and somewhere negative on the rest of the edges in the unit boundary layer for convex domains. Thus $\sigma(L)=0$ or~\eqref{eq_sdbCC}   for compatible elongations cannot hold.\end{proof}

\subsection{Sum of compatibility along a curve for the discrete linear problem}
Consider a triangulated structure.  
First, we show that the sum of all wagon wheel compatibility conditions vanishes on an interior edge.  

\begin{lemma}\label{lem_vanish} Let  $E_{02}$ be an interior edge that is bounded on  opposite sides by two nondegenerate triangles $V_0V_1V_2$ and $V_0V_2V_3$. Suppose further that all four $V_0$, $V_1$, $V_2$ and $V_3$ are interior vertices. Then the sum of the four wagon wheel conditions that involve $E_{02}$, the ones centered on $V_0$, $V_1$, $V_2$ and $V_3$,  has zero  $L_{02}$ coefficient.
\end{lemma}

\begin{proof}
Denote the lengths $\ell_1=|V_0V_1|$, $\ell_2=|V_0V_2|$, $\ell_3=|V_0V_3|$, $\ell_4=|V_1V_2|$ and $\ell_5=|V_2V_3|$.
Denote the angles $\alpha_1=\angle V_1V_0V_2$, $\alpha_2=\angle V_2V_0V_3$, $\beta_1=\angle V_0V_1V_2$, $\beta_2=\angle V_2V_3V_0$, $\delta_1=\angle V_1V_2V_0$ and $\delta_2=\angle V_0V_2V_3$. 

The wagon wheel conditions that involve $E_{01}$ are the ones centered at $V_0$ and $V_1$ where $E_{01}$ is a radial edge and those centered on $V_2$ and $V_3$ where $E_{01}$ is a concentric edge. Exactly one term in \eqref{eq_genWW} at each vertex appears. The sum of coefficients of $L_{01}$ is
\begin{align*}
\frac{\ell_2}{\ell_2\ell_3\sin\beta_1}+\frac{\ell_2}{\ell_3\ell_5\sin\beta_2}
-\left\{ \frac{\ell_2-\ell_3\cos\alpha_2}{\ell_2\ell_3\sin\alpha_2}
+ \frac{\ell_2-\ell_1\cos\alpha_1}{\ell_1\ell_2\sin\alpha_1}\right\}
-\left\{ \frac{\ell_2-\ell_4\cos\delta_1}{\ell_2\ell_4\sin\delta_1}
+ \frac{\ell_2-\ell_5\cos\delta_2}{\ell_2\ell_5\sin\delta_2}\right\}
\end{align*}
We observe that twice the areas of triangle $V_0V_2V_1$ and $V_0V_2V_3$ are, respectively,
\begin{align*}
2A_1&=\ell_1\ell_2\sin\alpha_1=\ell_1\ell_4\sin\beta_1=\ell_2\ell_4\sin\delta_1\\
2A_2&=\ell_2\ell_3\sin\alpha_2=\ell_3\ell_5\sin\beta_2=\ell_2\ell_5\sin\delta_2.
\end{align*}
The sum of coefficients of $L_{01}$ becomes
$$
\frac{-\ell_2+\ell_1\cos\alpha_1+\ell_4\cos\delta_1}{2A_1}
+\frac{-\ell_2+\ell_3\cos\alpha_2+\ell_5\cos\delta_2}{2A_2}=0.
$$
This is because the sum of the  lengths of the projections of the sides $V_0V_1$ and $V_2V_1$ onto the side $V_0V_2$ equals the length of $V_0V_2$, namely, $
\ell_1\cos\alpha_1+\ell_4\cos\delta_1=\ell_2$.
A similar equation holds for triangle $V_0V_2V_3$.
\end{proof}
   
   The union of the closed triangles that meet a vertex is called a \emph{closed star neighborhood.}
Since the interior compatibility conditions are supported on closed star neighborhoods, let us consider the union of interior stars whose boundary consists of a simple closed curve.  Let $\sigma$ be the sum of the compatibility conditions viewed as a linear functional on elongations. 
The value of $\sigma$ on different types of edges depends on which interior stars contain the edge. We prove that it vanishes for edges contained in four different stars which are sufficiently distant from a boundary.
\begin{thm}\label{th_bounlay} For the triangulated structure, let $\mathcal P$ be the union of closed star neighborhoods of all interior points. Let 
\begin{equation}\label{eq_dbCC}
\sigma(L)=\sum_{ij}\sigma(E_{ij})\, L_{ij}
\end{equation}
 be the sum of the compatibility conditions corresponding to the interior points. $\sigma(E_{ij})$ vanishes except for edges that either touch the boundary of $\mathcal P$ or both endpoints are one link away from the boundary. The coefficient $\sigma(E_{ij}), $ for such edges, is the sum of the $E_{ij}$ coefficient in all wagon wheel compatibility conditions whose star neighborhoods contain the edge $E_{ij}$ as either radial or circumferential edge. The values of the coefficients, which have expressions in terms of the geometry of the triangulation, are given in the proof.
\end{thm}
\begin{figure}[h]
      \begin{center}
          \scalebox{0.5}{\includegraphics{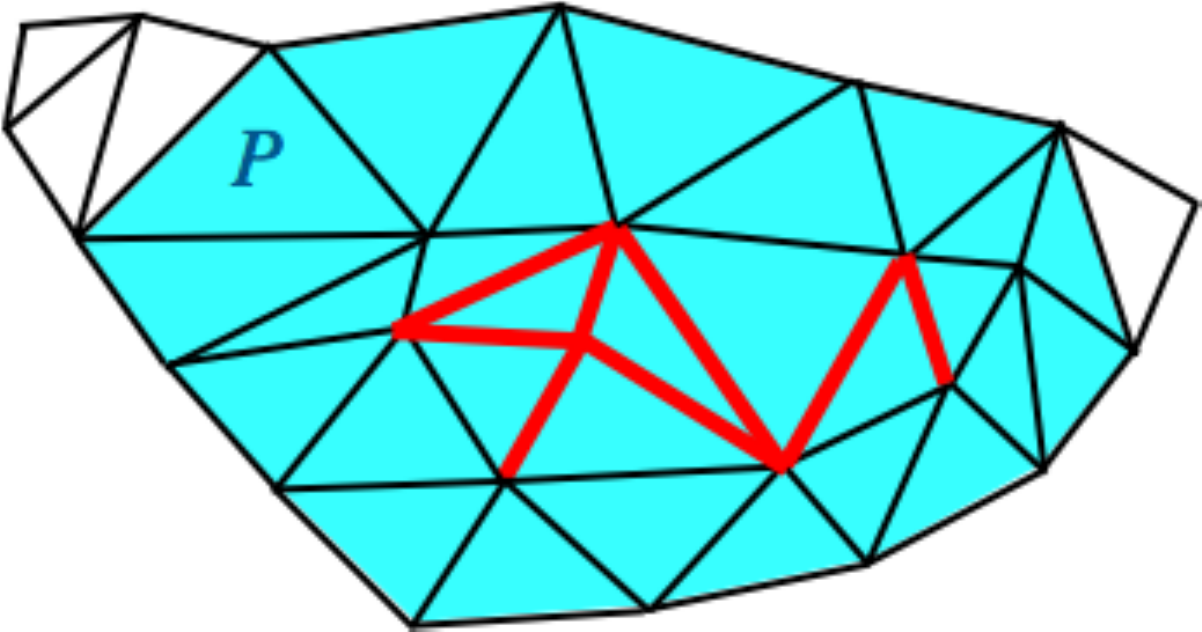}}
 \caption{Triangulated truss indicating plate $\mathcal P$ and  $\sigma(E_{ij})=0$ edges.}
 \end{center}
       \end{figure}

\begin{proof}
As before, consider an edge $E_{02}$ and its adjacent triangles $V_0V_2V_1$ and $V_0V_2V_3$. Let us fix lengths in this proof $\ell_1= \len(E_{01})$, $\ell_2=\len(E_{02})$, $\ell_3=\len(E_{03})$, $\ell_4=\len(E_{12})$ and $\ell_5=\len(E_{23})$ and  the angles  $\alpha_1=\angle V_1V_0V_2$, $\alpha_2=\angle V_2V_0V_3$, $\beta_1=\angle V_0V_2V_1$, $\gamma_1=\angle V_0V_2V_3$, $\delta_1=\angle V_0V_1V_2$ and $\delta_2=\angle V_0V_3V_2$.
By Lemma~\ref{lem_vanish}, $\sigma$ vanishes for edges with one endpoint farther than one link from $\partial \mathcal P$. There are seven combinatorial types of non-vanishing conditions: of the four vertices, there are several possibilities. (1) One vertex is interior which corresponds to a boundary edge or a unique incoming edge. (2)  two vertices are interior which corresponds to an isthmus edge, an extreme of multiple incoming edges or a spine edge; (3)  three vertices are interior which corresponds to the middle of multiple incoming edges or a parallel boundary edge.

{\bf Boundary edges.} Let $E_{02}$ and edge of the boundary $\partial \mathcal P$. This means that $\mathcal P$ is on one side of $E_{02}$, therefore we can take $V_1$  an interior vertex and $V_0$, $V_2$ and $V_3$  boundary vertices. Then the coefficient of $L_{02}$ has  only one term
$$
\sigma(E_{02})=\frac 1{h_1}.
$$
where 
$$
h_1=\ell_1\sin\alpha_1=\ell_4\sin\beta_1
$$
is the support distance of $V_1$ from $E_{02}$.

{\bf Unique incoming edge.} If both side triangles touch boundary, $V_0$ is interior but $V_1$, $V_2$ and $V_3$ are not, then $E_{02}$ is radial and
$$
\sigma(E_{02})= -\frac{\cos\beta_1}{h_1}-\frac{\cos\gamma_1}{h_2}
$$
where
\begin{equation}\label{eq_hs}
h_1=\ell_2\sin\beta_1,\qquad h_2=\ell_2\sin\gamma_1.
\end{equation}

{\bf Isthmus edges.} In case the edge straddles a neck of $\mathcal P$, $V_0,V_2\in\partial \mathcal P$  but $V_1$ and $V_2$ are interior vertices,
$$
\sigma(E_{02})=\frac 1{h_3}+\frac 1{h_4}
$$
where 
\begin{equation}\label{eq_h34}
h_3=\ell_1\sin\alpha_1=\ell_4\sin\beta_1,\qquad h_4=\ell_3\sin\alpha_2=\ell_5\sin\gamma_1.
\end{equation}

{\bf Extreme of multiple incoming edges.} If one side triangles touches the boundary and the other does not, say, $V_0$ and $V_1$ are interior but  $V_2$ and $V_3$ are not, then $E_{02}$ is both radial and circumferential so using~\eqref{eq_hs},~\eqref{eq_h34} and $2A_1=h_1\ell_4=h_3\ell_2$,
\begin{align*}
\sigma(E_{02})&= \frac 1{h_3}-\frac{\cos \beta_1}{h_1}-\frac{\cos\gamma_1}{h_2}
=\frac {\ell_2}{\ell_2h_3}-\frac{\ell_4\cos \beta_1}{\ell_4h_1}-\frac{\cos\gamma_1}{h_2}\\
&= \frac {\ell_1\cos\alpha_1}{\ell_1h_5}-\frac{\cos\gamma_1}{h_2}
= \frac {\cos\alpha_1}{h_5}-\frac{\cos\gamma_1}{h_2}.
\end{align*}

{\bf  Interior spine edges.} Both endpoints are interior but both triangles touch the boundary, say, $V_0$ and  $V_2$  are interior but  $V_1$ and $V_3$ are not, then $E_{02}$ is doubly radial  so using~\eqref{eq_hs},~\eqref{eq_h34}, $2A_1=h_1\ell_4=h_3\ell_2=h_5\ell_1$,
$2A_2=h_2\ell_5=h_4\ell_2=h_6\ell_1$ and the fact that the two projections $\ell_1\cos\alpha_1+\ell_4\cos\beta_1=\ell_3\cos\alpha_2+\ell_5\cos\gamma_1=\ell_2$, we have 
\begin{align*}
\sigma(E_{02})&= -\frac{\cos \alpha_1}{h_5}-\frac{\cos\alpha_2}{h_6}-\frac{\cos \beta_1}{h_1}-\frac{\cos\gamma_1}{h_2}
=-\frac{\ell_1\cos \alpha_1}{\ell_1h_5}-\frac{\ell_4\cos \beta_1}{\ell_4 h_1}-\frac{\ell_3\cos\alpha_2}{\ell_3h_6}-\frac{\ell_5\cos\gamma_1}{\ell_5 h_2}\\
&=-\frac{\ell_1\cos \alpha_1+\ell_4\cos \beta_1}{\ell_4 h_1}-\frac{\ell_3\cos\alpha_2+\ell_5\cos\gamma_1}{\ell_5 h_2}
=-\frac{\ell_2}{\ell_2 h_3}-\frac{\ell_2}{\ell_2h_4}
=-\frac{1}{h_3}-\frac{1}{h_4}
\end{align*}
where
\begin{equation}\label{eq_h56}
h_5=\ell_2\sin\alpha_1,\qquad h_6=\ell_2\sin\alpha_2.
\end{equation}
The remaining cases are computed similarly.

{\bf Middle of multiple incoming edges.} If neither side triangle touches the boundary, say, $V_0$, $V_1$ and $V_3$ are interior but  $V_2$ is not, then $E_{02}$ is both radial and doubly circumferential so using~\eqref{eq_hs},~\eqref{eq_h56}, $2A_1=h_1\ell_4=h_3\ell_2=h_5\ell_1$,
$2A_2=h_2\ell_5=h_4\ell_2=h_6\ell_1$  and the fact that the two projections $\ell_1\cos\alpha_1+\ell_4\cos\beta_1=\ell_3\cos\alpha_2+\ell_5\cos\gamma_1=\ell_2$,
\begin{align*}
\sigma(E_{02})&= \frac 1{h_3}+\frac 1{h_4}-\frac{\cos\beta_1}{h_1}-\frac{\cos\gamma_1}{h_2}
= \frac {\cos\alpha_1}{h_5}+\frac {\cos\alpha_2}{h_6}.
\end{align*}

{\bf  Boundary parallel  edges.} Both endpoints are interior but only one triangle touches the boundary, say, $V_0$, $V_1$ and  $V_2$  are interior but   $V_3$ is not, then $E_{02}$ is doubly radial and circumferential   so using~\eqref{eq_hs},~\eqref{eq_h56}, $2A_1=h_1\ell_4=h_3\ell_2=h_5\ell_1$,
$2A_2=h_2\ell_5=h_4\ell_2=h_6\ell_1$  and the fact that the two projections $\ell_1\cos\alpha_1+\ell_4\cos\beta_1=\ell_3\cos\alpha_2+\ell_5\cos\gamma_1=\ell_2$,
\begin{align*}
\sigma(E_{02})&= \frac 1{h_3}-\frac{\cos\alpha_1}{h_5}-\frac{\cos\alpha_2}{h_6}
-\frac{\cos\beta_1}{h_1}-\frac{\cos\gamma_1}{h_2}=-\frac 1{h_4}.
\end{align*}
\end{proof}
Note that for triangular trusses, the wagon wheel condition simplifies under another normalization ($\sqrt 3/2$ times the one for a general structure.) Thus in the triangular case, boundary edges have weight~$1$, unique incoming edges have weight~$-1$, isthmus edges have weight~$2$, extreme of multiple incoming edges have weight~$0$, interior spine edges have weight $-2$, middle of multiple incoming edges have weight~$1$ and boundary parallel edges have weight~$-1$, as in Figure~\ref{fig_hexwt}.

\subsection{Sum of compatibility conditions centered along a curve}
Consider a simple  closed curve $\gamma$  in a triangulated structure which doesn't surround a hole (is contractible) in the structure such that the points of $\gamma$  are interior points. Let $U$ be the union of the stars centered on $\gamma$ and suppose that the  part  $\tilde\gamma$ of the boundary  $\partial U$ outside $\gamma$ is also a simple closed curve.  Then  $U$  may be viewed as the union of two strips, the boundary layer $S$ inside $\gamma$ and the boundary layer $\tilde S$ inside  $\tilde\gamma$. The compatibility sums along the strips must be equal since their sums differ by the sum of stars along $\gamma$. Of course, we know them both to vanish. Let $C_i$ denote the wagon wheel functional at vertex $V_i$. Then
\begin{align*}
\sum_{E_{ij}\subset \tilde S}\tilde\sigma(E_{ij})L_{ij}&=
\sum_{\text{$V_i$ strictly inside $\tilde\gamma$}}C_i(L)\\
&=
\sum_{V_i\in \gamma }C_i(L)+\sum_{\text{$V_i$ strictly inside $\gamma$}}C_i(L)
=0+ \sum_{E_{ij}\subset  S}\sigma(E_{ij})L_{ij}.
\end{align*}
One notices that $\tilde\sigma(E_{ij})$ and $\sigma(E_{ij})$ have opposite signs for edges $E_{ij}\subset \gamma$ since they are radial edges in $\tilde S$ and circumferential edges of $S$. For triangular trusses, they add to zero. This sum of conditions along a curve limits to the compatibility condition at the vertex as the curve shrinks to the vertex.


\subsection{Sum along a curve of (ND) compatibility}\label{ss_NDGB}
The curve compatibility condition also holds for the nonlinear discrete problem. Let $\gamma$ be a contractible closed curve that bounds the subdomain $\mathcal P$. There is a boundary equation that holds for the double layer near the boundary that amounts to saying that the total angle change going around the outer boundary is $2\pi$. 
\begin{thm}  In a triangulated structure,   suppose that the union of stars~$\mathcal P$ is bounded by a single simple curve $\gamma$. Then the total turning angle of the $\gamma$ may be expressed in terms of  the prescribed lengths of edges on or within one link of the boundary edge
\begin{equation}\label{eq_BBanglesum}
2\pi=\sum_{V_i\in\gamma}\left[ \pi-\sum_{\bigtriangleup(V_j,V_i,V_k)\in\mathcal F}\alpha(V_j,V_i,V_k)\right]
\end{equation}
where $\mathcal F$ are the triangular faces of $\mathcal P$ and
$$
\alpha(V_j,V_i,V_k)=\cos^{-1}\left( \frac {\ell_{ij}{}^2+\ell_{ik}{}^2-\ell_{jk}{}^2}{2\ell_{ij}\ell_{jk}}\right)
$$
is the angle of the triangle at $V_i$.
\end{thm}
\begin{proof} The inner sum in \eqref{eq_BBanglesum} is the interior angle, the sum of the angles of triangles adjacent to the boundary vertex $V_i$. Thus the bracket is the outer turning angle of $\gamma$ at $V_i$. The outer sum is the total over boundary vertices of the turning angles, which adds up to $2\pi$ for planar domains.
\end{proof}

\subsection{Integral of  compatibility along a curve for  (LC)}\label{ss_doublay}

In this section, we deduce the continuum compatibility integral along a contractible curve $\gamma$. Let $\Omega$ be the domain bounded by $\gamma$. We begin with a derivation of the compatibility conditions in curvilinear coordinates using Cartan's moving frames (see {\it, e.g.,} \cite{O:1966}) and Chern's computation of the variation (as in \cite{C:1962}). A computation in local coordinates is given by Brown~\cite{B:1957}.

\paragraph{Moving Frame derivation of compatibility of linearized prescribed strain.}
Suppose that  $\{ {\mathbf e}_i\}$ is an orthonormal frame and $\{ \omega^i\}$ such that $\omega^i({\mathbf e}_j)=\delta^i_j$ is the dual orthonormal co-frame for $\mathcal S$. In the flat plane, they satisfy the structure equations
\begin{align*}
d\, \omega^1 &= \omega^j \wedge \omega_j^i,\qquad
d\, \omega_i^j= \omega_i^p\wedge \omega_p^j
\end{align*} 
where $\omega_i^i=-\omega_j^i$ is the connection form.
  Assume that  $t$ is the deformation parameter where the deformation is given by $\phi(X,t):\mathcal D\times(-\delta,\delta)\to\mathcal S$.
The infinitesimal displacement vector field is
$$
A=\left.\frac{d\phi}{dt}\right|_{t=0}=d\phi\left( \frac d{dt}\right)=a^i\, {\mathbf e}_i.
$$
The covariant derivative of this vector field is the vector valued one-form
$$
DA = da^i\, {\mathbf e}_i + a^i D{\mathbf e}_i=\left(da^i+a^j \omega_j^i \right){\mathbf e}_i=a^i_{,j}\omega^j\, {\mathbf e}_i.
$$
Pulling back the co-frame and connection form gives
\begin{align*}
\theta^i+ a^i\, dt &= \phi^*\omega^i,\qquad
\theta_i^j+ a_i^j\, dt = \phi^*\omega_i^j,
\end{align*}
where $\theta^i$ are forms on $\mathcal D$ and $a_i^j=-a_j^i$. Exterior differentiation on $D\times(-\delta,\delta)$ is given by
$$
d=d_{\mathcal B} + dt \wedge \frac d{dt}
$$
The time derivative is computed by computing $d$ and picking off the ``$dt$'' part.
Differentiating,  dropping ``$\phi^*$,'' and calling $\frac d{dt}\theta^i=\dot\theta^i$,
\begin{align*}
d\theta^i + dt\wedge \dot\theta^i + da^i\wedge dt = d\omega^i&=\omega^j\wedge\omega_j^i=\left(\theta^j+ a^j\, dt\right)\wedge\left(\theta_j^i+ a_j^i\, dt \right)
=\theta^j\wedge \theta_j^i+dt\wedge\left(a^j\theta_j^i -a_j^i\theta^j\right)
\end{align*}
Equating the $\mathcal D$ parts and the $(-\delta,\delta)$ parts, gives
\begin{align*}
d\theta^i&=\theta^j\wedge \theta_j^i,\qquad
\dot\theta^i=d\,a^i+a^j\theta_j^i -a_j^i\theta^j=\left( a^i_{,j}-a_j^i\right)\theta^j.
\end{align*}
The derivative of the metric is therefore
\begin{align*}
\left.\frac{d\phi}{dt}\right|_{t=0}ds^2&=\left.\frac{d\phi}{dt}\right|_{t=0}\sum_{i=1}^2\omega^i\otimes\omega^i=\sum_{i=1}^2\dot\theta^i\otimes\theta^i+\theta^i\otimes\dot\theta^i\\
&=\sum_{i=1}^2\left( a^i_{,j}-a_j^i\right)\theta^j
\otimes\theta^i+\theta^i\otimes\left( a^i_{,j}-a_j^i\right)\theta^j
=\sum_{i=1}^2\sum_{j=1}^2\left( a^i_{,j}+a^j_{,i}\right)\theta^j
\otimes\theta^i
\end{align*}
where we have used the skew symmetry of $a_j^i$. Thus the linearized prescribed strain equation becomes
\begin{equation}\label{eq_fel}
a^i_{,j}+a^j_{,i}=2\epsilon_{ij}
\end{equation}
where $\epsilon_{ij}=\epsilon_{ji}$ is the prescribed strain. The second and third covariant derivatives of the deformation are defined by
\begin{align*}
a^i_{,jk}\omega^k&=da^i_{,j}+a^p_{,j}\omega_p^i-a^i_{,p}\omega_j^p\\
a^i_{,jk\ell}\omega^{\ell}&=da^i_{,jk}+a^p_{,jk}\omega_p^i-a^i_{,pk}\omega_j^p-a^i_{,jp}\omega_{\ell}^p
\end{align*}
where $a^i_{,jk}=a^i_{,kj}$ and $a^i_{,jk\ell}=a^i_{,j\ell k}=a^i_{,k\ell j}=a^i_{,kj\ell}=a^i_{,\ell jk}=a^i_{,\ell kj}$.
The compatibility condition is obtained by cyclicly permuting second covariant derivatives of \eqref{eq_fel}, alternating signs and adding,
\begin{align*}
a^i_{,jk\ell}+a^j_{,ik\ell}&=2\epsilon_{ij,k\ell},\qquad
-a^j_{,k\ell i}-a^k_{,j\ell i}=-2\epsilon_{jk,\ell i},\\
a^k_{,\ell ij}+a^{\ell}_{,kij}&=2\epsilon_{k\ell,ij},\qquad
-a^{\ell}_{,ijk}-a^i_{,\ell jk}=-2\epsilon_{\ell i,jk}
\end{align*}
The resulting compatibility equations are
$
0=\epsilon_{ij,k\ell}-\epsilon_{jk,\ell i}+\epsilon_{k\ell,ij}-\epsilon_{\ell i,jk}
$.
In two dimensions this amounts to a single equation, namely $i=j\ne k=\ell$ so  (no sum)
\begin{equation}\label{eq_deriveCC}
0=\operatorname{Ink}(\epsilon)=\epsilon_{ii,kk}-2\epsilon_{ik,ki}+\epsilon_{kk,ii}.
\end{equation}

\paragraph{Integral compatibility condition}
The equivalence of the  closedness of the one form
$\beta=\beta_i\omega^i$, where $\beta_i=\epsilon_{1i,2}-\epsilon_{2i,1}$,
and  the vanishing of $\operatorname{Ink}(\epsilon)$
was observed by Weingarten~\cite{W:1901} and used by Ces\`aro and Volterra to study dislocations along cracks~\cite{Ac:2017},~\cite[p.~223]{Lo:1944} and by Michell~\cite{Mi:1899} and Yavari~\cite{Y:2013} to study solvability in non-simply connected bodies.
\begin{align*}
d\beta&= d\beta_i\wedge \omega^i+ \beta_i\omega^p\wedge\omega_p{}^i=\left( d\epsilon_{1i,2}-d\epsilon_{2i,1}\right)\wedge\omega^i+ \beta_i\omega^p\wedge\omega_p{}^i\\&=
\left( \epsilon_{1i,2p}\omega^p+ \epsilon_{pi,2}\omega_1{}^p +\epsilon_{1p,2}\omega_i{}^p +\epsilon_{1i,p}\omega_2{}^p-\epsilon_{2i,1p}\omega^p-\epsilon_{pi,1}\omega_2{}^p-\epsilon_{2p,1}\omega_i{}^p-\epsilon_{2i,p}\omega_1{}^p\right)\wedge\omega^i\\
&\qquad 
+\left(\epsilon_{1i,2}-\epsilon_{2i,1} \right)\omega^p\wedge\omega_p{}^i= \left(\epsilon_{12,21}-\epsilon_{11,22}-\epsilon_{22,11}+\epsilon_{21,12} \right) \omega^1\wedge\omega^2=0.
\end{align*}
Note that $\beta$ is a contraction of the covariant derivative of the prescribed strain, thus is a globally defined one-form depending on the prescribed strain. We have
$
\beta_i=\eta^{pq}\epsilon_{pi,q}$,
where the $\eta^{pq}$ is the usual skew rotation tensor
\begin{equation}\label{eq_RotM}
\eta^{pq}=\binom{\ 0\quad \ 1}{-1\quad 0},
\end{equation}
which is invariant like the identity tensor. Upon changing frame
$$
\eta^{pq}=\eta^{rs}R_r{}^pR_s{}^q\qquad\text{for all $R=\binom{\ \cos\alpha\ \ -\sin\alpha}{\sin\alpha\quad\ \cos\alpha}$}
$$
where $\alpha$ is a function. The tensorial nature of $\epsilon_{ij,k}$ and $\beta_j$ dictates how components change under change of frame. For a frame change
\begin{equation}\label{eq_changeframe}
\tilde {\mathbf e}_i = a_i{}^p\mathbf e_p;\qquad \tilde \omega^k{}=R_q{}^k\omega^q
\end{equation}
for some special orthogonal matrix function $R_q{}^k$ and its inverse $R_q{}^ka_k{}^p=\delta_q{}^p$. Then
$$
\tilde \epsilon_{ij,k}=a_i{}^pa_j{}^qa_k{}^r\epsilon_{pq,r};\qquad
\tilde \beta_i=a_i{}^q\beta_q=\tilde\epsilon_{1i,2}-\tilde\epsilon_{2i,1}.
$$

Let us introduce some notation as in  the derivation of the Gauss Bonnet Theorem~\cite[p. 375.]{O:1966}. Let $r$  denote arclength along the boundary, $\gamma(r)$ parameterize the boundary curve with positive orientation and let
\begin{align*}
\mathbf t&=\dot \gamma(r) = c(r) \mathbf e_1(\gamma(r)) + s(r) \mathbf e_2(\gamma(r)),\qquad
{\text{\boldmath$\nu$}}=-s(r) \mathbf e_1(\gamma(r)) + c(r) \mathbf e_2(\gamma(r))
\end{align*}
be the unit tangent and inner normal vectors along the boundary where 
$c(r)=\cos\phi(r)$, $s(r)=\sin\phi(r)$
and $\phi(r)=\angle(\mathbf e_1(r),\mathbf t(r))$ is the angle of the tangent vector relative to the coordinate chart. Thus, if $\partial\Omega$   has length $L$, then the total change in angle going around the boundary is $\phi(L)-\phi(0)=2\pi$.
For  $\mathcal C^2$ boundary, the Frenet equation gives the geodesic curvature in terms of the covariant derivative of the tangent vector
\begin{equation}\label{eq_kappag}
\begin{aligned}
\kappa_g\, {\text{\boldmath$\nu$}}&=\nabla_{\mathbf t}\mathbf t.
\end{aligned}
\end{equation}
In the neighborhood of a boundary point in Fermi coordinates $X(u,v)=\gamma(u)+v$\mbox{\boldmath$\nu$}$(u)$ where $u$ is arclength along $\partial\Omega$ and \mbox{\boldmath$\nu$}$(u)$ is the inner unit normal to $\partial\Omega$. The corresponding new frame on $\partial\Omega$ has  $\tilde{\mathbf e}_1={\mathbf t}$ and $\tilde{\mathbf e}_2=\bnu$.  $v\mapsto X(u,v)$ is a straight line so $\nabla_{\bnu}\bnu=0$.  The geodesic curvature $\kappa_g$ of the boundary is given by 
$$
\nabla_{\tilde{\mathbf e}_1}\tilde{\mathbf e}_1=\tilde\omega_1{}^2(\tilde{\mathbf e}_1)\tilde{\mathbf e}_2=\kappa_g \, \tilde{\mathbf e}_2.
$$

Recalling the definition of covariant derivative,
$$
\tilde\epsilon_{ij,k}\omega^k=d\tilde\epsilon_{ij}-\tilde\epsilon_{pj}\tilde\omega_i{}^p-\tilde\epsilon_{ip}\tilde\omega_j{}^p,
$$
and using skewness of $\tilde\omega_i{}^j$ and $\tilde\epsilon_{ij}=\tilde\epsilon_{ji}$,
$$
\tilde\epsilon_{21,1}=\tilde\epsilon_{21,k}\tilde\omega^k(\tilde{\mathbf e}_1)=\tilde{\mathbf e}_1\tilde\epsilon_{21}-\tilde\epsilon_{11}\tilde\omega_2{}^1(\tilde{\mathbf e}_1)-\tilde\epsilon_{22}\tilde\omega_1{}^2(\tilde{\mathbf e}_1).
$$
Similarly, using Fermi coordinates where straight lines have zero geodesic curvature
$$
\tilde\epsilon_{11,2}=\tilde\epsilon_{11,k}\tilde\omega^k(\tilde{\mathbf e}_2)=\tilde{\mathbf e}_2\tilde\epsilon_{11}-2\tilde\epsilon_{12}\tilde\omega_1{}^2(\tilde{\mathbf e}_2)=\tilde{\mathbf e}_2\tilde\epsilon_{11}\tilde\epsilon_{11}.
$$
Thus we have an expression for the integrand 
\begin{equation}\label{eq_beta}
\begin{aligned}
\beta(\mathbf t)&=\tilde\beta_i\tilde\omega^i(\tilde{\mathbf e}_1)=\tilde\beta_1= \tilde\epsilon_{11,2}-\tilde\epsilon_{21,1}=\tilde{\mathbf e}_1\tilde\epsilon_{21}+\left(\tilde\epsilon_{11}-\tilde\epsilon_{22}\right)\tilde\omega_1{}^2(\tilde{\mathbf e}_1)-\tilde{\mathbf e}_2 \tilde\epsilon_{11}\\
&=\mathbf t\tilde\epsilon_{21}+\left(\tilde\epsilon_{11}-\tilde\epsilon_{22}\right)\tilde\omega_1{}^2(\mathbf t)-\bnu \tilde\epsilon_{11}
\end{aligned}
\end{equation}

Integrating gives the boundary compatibility equation.

 \begin{thm}\label{th_boundaryintCL}
 Let $\Omega$ be a simply connected domain with $\mathcal C^2$ boundary and $\epsilon_{ij}$ be a $\mathcal C^1$ prescribed strain field satisfying the local compatibility conditions and  defined in the neighborhood of the closure of $\Omega$. Then
\begin{equation}\label{eq_bounintegrand}
0=
\int_{\partial\Omega}-\frac{\partial}{\partial\bnu} \tilde\epsilon_{11}+\left(\tilde\epsilon_{11}-\tilde\epsilon_{22}\right)\kappa_g\, ds.
\end{equation}
The components $\tilde \epsilon_{ij}$ near each boundary point are expressed in a local frame where $\tilde{\mathbf e}_1=\mathbf t$ and $\tilde {\mathbf e}_2=\bnu$ are the unit tangent and inner normal vectors on $\partial\Omega$.
\end{thm}
\begin{proof} The function $\tilde\epsilon_{21}=\bnu^T\epsilon\, \mathbf t$ is periodic around $\partial \Omega$ so the integral of its tangential derivative vanishes.
Applying Stokes's Theorem and~\eqref{eq_beta}  to $d\beta=0$ completes the proof of  the  boundary compatibility condition
$$
0=\int_{\Omega}d\beta=\int_{\partial \Omega}\beta=
\int_{\partial\Omega}\beta({\mathbf t})\, dr.
$$
\end{proof}


\paragraph{Compatibility integral for curve with corners.}
 If the boundary is piecewise $\mathcal C^2$, then there are finitely many corner points $\gamma(r_i)$ with $0\le r_1<r_2<\cdots < r_k<L$ such that $\gamma$  can be extended to a $\mathcal C^2$ function on each interval $[r_i,r_{i+1}]$. Moreover, $\gamma$ has  limiting directions $\dot\gamma(r_i+)$ and $\dot\gamma(r_{i+1}-)$ at  endpoints. The angle change in direction at the corner $r_i$ is $\alpha_i=\angle(\dot\gamma(r_i-),\dot\gamma(r_i+))=\phi(r_i+)-\phi(r_i-)$. As in the Gauss Bonnet theorem, the
formula~\eqref{eq_bounintegrand} may be generalized to include corners. The idea is to round off the corner with a circular arc $C_i(\delta)$ with arbitrarily small radius $\delta$  which will tend to zero. We assume $\epsilon_{ij}$ is $\mathcal C^2$ and assume that the fillet $C_i(\delta)$ of radius $\delta$ rounding out the corner at $\gamma(r_i)$ osculates the boundary at $r_i-\rho(\delta)$ and $r_i+\sigma(\delta)$. Hence $C_i(\delta)$ is in the $c_i\delta$ disk about the corner point  $\gamma(r_i)$, where we may take $c_i=1+\sec\alpha_i$. The change of angle along the fillet is $\theta_i=\phi(r_i+\sigma)-\phi(r_i-\rho)$. We have
$$
\lim_{\delta\to 0+}{\theta_i}=\alpha_i.
$$
Pick some points in the middle of each segment $r_i<m_i<r_{i+1}$. The curve near the corner is approximated
$$
\gamma([m_{i-1},m_i])\approx  \gamma([m_{i-1},r_i-\rho])\cup C_i(\delta)\cup \gamma([r_i+\sigma,m_i]).
$$
The length of $C_i(\delta)$ is $\delta\theta_i$ and the curvature is $\kappa_g=\frac 1{\delta}$. The corner contribution at $r_i$ is $A_i$ where
$$
A_i+\int_{\gamma([m_{i-1},m_i])}\beta=\lim_{\delta\to 0+}\left\{\int_{C_i(\delta)}\beta+
\int_{\gamma([m_{i-1},r_i-\rho])}\beta+\int_{\gamma([r_i+\sigma,m_i])}\beta\right\}.
$$

For simplicity, we may  assume $\gamma(r_i)=0$ and that the corner is symmetric about the $y$-axis $\dot\gamma(r_i+)=(\cos\psi,\sin\psi)$ and $\dot\gamma(r_i-)=(\cos\psi,-\sin\psi)$, so that the change in angle is $\alpha_i=2\psi$. Hence
\begin{equation}\label{eq_anglelimit}
\lim_{\delta\to 0+} \phi(r_i-\rho)=-\psi,\qquad \lim_{\delta\to 0+} \phi(r_i+\sigma)=\psi.
\end{equation}
Let us approximate $\epsilon_{ij}$ by its Taylor expansion at  the origin. Call the first order Taylor polynomial
$$
\epsilon^1_{ij}=f_{ij}+xg_{ij}+yh_{ij}
$$
where $f_{ij}$, $g_{ij}$ and $h_{ij}$ are constants so that the strain and its derivative
$$
\epsilon_{ij}-\epsilon^1_{ij}=\mathbf O(x^2+y^2)
,\qquad\qquad
\epsilon_{ij,k}-\epsilon^1_{ij,k}=\mathbf O\left(\sqrt{x^2+y^2}\right)
$$
as $(x,y)\to (0,0)$. 

Let the circular curve  $C_i(\delta)$ be parameterized by $\chi(u)$ where $u$ is arclength such that $\delta\phi(r_i-\rho)\le u \le \delta\phi(r_i+\sigma)$ so that $\dot\chi(\delta\phi(r_i-\rho))=
\dot\gamma(r_i-\rho)$ and $\dot\chi(\delta\phi(r_i+\sigma))=
\dot\gamma(r_i+\sigma)$. Observe that by~\eqref{eq_anglelimit},
$$
\lim_{\delta\to 0+}\int_{\delta\phi(r_i-\rho)}^{-\delta\psi} \beta  =\lim_{\delta\to 0+}\int^{\delta(\phi(r_i+\sigma))}_{\delta\psi} \beta =0
$$
so that
$$
A_i=\lim_{\delta\to 0+}\int_{C_i(\delta)}\beta
=\lim_{\delta\to 0+}\int_{-\delta\psi}^{\delta\psi} \beta .
$$
To approximate $\beta$, observe that the tangent and normal vectors along $C_i{\delta}$ are
$$
\mathbf t(u)=\binom{c(u)}{s(u)},\qquad
\bnu(u)=\binom{-s(u)}{c(u)}
$$
where $c(u)=\cos \frac u{\delta}$ and $s(u)=\sin\frac u{\delta}$.
Thus, we may approximate
\begin{equation}\label{eq_cornerz11}
\tilde \epsilon_{11}-\tilde\epsilon_{22}= \mathbf t^T\epsilon \mathbf t -\bnu^T \epsilon \bnu+\mathbf O(\delta)=(f_{11}-f_{22})(c^2-s^2)+4f_{12}cs+\mathbf O(\delta).
\end{equation}
Also
$$
\bnu \tilde\epsilon_{11}=\left(-s\tfrac{\partial}{\partial y}+c\tfrac{\partial}{\partial y}\right)\mathbf t^T\epsilon \mathbf t+\mathbf O(\delta)=h_{11}c^3+(2h_{12}-f_{11})c^2s+(h_{22}-2f_{12})cs^2-f_{22}s^3+\mathbf O(\delta).
$$
Hence
\begin{equation}\label{eq_cornerlimit}
\begin{aligned}
\int_{-\delta\psi}^{\delta\psi} \beta&=\int_{-\delta\psi}^{\delta\psi} \Bigl[\tilde\epsilon_{11}-\tilde\epsilon_{22}\Bigr]\kappa_g-\frac{\partial}{\partial\bnu} \tilde\epsilon_{11}\, du\\
&=\int_{-\delta\psi}^{\delta\psi} \Bigl[(f_{11}-f_{22})(c^2-s^2)+4f_{12}cs\Bigr]\frac 1{\delta}\\
&\qquad\qquad-\Bigl[ h_{11}c^3+(2h_{12}-f_{11})c^2s+(h_{22}-2f_{12})cs^2-f_{22}s^3\Bigr]\, du+\mathbf O(\delta)\\
&=(f_{11}-f_{22})\sin 2\psi + \mathbf O(\delta).
\end{aligned}
\end{equation}
Adding over all segments, we have proved the boundary compatibility condition with corners.

 \begin{thm}
 Let $\Omega$ be a simply connected domain with piecewise $\mathcal C^2$ boundary and $\epsilon_{ij}$ be a $\mathcal C^2$ prescribed strain field satisfying the local compatibility conditions and defined in the neighborhood of the closure of $\Omega$. 
 Suppose there are $k$ corners of $\partial\Omega$ at the vertices $\gamma(r_i)$, where $\gamma$ is a parameterization of $\partial\Omega$ by arclength and $0\le r_1<r_2<\cdots < r_k<L$ where $L$ is the length of the boundary. Then
\begin{equation}\label{eq_curveCCcorner}
0=
\int_{\partial\Omega}\left(\tilde\epsilon_{11}-\tilde\epsilon_{22}\right)\kappa_g-\frac{\partial}{\partial\bnu} \tilde\epsilon_{11}\, ds+\sum_{j=1}^m \left(\hat\epsilon_{11}-\hat\epsilon_{22}\right)\sin\alpha_i.
\end{equation}
The components $\tilde \epsilon_{ij}$ near each boundary point are expressed in a local frame where $\tilde{\mathbf e}_1=\mathbf t$ and $\tilde {\mathbf e}_2=\bnu$ are the unit tangent and inner normal vectors. The angle change at the corners is given by 
$\alpha_i=\angle(\dot\gamma(r_i-),\dot\gamma(r_i+))$. At the corners, the components $\hat \epsilon_{ij}$ near each boundary point are expressed in a local frame where 
\begin{equation}\label{eq_cornerframe}
\hat{\mathbf  e}_1=\frac{\dot\gamma(r_i+)+\dot\gamma(r_i-)}{|\dot\gamma(r_i+)+\dot\gamma(r_i-)|},\qquad\hat {\mathbf e}_2=\frac{\dot\gamma(r_i+)-\dot\gamma(r_i-)}{|\dot\gamma(r_i+)-\dot\gamma(r_i-)|}
\end{equation}
are the unit vectors halfway between the tangent directions at the corner and the angle bisector.
\end{thm}
This theorem matches the  discrete curve sum~\eqref{eq_dbCC}.
 $\tilde\epsilon_{11}-\tilde\epsilon_{22}$ corresponds to the strain of the links along the boundary. $\partial \tilde\epsilon_{11}/\partial \mathbf\nu$ corresponds to the difference between elongations on the curve parrallel to the boundary and on the boundary curve. $\kappa_g$ is a delta function which is nonzero at the corners and $\alpha=60^{\circ}$ at convex exterior corners of a triangular domain $\mathcal P$.

\subsection{Integral of compatibility along curve for (NC)}\label{ss_NCdoublay}

Let $\zeta$ be the prescribed right Cauchy-Green tensor.
The compatibility condition that the curvature of $\zeta$ vanishes implies the Gauss-Bonnet Formula~\eqref{eq_GaussBonnet} which says that the total change in angle going around of the boundary of a domain is $2\pi$ ({\it, e.g.,}~\cite{O:1966}).  This is easy to see if $\zeta$ is transformed to Euclidean coordinates, but here we express the angle change in the background coordinates in which $\zeta$ was initially given to compare with Theorem~\ref{th_boundaryintCL}.

 Take a local $\zeta$-orthonormal frame $\mathbf f_i$ and let $\theta^j$ be its dual coframe. Then the structure equations 
$$
d\theta^i=\theta^j\wedge\theta_j^i,\qquad \theta_i^j+\theta_j^i=0,
$$
define the connection form. Vanishing  of the curvature of $\zeta$ is given by the equation
$
d\theta_1^2=0$.
Recall the  change of coordinates formula for the connection form under~\eqref{eq_changeframe}
$$
\tilde\theta_t^j=-a_t^r da_j^r+  a_t^ra_j^q\theta_r^q
$$
so that $\theta_i^j$ is not a tensor.
In our case, the frame has rotated by an angle $\phi(r)$ at $\gamma(r)$ so that at least along the curve the coordinate change is given by an orthonormal matrix
$$
\binom{a_1^1\  a_1^2}{a_2^1\  a_2^2}
=\binom{c(r) \ \  s(r)}{-s(r)\  c(r)}
$$
where $c(s)=\cos\phi(s)$ and $s(s)=\sin\phi(s)$. From this we can compute
\begin{align*}
\tilde\theta_1^2&=-a_1^r da_2^r+  a_1^ra_2^q\theta_r^q=-a_1^1 da_2^1-a_1^2 da_2^2+  a_1^1a_2^2\theta_1^2+  a_1^2a_2^1\theta_2^1
\end{align*}
so that 
\begin{equation}\label{eq_omegae1}
\begin{aligned}
\tilde\theta_1^2(\tilde{\mathbf f}_1)&=-a_1^1 \tilde{\mathbf f}_1a_2^1-a_1^2 \tilde{\mathbf f}_1a_2^2+  \left(a_1^1a_2^2-  a_1^2a_2^1\right)\theta_1^2(\tilde{\mathbf f}_1)\\&=c^2\dot\phi +s^2\dot\phi +(c^2+s^2)\theta_1^2(\tilde{\mathbf f}_1)=\dot\phi +\theta_1^2(\tilde{\mathbf f}_1).
\end{aligned}
\end{equation}

To see this in terms of derivatives of $\zeta$,  we derive an expression for the connection form of the $\zeta$ metric in terms of background curvilinear coordinates.   Let $\mathbf e_i$  be an orthonormal background frame, $\omega{}^j$ its dual coframe and $\omega_i^j$ its connection form.  The $\zeta$-metric is
$
\zeta= \zeta_{ij}\, \omega^i\otimes\omega^j$.
Take a $\zeta$-orthonormal frame  with $\mathbf f_1$ is in the $\mathbf e_1$ direction given by
$$
\mathbf f_1=\frac 1{\sqrt \zeta_{11}}\mathbf e_1,\qquad
\mathbf f_2=-\frac { \zeta_{12}}{\sqrt{D\zeta_{11}}}\mathbf e_1+\frac{\sqrt{\zeta_{11}}}{\sqrt{ D}}\mathbf e_2
$$
where $D=\zeta_{11}\zeta_{22}-\zeta_{12}{}^2$ is the determinant. The dual coframe is 
$$
\theta^1=\sqrt{\zeta_{11}}\omega^1 + \frac{\zeta_{12}}{\sqrt{\zeta_{11}}}\omega^2,\qquad
\theta^2=\frac{\sqrt D}{\sqrt{\zeta_{11}}}\omega^2.
$$
Splitting $\omega_1^2=\Gamma_{11}^2\omega^1+\Gamma_{12}^2\omega^2$, the connection form satisfies 
$$
\theta^2\wedge\theta_2^1=d\theta^1=\left[
(\sqrt{\zeta_{11}})_2 - \left(\frac{\zeta_{12}}{\sqrt{\zeta_{11}}}\right)_1\right]\omega^2\wedge\omega^1+
\sqrt{\zeta_{11}}\omega^2\wedge\omega_2^1 + \frac{\zeta_{12}}{\sqrt{\zeta_{11}}}\omega^1\wedge\omega_1^2
$$
so 
$$
\theta_2^1=\frac{\sqrt{\zeta_{11}}}{\sqrt D}
\left[
(\sqrt{\zeta_{11}})_2 - \left(\frac{\zeta_{12}}{\sqrt{\zeta_{11}}}\right)_1+
\sqrt{\zeta_{11}}\Gamma_{21}^1 - \frac{\zeta_{12}}{\sqrt{\zeta_{11}}}\Gamma_{12}^2\right]\omega^1
\mod \omega^2.
$$
In case of Fermi coordinates where $\tilde{\mathbf e}_1$ is tangent to the boundary and $\tilde{\mathbf e}_2$ is the inner normal, since $\Gamma_{ij}^k=-\Gamma_{kj}^i$,
$$
\kappa_g\tilde{\mathbf e}_2=\nabla_{\tilde{\mathbf e}_1}\tilde{\mathbf e}_1=\tilde{\omega}_1^2(\tilde{\mathbf e}_1)\tilde{\mathbf e}_2=-\tilde{\Gamma}_{21}^1\tilde{\mathbf e}_2,\qquad
0=\nabla_{\tilde{\mathbf e}_2}\tilde{\mathbf e}_2=\tilde{\omega}_2^1(\tilde{\mathbf e}_2)\tilde{\mathbf e}_2=-\tilde{\Gamma}_{12}^2\tilde{\mathbf e}_1.
$$
Hence
$$
\tilde{\theta}_1^2(\tilde{\mathbf e}_1)=\frac{1}{\sqrt{ \tilde D}}
\left[-\frac 12\tilde{\zeta}_{11,2}+\tilde{\zeta}_{12,1}-\frac{\tilde{\zeta}_{12}\tilde{\zeta}_{11,1}}{2\tilde{\zeta}_{11}}+\tilde{\zeta}_{11}\tilde{\kappa}_g
\right].
$$
Applying Stokes's Theorem and \eqref{eq_omegae1}, 
\begin{align*}
0&=\int_{\Omega}d\theta_1^2=\int_{\partial\Omega}\theta_1^2=
\int_0^L \theta_1^2(\mathbf t)\, d\sigma =\int_0^L \tilde{\theta}_1^2(\tilde{\mathbf f}_1) -\dot\phi\, d\sigma=\int_0^L \tilde{\theta}_1^2(\tilde{\mathbf e}_1)\, ds-2\pi\\
&=\int_0^L\frac{1}{\sqrt {\tilde D}}
\left[-\frac 12\tilde{\zeta}_{11,2}+\tilde{\zeta}_{12,1}-\frac{\tilde{\zeta}_{12}\tilde{\zeta}_{11,1}}{2\tilde{\zeta}_{11}}
+\tilde{\zeta}_{11}\tilde{\kappa}_g
\right]\, ds - 2\pi
\end{align*}
where $\sigma$ is $\zeta$-arclength, $s$ is background arclength so $d\sigma=\sqrt{\tilde{\zeta}_{11}}\, ds$.
Arguing as before at the corners, we obtain

 \begin{thm}
 Let $\Omega$ be a simply connected domain with piecewise $\mathcal C^2$ boundary and $\zeta_{ij}$ is a $\mathcal C^2$ prescribed right Cauchy-Green tensor field satisfying the local compatibility conditions and defined in the neighborhood of the closure of $\Omega$. 
 Suppose there are $k$ corners of $\partial\Omega$ at the vertices $\gamma(r_i)$, where $\gamma$ is a parameterization of $\partial\Omega$ by arclength and $0\le r_1<r_2<\cdots < r_k<L$ where $L$ is the length of the boundary. Then
$$
2\pi=
\int_0^L
\left[-\frac 12\tilde{\zeta}_{11,2}+\tilde{\zeta}_{12,1}-\frac{\tilde{\zeta}_{12}\tilde{\zeta}_{11,1}}{2\tilde{\zeta}_{11}}
+\tilde{\zeta}_{11}\tilde{\kappa}_g
\right]\frac{ds}{\sqrt{\tilde D}}+\sum_{j=1}^m\frac{\hat{\zeta}_{11}(\alpha_i+\sin\alpha_i)+\hat{\zeta}_{22}(\alpha_i-\sin\alpha_i)}{2\sqrt{ \hat D}}.
$$
The components $\tilde \zeta_{ij}$ near each boundary point are expressed in a local frame where $\tilde{\mathbf e}_1=\mathbf t$ and $\tilde {\mathbf e}_2=\bnu$ are the unit tangent and inner normal vectors. $\kappa_g$ is the geodesic curvature of the boundary in the background metric. The angle change at the corners is given by 
$\alpha_i=\angle(\dot\gamma(r_i-),\dot\gamma(r_i+))$. At the corners, the components $\hat \epsilon_{ij}$ near each boundary point are expressed in a local frame~\eqref{eq_cornerframe} in unit vectors halfway between the tangent directions at the corner and the angle bisector.
\end{thm}
\begin{proof}
Expading~\eqref{eq_cornerz11} at the corner,
\begin{equation}\label{eq_NDcornerlim}
\frac{\tilde{\zeta}_{11}}{\sqrt {\tilde D}}=\frac{f_{11}c^2+2f_{12}sc+f_{22}s^2}{\sqrt {f_{11}f_{22}-f_{12}{}^2}}+\mathbf O(\delta).
\end{equation}
The corresponding corner limit~\eqref{eq_cornerlimit} gives the result.
\end{proof}


\section{The genericity of trusses.}

Mechanical intuition suggests that the trusses we consider are infinitesimally rigid, thus are generic, and the number of compatibility conditions is given by the Maxwell count. 
In this section, we prove that many planar trusses are generic. For triangulated trusses without holes, this number is the number of interior vertices~\eqref{eq_gen}.
The independence of wagon wheel conditions centered at different interior vertices shows that they form a basis for compatibility conditions.

\subsection{A geometric basis for the  compatibility conditions of  triangular structures.}\label{ss_basis}

A basis for the compatibility conditions will be determined for
a triangular structure $X$ in the hexagonal lattice.
 We shall decompose $X$ into pieces, thick regions, \emph{plates}, which are connected by thin parts, \emph{girders}.  Consider the collection of nodes $\mathcal V_H$  that are centers of hexagons contained in $X$.  Consider the graph $\mathcal G_H$ whose vertices are $\mathcal V_H$ and whose edges are any pair of nodes in $\mathcal V_H$ that are a unit apart. In general $\mathcal G_H$ is not connected. If $\mathcal G_i$ is a connected component of $\mathcal G_H$, let $P_i$ be the union of hexagons whose centers are $\mathcal G_i$. Let us call $P_i$ a {\it plate}. Let us call a simple truss that contains no hexagons a {\it girder.}
The plates may not make up all of $X$, what remains is a collection of girders that attach to the plates. 

It will be shown that the plates  support the compatibility conditions and the girders are statically determined and don't support any compatibility conditions. Unlike for continuous materials, the girders bounded by a single simple curve are examples of rigid structures without compatibility conditions. If  such a girder   loops around and a single new edge is attached  connecting the ends of the girder, then the new structure gains a single compatibility condition. A girder is bounded by two simple disjoint curves,  a \emph{ring girder}, has three compatibility conditions.

  For a structure $X$ in the hexagonal lattice,  the compatibility conditions consist of wagon wheel equations centered on the interior vertices and ring girders around the holes.  
    We begin with simply connected domains.

\begin{thm} \label{th_simple} Let $X$ be the union of finitely many unit triangles of the triangular lattice. Suppose that the boundary $\partial X$ consists of a single simple closed curve. Then the truss 
$X$ is generic: the number of compatibility conditions for (LD) is given by the Maxwell number which also equals the number of interior vertices.
\begin{equation}\label{eq_gen}
c=\CCM=v_i.
\end{equation}
Moreover, a basis for the compatibility conditions is given by the wagon wheels centered at the interior vertices which are supported on the hexagon neighborhood of the vertex.
\end{thm}
The simplicity of the boundary means that the boundary edge path has no self-intersections, thus $X$ cannot be pinched together at a hinge vertex. The proof of Theorem~\ref{th_simple} is given in Appendix, Section~\ref{ss_hex}.

We expect that wagon wheels form a compatibility basis in an arbitrary triangulated structure.
 Since the triangulated truss is generic by Corollary~\ref{cor_BTP}, we know that the dimension of the compatibility conditions is the number of interior vertices~\eqref{eq_CCmholes}. They would form a basis if we knew the wagon wheel conditions are linearly independent. The independence of wagon wheels in (ND) can be seen geometrically. The total angle at $V_i$ is not determined by the flatness of the surrounding vertices. To see this, imagine that the truss lived on a cone that is flat except at the vertex $V_i$ where the curvature atom might not vanish. Just imagine rolling a piece of paper into a cone with the vertex at $V_i$. Since the stars not centered at $V_i$ do not surround $V_i$, they are Euclidean, and the compatibility condition holds for lengths in the cone. However, they do not determine the cone angle at $V_i$ which may be arbitrary.

\subsection{The Number of compatibility conditions in a multiply connected truss.}

Next, we consider multiply connected planar domains. Thus we imagine $X$ is a  triangulated truss with $g$ holes removed. So $X$ is bounded by $g+1$ pairwise disjoint simple closed curves $\gamma_i$ such that one of them, say $\gamma_0$, contains the others within. 
The rest of the curves have at least four links so do not bound a single triangle and do not surround other components $\gamma_i$.
Then holes are the regions bounded by the interior curves $\gamma_1,\ldots,\gamma_g$.

\begin{thm} \label{th_multiple} Let $X$ be a triangulated truss with holes. Suppose that the boundary $\partial X$ consists of  $g+1$ pairwise disjoint, simple closed curves. Then
$X$ is generic: the number of compatibility conditions equals the Maxwell number
\begin{equation}\label{eq_CCM3}
c=\CCM=3g+v_i.
\end{equation}
\end{thm}

\paragraph{Intuitive explanation by a ``hole filling'' argument.}
Consider a triangulated truss $\mathcal T$  bounded by $g+1$ disjoint simple closed curves.
Theorem~\ref{th_multiple} can be seen by an inductive argument,   removing the holes one after another starting from a simply connected truss. The argument  assumes that wagon wheel   conditions used are independent, an assumption that will not be required for the proof. Suppose that $\gamma_i$, $i=0,1,\ldots,g$ are the boundary curves with $n\ge 4$ links each such that $\gamma_0$ contains the others. Fill in the holes with a continuation of the triangulation of $\mathcal T$. Let $\mathcal H_i$ be the disk region inside $\gamma_i$, a hole to be removed, and $\mathcal T_i=\mathcal T\cup(\mathcal H_i\cup\cdots\cup\mathcal H_{g})$ the completely filled region $\mathcal T_1$ with $i-1$ holes removed.  $\mathcal T_1$ is simply connected so $CC(\mathcal T_1)= v_i(\mathcal T_1)$,
the number of interior vertices, by Theorem~\ref{th_simple}. Removing one hole at a time, suppose that $i-1$ holes have been removed and that
$$
CC(\mathcal T_i)= v_i(\mathcal T_i)+3(i-1).
$$
To find $CC$ after removing the next hole, suppose
 $\mathcal H_i$ has $v_h$ interior vertices, $e_h$ interior edges, and $\gamma_i$ has $f_h$ triangles and $n\ge 4$ links. Then the number of vertices and edges on $\gamma_i$ is $n$. The Euler Characteristic of $\mathcal H$ is $1=v_i-e_i+f_i$.
Being a triangulation implies $3f_i=n+2e_i$ so that 
$$
e_i=3v_i+n-3.
$$

Now proceed inductively on the number of interior vertices. If there are no interior vertices, then there are $n-3$ interior edges and $n$ vertices on $\gamma$. The number of compatibility conditions of $\mathcal T_i$ is the difference 
between the number of equations $CC(\mathcal T_i)+2e(\mathcal T_i)$  minus the number of variables $2e(\mathcal T_i)$ for the generic  system~\eqref{eq_tildeAU} augmented by annihilators of rigid motions. Removing $n$ wagon wheel conditions centered on $\gamma_i$ and $n-3$ interior edges gives the compatibility count for $\mathcal T_{i+1}$,
$$
CC(\mathcal T_{i+1})=\Bigl[ CC(\mathcal T_i)+2e(\mathcal T_i) -(n-3)\Bigr] - 2\Bigl[ e(\mathcal T_i)-n\Bigr]=CC(\mathcal T_i)+n+3.
$$
$\mathcal T_{i+1}$ has $n$ fewer interior vertices, thus
\begin{equation}\label{eq_holefill}
 CC(\mathcal T_{i+1})= 3g+v_i(\mathcal T_{i+1}).
\end{equation}

If there are interior vertices,  then we may replace the hole with one with fewer interior vertices and with the same number of compatibility conditions.
Suppose~\eqref{eq_holefill} holds for $j$ interior vertices. Arguing inductively, if there are $j+1$ interior vertices in $\mathcal H_i$, choose a triangle of the triangulation $\tau\subset\mathcal H_i$ such that one side is an edge of $\gamma_i$ and the other edges are interior.  Cut $\tau$ from $\mathcal H_i$ and glue it to $\mathcal T_i$. Now the region $\mathcal T_i\cup\tau$ has the same number of compatibility conditions as $\mathcal T_i$, and its new hole $\mathcal H_i-\tau$ has $n+1$ boundary links, but $j$ interior vertices. Thus the induction holds, and the new hole may be removed from $\mathcal T_i\cup\tau$, completing the induction.

\paragraph{Proof using ``branch cut'' argument.}
\begin{proof}
The statement is proved by induction on the number of holes. The base case $g=0$ is proved by Theorem~\ref{th_simple}. Let $X_{g+1}$ be a truss with $g$ holes. Again, we shall construct a maximal subtruss $Z_{g+1}$ containing all vertices of $X_{g+1}$ which is statically determined and is obtained from $X_{g+1}$ by removing $v_i+3g$ edges.

The idea is to make a ``branch cut.'' From the hole bounded by $\gamma_g$ in $X_{g+1}$ closest to the outside, draw a line segment from the hole to the outside that meets only some edges $B$ between the hole and outside. There must be $b\ge 3$ of them. The truss $X_g=X_{g+1}\backslash B$ now has $g-1$ holes. The outside curve and $\gamma_g$ have been combined to a single closed curve (their connected sum) by replacing the two $B$- edges on the curves $\gamma_0$ and $\gamma_g$ by two connecting segments from $\gamma_0$ to $\gamma_g$. By the induction hypothesis, there is a statically rigid $Z_g$ in $X_g$ obtained by removing $\tilde v_i+3(g-1)$ edges of $X_g$ where $\tilde v_i$ is the number of interior vertices of $X_g$.  Notice that there were $b-3$ interior vertices of $X_{g+1}$ lost by making the branch cut. Also, notice that $Z_{g+1}=Z_g$ is also a maximal statically determined subtruss in $X_g$ obtained by removing 
$$
[\tilde v_i+3(g-1)] + b= [\tilde v_i + b-3]+3g=v_i+3g
$$
of its edges, where $v_i$ is the number of interior vertices of $X_{g+1}$. Thus the induction is complete.
\end{proof}

\paragraph{Example of infinitesimally rigid truss with maximally many interior links removed}
In general, one can remove a link from every wagon wheel in a simply connected truss and still maintain static determinacy.
Consider a   rhombus $\mathcal P_n$ consisting of the union of hexagons centered on $ke_1+\ell e_2$ where $k,\ell=1,\ldots,n$,  $e_1=(1,0)$, $e_2=\left(\frac 12,\frac{\sqrt 3}2\right)$. As in Theorem~\ref{th_BTP}, for each hexagon we may remove a link (the NE link) and still keep rigidity. In fact, the resulting figure is statically determined with no compatibility condition. Thus it has no material points according to our definition. removing one more link makes the structure flexible.
There are $n^2$ hexagons and $3n^2+8n+1$ total links in $\mathcal P_n$. Thus the maximal number that can be removed and still maintain rigidity approaches one third of the links as $n\to\infty$.

Compare this to minimal number of links that need to be removed from  $\mathcal P_n$  to make the structure flexible but keeping it a period cell. If all $n+2$ NE links of $\mathcal P_n$ were removed from the $\ell=\text{const.}$ row, then the structure would flex along that row. The proportion of number of links removed would tend to zero as $n\to\infty$.

\subsection{The definition of  BTP Trusses.}\label{ss_defBTP}

We establish the genericity of a more general class of trusses built up by assembling rigid pieces, which we call BTP Trusses which provides a second argument for the genericity of trusses that are subdomains of the triangular lattice. 
BTP trusses are built up erector set fashion by assembling rigid pieces to form a larger composite rigid piece. Besides, we determine the number of compatibility conditions of the composite truss in terms of its components and attaching procedure.

The BTP-Trusses (Bigon-Triangle-Prism trusses) are finite trusses built by assembling subunits of smaller BTP-Trusses according to some rules.  The following constructions define BTP-Trusses. A single edge with two ending vertices is the basic BTP truss. A pair of edges attached to the same two vertices form a {\it bigon}, which is also rigid. Three edges connected in a triangle also make a rigid truss. We observe that a rigid truss with two labeled vertices behaves like a single edge: two rigid trusses may be attached bigon or triangle fashion to make a larger rigid truss. Two distinct nodes at the same coordinates may be pinned together to make a single node. In addition, there is the prism construction which uses three edges to connect two rigid trusses to make a larger rigid truss not gotten by forming bigons or triangles. Of course, our prisms are projections into two dimensions! Since the third connecting edge may be far separated from the other edges,  determining the rigidity of a truss is not a local problem.

The composition rules of BTP trusses are as follows.
\begin{enumerate}
\item Single links. These consist of two different points and the edge connecting them.
\item Bigons. Suppose $S$ and $T$ are two BTP-Trusses, each containing at least two distinct points $z_1,z_2\in S$ and $z_3,z_4\in T$ such that the coordinates  $z_1=z_3$ and $z_2=z_4$. The bigon is the disjoint union ``$\amalg$'' of $S$ and $T$ whose two points are identified.
$$
T_{\text{bigon}}=(S \amalg T)/\{ a_1\sim z_3, z_2\sim z_4\}
$$
The bigon is assembled by pinning two points together in each of the subassemblies.
\item Triangles. 
 Suppose $S$, $T$ and $U$ are three BTP-Trusses, each containing at least two distinct points $z_1\ne z_2$ in $S$, $z_3\ne z_4$ in $T$ and $z_5\ne z_6$ in $ U$ such that the coordinates  $z_2=z_3$, $z_4=z_5$ and $z_6=z_1$ and such that $z_1z_2z_4$ is non-degenerate (the three points are not collinear.) The triangle is the disjoint union of three sides whose three points are identified pairwise.
$$
T_{\text{triangle}}=(S \amalg T\amalg U)/\{ z_1\sim z_3, z_2\sim z_5, z_4\sim z_6\}
$$
The triangle is assembled by pinning two points together in each of the three subassemblies to form a triangle.

\begin{figure}[h]\label{fig_BTP}
      \begin{center}
          \scalebox{0.5}{\includegraphics{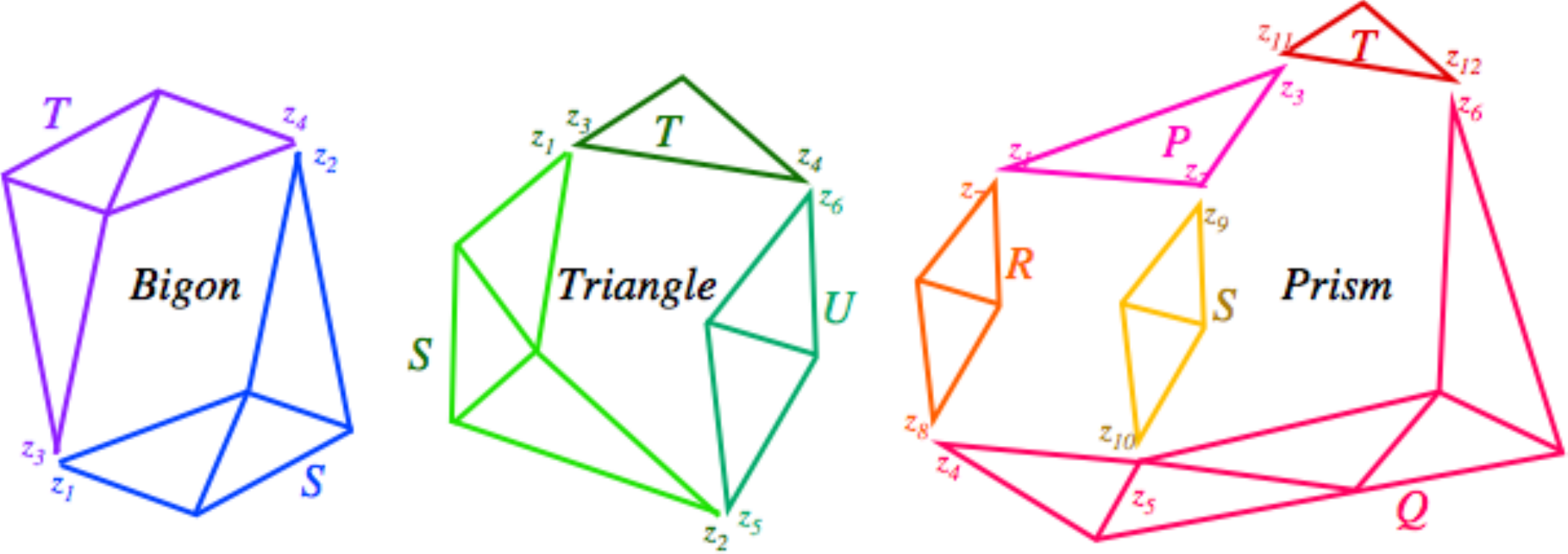}}
 \caption{Bigon-Triangle-Prism Constructions. The parallelogram $z_1,z_2,z_4,z_5$ would flex if it were not for $z_3,z_6$.}
 \end{center}
       \end{figure}

\item (Planar) Prisms. Suppose $P,Q,R,S,T$  are five BTP-Trusses, two of which contain at least three distinct points $z_1,z_2,z_3\in P$ and  $z_4,z_5,z_6\in Q$ satisfying the non-degeneracy condition \eqref{eq_prismgen} and the others contain at least two distinct points $z_7\ne z_{10}$ in $R$, $z_8\ne z_{11}$ in $ S$ and $z_{9}\ne z_{12}$ in $ T$  such that the coordinates  $z_1=z_7$, $z_2=z_8$, $z_3=z_{9}$, $z_4=z_{10}$, $z_5=z_{11}$ and $z_6=z_{12}$. The prism is the disjoint union with six pairs of points  identified.
$$
T_{\text{prism}}=(P \amalg \cdots\amalg T)/\{ z_1\sim z_7,z_2\sim z_8,z_3\sim z_{9},z_4\sim z_{10},z_5\sim z_{11},z_6\sim z_{12}\}
$$
The prism is assembled by connecting the vertices of the triangles $z_1z_2z_3$ of $P$ and $z_4z_5z_6$ of $Q$  with edges $z_1z_4$ of $R$, $z_2z_5$ of $S$ and $z_3z_6$ of $T$.
\item Pin a vertex. Suppose that $T$ is a truss that has two distinct vertices $z_1,z_2\in T$ 
with the same coordinates. The new truss is built by pinning the vertices
$$
T_{\text{pin}}=T/\{ z_1\sim z_2\}.
$$
\end{enumerate}
For example, three single links may be assembled to a simple nondegenerate triangle. Another identical copy of this triangle may be attached to the first at two vertices and overlapping the first, forming a ``bigon.'' The third vertices from each triangle are distinct nodes but have the same coordinates. Finally, these vertices may be pinned together. The BTP-Truss structure is not unique. The same double triangle truss also results from attaching the second edge to each of the three original edges of a triangle.


\subsection{The genericity of BTP Trusses.}\label{ss_BTP}

The triangle and prism constructions require that a nondegeneracy condition be satisfied. For example, in a triangle, the three edges cannot be collinear. In a prism, if the upper and lower triangles are connected by three parallel line segments, then the resulting truss is not infinitesimally rigid because it has a shearing flex. Similarly, if the line segments have a common point of intersection then the prism isn't infinitesimally rigid it will have a rotational flex about the common point.
The nondegeneracy condition~\eqref{eq_prismgen} will be stated in the proof of the theorem.

\begin{thm} \label{th_BTP} BTP-Trusses are infinitesimally rigid, hence generic. The number of compatibility conditions under a BTP combination is determined from the compatibility conditions of its parts. Let  $c_i$ be the number of compatibility conditions for the part $T_i$.
\begin{itemize}
\item Segments have $c=0$.
\item Bigons have $c_{\mathrm{bigon}}=c_1+c_2+1$.
\item Triangles have $c_{\mathrm{triangle}}=c_1+c_2+c_3$.
\item Prisms have $c_{\mathrm{triangle}}=c_1+\cdots +c_5$.
\item Pinning a vertex has $c_{\mathrm{pin}}=c_1+2$.
\end{itemize}
\end{thm}
The proof is given in the Appendix Section~\ref{ss_aBTP}.
It is unknown to the authors whether all infinitesimally rigid trusses are BTP-trusses.

An immediate consequence is that the trusses of triangulated domains are infinitesimally rigid.

\begin{cor}\label{cor_BTP}  Let $T$ be a triangulated truss such that all triangles are non-degenerate. Then $T$ is generic. 
Suppose that $T$ is built up starting from a single edge one step at a time by attaching two connected edges to form a triangle, such as gluing on a triangle to an outer edge, or by attaching a single boundary edge to two existing vertices, such as   gluing on a triangle to two existing edges, or such as connecting two vertices to surround a hole. 
The number of compatibility conditions is
 $n_b$, the number of times a single edge is glued to two vertices.
\end{cor}

\begin{proof}
The process of building the truss is just the BTP construction where triangles are made from the previous stage and two segments, and bigons are made from the previous stage and one segment. Each bigon increases the compatibility count by one.
\end{proof}

\section{Asymptotic Compatibility Conditions.}\label{ss_AC}

How do the relative area and the relative number of holes influence the asymptotic compatibility condition? For simplicity, we restrict consideration to periodic triangular structures.
Let the basic periodicity cell $\Upsilon$ by a $k\times k$ union of hexagons centered on $ae_1+be_2$ where $a,b=1,\ldots,k$.  Suppose there are $h$ holes per cell and $m$ interior vertices taken by each hole. For simplicity, we assume that cells are bounded by $h+1$ pairwise disjoint simple closed curves. Let $\Omega_n$ be the   $n\times n$ union of cells slightly overlapping, centered on $ake_1+bke_2$ where $a,b=1,\ldots,n$.

The asymptotic compatibility density is defined to be
\begin{equation}\label{eq_defAC}
AC =\lim_{n\to\infty}\frac {c(\Omega_n)}{ \Area(\Omega_n)}.
\end{equation}
The total number of holes is $g=n^2h$. The total number of interior vertices is
$$
v_i=k^2n^2-hmn^2.
$$
The area is base times height minus corner triangles, thus
\begin{equation}\label{eq_compAC}
AC =\lim_{n\to\infty}\frac {v_1+3g}{ \Area(\Omega_n)}
=\lim_{n\to\infty}\frac {[k^2n^2-hmn^2]+3n^2h}{ nk(nk+1)\frac{\sqrt 3}2}=\frac{k^2-hm+3h}{\frac{\sqrt 3}2 k^2}.
\end{equation}
This shows that the asymptotic compatibility depends not just on the total area of the holes removed from the cell. Taking out more holes of the same total are increases $AC$, a proxy for material resilience. What is the same, the influence of the holes in a triangular truss depends on the number of interior vertices removed.

For example, if the hole is a $p\times p$ rhombus, then there are $m=p^2+4p+2$ interior vertices removed, but the hole has area $2(p-1)^2$ triangles. If the hole is a $(p-1)^2\times 1$ trapezoid, it has the same number of triangles, but it has $m=2p^2-4p+4$ interior vertices removed.

\subsection{A Single Hole}

Suppose $X$ is an annular domain in the triangular lattice bounded by two disjoint, simple closed curves, an inner one $\gamma_1$ and an outer one $\gamma_0$.
The number of compatibility conditions is $v_i+3$ where $v_i$  is the number of interior vertices of $X$. We address here how much weaker the region with the hole is than the region bounded by just by $\gamma_0$ without the hole.

For sake of argument, let $Y$ denote the hole-region bounded by $\gamma_1$ and $Z=X\cup Y$ the region bounded by just $\gamma_0$. Let $\ell_0$ and $\ell_1$ be the number of boundary vertices on $\gamma_0$ and $\gamma_1$, resp., $v_0$, $v_1$ and $v_2$ the number of interior vertices of $X$, $Y$ and $Z$, resp.  Neighboring the boundary curve $\gamma_1$ are the inside and outside unit neighborhoods
$$
G_0=\{ v\in X: \dist(v,\gamma_1)\le 1\},\qquad G_1=\{ v\in Y: \dist(v,\gamma_1)\le 1\}
$$
A {\it collar}\rm\  of $\gamma_1$ is the domain $G_0\cup G_1$.
If $\gamma_1$ is smooth enough, then the $1$-neighborhoods are a sequence of triangles which form a girder-ring.  However, in general, a $1$-neighborhood may not even be an annular domain; it may contain additional girders and even hexagons.

Assume that $\gamma_1$ is a curve whose outer $1$-neighborhood is a girder-ring. This would be the case if the curvature exceeds $- 60^{\circ}$ at each vertex and no three consecutive vertices have angle  $-60^{\circ}$ with respect to the outer normal. The outer curve must be longer and must have $\ell_1+6$ edges. This description fits Steiner's formula for a parallel curve. Since $\gamma_1$ is a closed curve in the plane, if we traverse it in the counter-clockwise direction, then the total curvature is
$$
T=\sum_{v_i\in\gamma_1} \angle_{\gamma_1}(v_i)=360^{\circ}.
$$
Now imagine that $\gamma_1$ is a bicycle chain. Attach an outer triangle to each link. Then the angles between consecutive triangles are $\angle_{\gamma_1}(v_i)+60^{\circ}$. It follows that we can insert one triangle whenever $\angle_{\gamma_1}(v_i)=0$ for each link in the chain. In total, we can insert $\ell_1+T/60$ triangles to fill in the girder-ring.

Similarly, if the inner $1$-neighborhood is a girder-ring, then its inner edge has $\ell_1-6$ edges. This would be the case if the curvature does not exceed $ 60^{\circ}$ at each vertex and no three consecutive vertices have angle  $60^{\circ}$ with respect to the outer normal.

Note that if both $G_0$ and $G_1$ are girder rings, then an open hexagon about a vertex either lies entirely in the inside $\gamma_1$, entirely in the outside of $\gamma_1$  or its center is on a simple closed curve $\gamma_1$. 
This means that in the collared case,
$$
v_2=v_0+v_1+\ell_1.
$$

Now we can compare the number of compatibility conditions before and after drilling out the hole.
$$
c(X)=v_0+3,\qquad c(Y)=v_1,\qquad C(Z)=v_2.
$$
Thus, the number of compatibility conditions lost by drilling the hole is
\begin{equation}\label{eq_hole}
c(Z)-c(X)=v_1+\ell_1-3=c(Y)+\ell_1-3.
\end{equation}
Strength is not only lost by removing the interior particles of the hole but from the boundary points. The drop in the number of comptiblity conditions is the number of interior and boundary points lost to the hole plus three~\eqref{eq_hole}.

\paragraph{Small holes.}
Let us work out some examples. As a reality check, suppose that the hole is a single triangle $Y$. This hole should not change the truss since it is made of triangular holes. Its  removal results in the same number of compatibility conditions
$$
c(Z)-c(X)=c(Y)+\ell_1-3=0+3-3=0. 
$$

The smallest nontrivial inner boundary $\gamma_1$ is for the unit rhombus. The loss of compatibility dimensions is
$$
c(Z)-c(X)=c(Y)+\ell-3=0+4-3=1.
$$
For holes consisting of 3, 4 or 5  adjacent triangles, there are
 eight (up to reflection and rotation) different shapes with a simple boundary curve. They  have $\ell=5$, $6$ or $7$, resp., but $v_1=0$ for all of them so that $c(Z)-c(X)=2$, $3$, and $4$, resp. 

Of the twelve simple curves about six triangles,  the hexagon is unique. For the others $\ell=8$ and $c(Y)=0$ so that $c(Z)-c(X)=5$.  However, for the hexagon, $\ell=6$ and $v_1=1$. Hence, $c(Z)-c(X)=4$. Thus the shape of the hole influences how many compatibility conditions are lost.

\paragraph{Isoperimetric bound on the number of compatibility conditions lost by a single hole.}
For the  girder $Y$ with $p$ triangles, $c(Y)=0$, $\ell_1=p+2$ so 
$c(Z)-c(X)=p-1$. For a regular hexagon with side $p$, $\ell_1=6p$, $v_1=3p^2-3p+1$ so
$c(Z)-c(X)=3p^2+3p-2$. For a girder with the same number of triangles $6p^2$ as the hexagon,  we have $c(Z)-c(X)=6p^2-1$ which is a much weaker.

For a given length $\ell_1$, the range of losses of strength \eqref{eq_hole} is bounded on the one side by $v_1=0$ and on the other side by an isoperimetric inequality for domains in the triangular lattice.

\begin{thm} \label{th_isop} Let $\Omega\subset \mathcal H$ be a connected region bounded by (not necessarily simple) boundary curve of length $\ell$. Then $\Omega$ contains at most
$$
v_i\le \frac{\ell_1^2}{12}-\frac{\ell_1}2+1.
$$
hexagon centers.
\end{thm}
\begin{proof}
We find the region $\Omega$ with maximal number of interior vertices subject to  boundary length at most $\ell_1$.
First, observe that the maximizing region is a hexagon.  $\Omega\subset \Theta$ where $\Theta$ is a hexagon with not necessarily equal sides and boundary length at most $\ell$. To see it, consider the function $\wp_i(x)=n_i\dot x$, where $n_i$ is outward normal vector obtained by a $90^{\circ}$ rotation of the direction vector $e_i$ of the $i$th side. Let the support distance in the $n_i$ direction  be
$$
h_i=\max_{x\in\Omega} \wp_i(x)=\wp_i(z_i)
$$
 and $z_i$ be a point on the side that achieves this distance. Fill in the consecutive points by geodesics, for example, take the geodesic from $z_i$ to $z_{i+1}$ which goes in the $e_i$ direction first, and then the $e_{i+1}$ direction. The resulting figure is the desired hexagon $\Theta$. 

Now maximize the number of interior points of a hexagon.
Let us call the lengths of the sides $a,b,\ldots,f$ corresponding to the $e_1,e_2,\ldots$ directions. Observe that the length of any pair of adjacent sides equals the length of the opposite pair because both pairs connect  two parallel lines.
\begin{equation}\label{eq_oppo}
\begin{aligned}
a+b&=d+e\\
b+c&=e+f\\
c+d&= f+a.
\end{aligned}
\end{equation}
The last equation is the difference of the first two.
The centers of hexagons within $\Theta$ form a hexagonal pattern bounded by points just inside each edge, thus having $a,b,\ldots$ points on a side.
The objective function is the total number of points inside, which equals the area of the parallelogram between the $e_1$-$e_4$  and the $e_3$-$e_6$ parallel lines, minus the triangles in the $e_2$ and $e_5$ corners
$$
N=(a+b-1)(b+c-1)-\frac 12 b(b-1)-\frac 12(a+b-d)(a+b-d-1).
$$
The number of points an the bottom are the number of points on the ``$a$'' side plus those on the ``$b$'' side projected to the ``$a$'' line. The point in the corner is counted twice. The number of lattice points in an equilateral triangle with $b-1$ points on a side is the triangular number $\frac 12 b(b-1)$. Replacing $e$ and $f$ from \eqref{eq_oppo}, we obtain the total length 
$$
L=a+2b+2c+d.
$$
Maximizing $N$ subject to fixed length yields all sides equal length, so the area maximizing figure is the hexagon.

A hexagon with $a$ points on a side is made up of six triangles with side $a-1$ plus the point in the middle, yielding the formula.
\end{proof}

\begin{thm} Let $X$ be an annular domain bounded by disjoint simple closed curves. If $\ell_1\ge 4$ denotes the length of the inner boundary, then the compatibility loss may be estimated
$$
\ell_1-3\le c(Z)-c(X)=\frac{\ell_1^2}{12}+\frac{\ell_1}2-2.
$$
\end{thm}
The result follows from applying the isoperimetric estimate, Theorem~\ref{th_isop} to \eqref{eq_hole}.

\subsection{How geometry of a hole influences asymptotic compatibility}

How much do holes weaken a hexagonal lattice? We assume that the lattice is periodic and compute the large-scale average compatibility condition density for damaged material relative to the undamaged material.

  Note that removing a single edge reduces the number of interior vertices by four, but introduces a ring girder which supports three compatibility conditions. Thus $m-3\ge 1$ compatibility conditions are lost  for each hole. \eqref{eq_compAC} becomes
$$
AC =\frac{k^2-h(m-3)}{\frac{\sqrt 3}2 k^2},\qquad
AC_{\text{many $1$-link holes}}=\frac{k^2-p^2+2p-1}{\tfrac{\sqrt 3}2 k^2}.
$$
The asymptotic compatibility is $AC=\frac 2{\sqrt 3}\cong 1.15447$ for the triangular material without holes. Now we compare the removal of a total area of $18$ triangles with different hole configurations: nine separated $1\times 1$ rhombi, one $3\times 3$ rhombus or one $1\times 9$ rhombus in a $13\times 13$ period cell. Thus  $(h,m)=(9,4),(1,16),(1,20)$ and $AC\cong 1.0932$, $1.0659$ and $1.0385$, resp.  If we remove the largest $11\times 11$ hole we have $(h,m)=(1,144)$ and $AC\cong 0.1913$. Thus $AC$ decreases as our intuition of the strength of these damaged materials decreases.
\begin{figure}[h]
      \begin{center}
          \scalebox{0.5}{\includegraphics{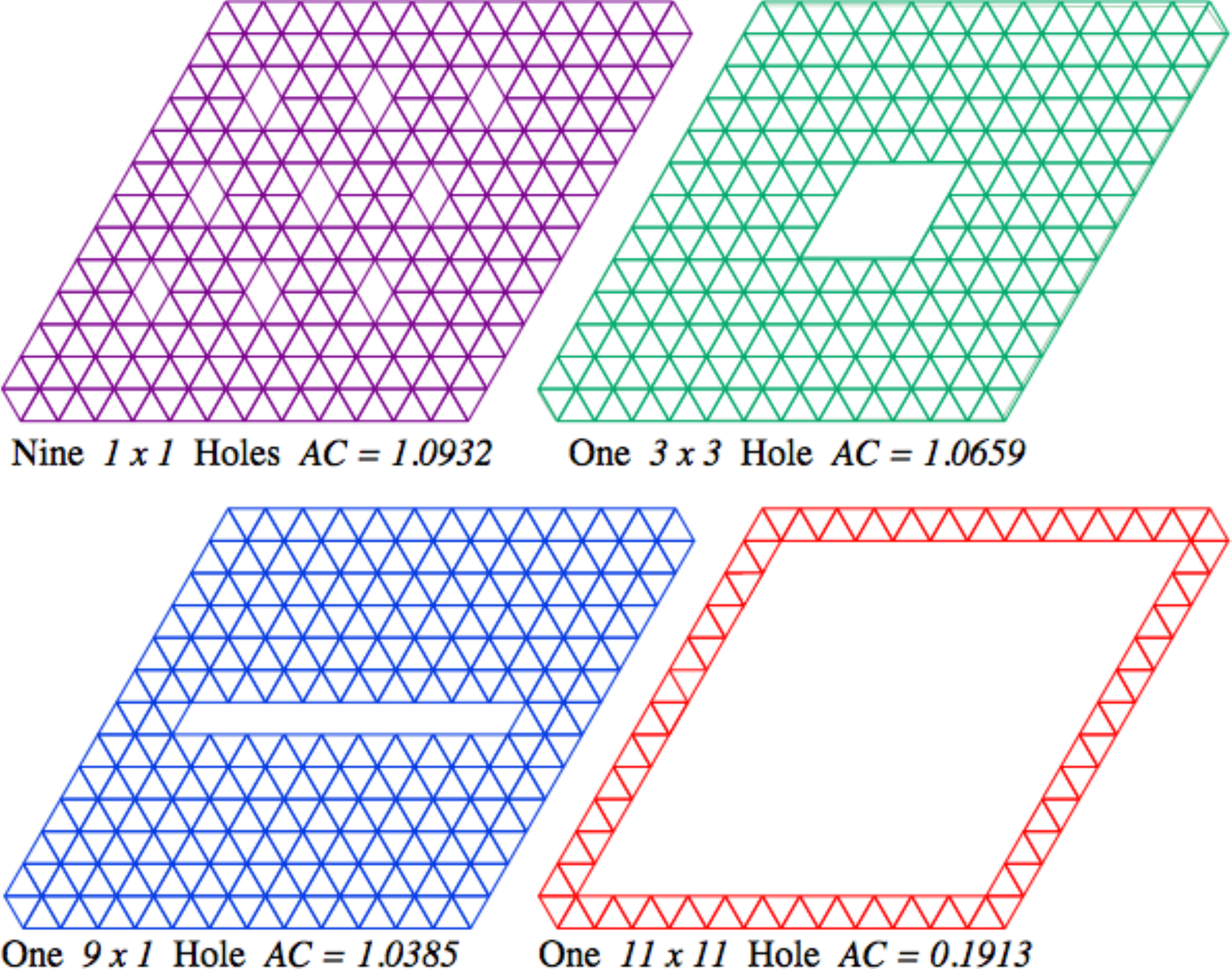}}
 \caption{$13 \times 13 $  Cell with Holes of Area $18$ Triangles, $(h,m)=(9,4),(1,16),(1,20),(1,144)$.}
 \end{center}
       \end{figure}

In this example, if one link is removed, $m=4$ and triangle has area $2$ triangles. Removing $h=(p-1)^2$ one-links removes $m=4$ interior vertices per hole and has the same area $2(p-1)^2$ triangles as the  $(p-1)\times (p-1)$ rhombus, which removes $h=1$ hole and  $m=p^2$ interior vertices. If the hole is a $(p-1)^2\times 1$ trapezoid, it also has the same  number of triangles, $h=1$  but  removes $m=2p^2-4p+2$ interior vertices.
$$
AC_{\text{rhombus}}=\frac{k^2-p^2+3}{\frac{\sqrt 3}2 k^2},\qquad
AC_{\text{trapezoid}}=\frac{k^2-2p^2+4p-5}{\frac{\sqrt 3}2 k^2}.
$$
The asymptotic compatibility depends not just on the total area  removed from the cell. Taking out more holes of the same total area has larger $AC$, a proxy for material resilience.   In other words, one large hole weakens the material more than a number of small holes of the same total area. A similar phenomenon was observed by Cherkaev and Ryvkin~\cite{CRa},~\cite{CRb}.

\section{Discretization of the nonlinear continuum problem (NC).}
\label{ss_nc} 

We consider the approximation of the prescribed Green Tensor problem by discrete structures in two-dimensional materials. Discrete approximations of the prescribed strain equation and other elasticity problems are sometimes justified as discretizations of partial differential equations  via  finite elements\cite{Bd:2007},~\cite{BS:2010} or discrete differential forms\cite{AFW:2006}. In this section, we recall another justification using the theory of approximating Riemannian metrics by piecewise linear metrics, first developed by A.~D.~Alexandrov\cite{AZ:1962}.

 The prescribed Green (deformation) tensor $\zeta$ gives a Riemannian metric. The unknown is a mapping to Euclidean space which pulls back the Euclidean metric to the prescribed metric $\zeta$. Its compatibility condition is that  Riemannian curvatures of both metrics agree, that is they both vanish. We discuss its approximation by the discrete prescribed length problem an approximating triangulated discrete structure whose lengths are determined from $\zeta$.   Associated to a triangulated structure is a piecewise Euclidean surface made by filling in triangles with pieces of Euclidean planes in such a way that the lengths of links correspond to Euclidean distances.  It is also the continuous piecewise linear finite element map on the triangulation. The resulting piecewise Euclidean metric approximates $\zeta$.
  The global existence problem for prescribed Green tensor may be solved by solving the discrete prescribed length problem and taking the limiting configuration.
 In a similar vein,     configurations of sequences of converging discrete structures converge to a continuous material. The compatibility condition, the vanishing of curvature atoms of the structures, also converges to the continuum compatibility condition of the limiting structure.

\subsection{Nonlinear Prescribed Green Deformation, Cauchy Tensor, Compatibility.}\label{ss_NPGDCTC}
The geometric interpretation of the prescribed Green Deformation equation allows an interpretation of the discrete prescribed length problem as an approximation.
For (NP) to be solvable, we assume first the nondegeneracy condition that the prescribed $\zeta$ is a symmetric positive definite function. Hence it is a Riemannian metric for $\mathcal B$. Viewing $\phi:\mathcal B\to\mathcal S\subset \mathbf R^2$  as curvilinear coordinates for $\mathcal S$,  then $\sum_iF_A^i F_B^i\, dX^A\otimes dX^B$ is the Euclidean metric of $\mathcal S$. The geometric interpretation of~\eqref{eq_pullba}  is that $\phi$ is an isometric immersion of a Euclidean domain to the Euclidean plane which pulls back the Euclidean metric to  $\zeta$. The fact that $\phi$ pulls back a Euclidean metric to a Euclidean metric implies the integrability condition (compatibility condition for prescribed Green Deformation tensor equation)  that the Riemannian curvature tensor of $\zeta$ vanish on $\mathcal B$~\cite[p.~25]{E:1925},  \cite[p.~81]{MH:1983}. 
Equivalently in two dimensions, the  scalar Gauss Curvature of
$\zeta$  vanishes. The Gauss Curvature may be expressed from known formulas \cite[p.~114]{S:1961}. Putting $D^2=\zeta_{11}\zeta_{22}-\zeta_{12}{}^2$, the Gauss curvature is
\begin{align*}
\mathcal K&=\frac 1{2D^2}\left( - \zeta_{22,11}+2\zeta_{12,12}-\zeta_{11,22}\right)\\
&\quad +\frac {\zeta_{22}}{4D^4} \left( \zeta_{11,1}\zeta_{22,1} -2\zeta_{11,1}\zeta_{12,2}+ \zeta_{11,2}{}^2\right)+\frac {\zeta_{11}}{4D^4} \left( \zeta_{11,2}\zeta_{22,2}-2\zeta_{12,1}\zeta_{22,2}+\zeta_{22,1}{}^2\right)\\
&\quad+\frac {\zeta_{12}}{4D^4} \left(  -2\zeta_{12,1}\zeta_{22,1}-2\zeta_{11,2}\zeta_{12,2}+ 4\zeta_{12,1}\zeta_{12,2}- \zeta_{11,2}\zeta_{22,1}+ \zeta_{11,1}\zeta_{22,2}\right).
\end{align*}
Note that the compatibility equation $\mathcal K\equiv 0$ has linear highest order term which is the compatibility condition of the linearized equation.

The integrability condition implies local solvability of the prescribed metric equations. A nice exposition  is in \cite[p.~242]{L:1926}. The existence of a deformation with prescribed right Cauchy Green tensor is further discussed by Shield~\cite{Sh:1973} and by Blume~\cite{Bl:1989}. Integrability condition of the linearized prescribed strain equation implies locally solvability similarly ~\cite{So:1956}.
The global solubility of (LC) may be deduced from the discrete approximation.

\subsection{Approximation (NC) equation by (ND) the nonlinear discrete prescribed length problem.}\label{ss_NCapprox}

Of several ways to see how the discrete nonlinear problem is approximating the nonlinear problem, we exploit the geometric interpretation of the prescribed Green deformation problem to give a geometric notion of the weak convergence, due to A.~D.~Alexandrov.

\paragraph{Triangulated Structures.}
Recall that a triangulated structure $(\mathcal D,\ell_{ij})$ is combinatorially the nodes and links of a piecewise linear triangulation of a simple connected planar domain $\mathcal D$ with an assignment of a positive length $\ell_E$ to each edge $E$ of $S$ (Section~\ref{def_str}). 

An approximation to $\phi:\mathcal B \to \mathcal S$ is obtained by taking a closed approximating triangulated disk $\mathcal D\subset (\mathcal B, \zeta)$ which is subdivided into sufficiently fine straight line segments, which is sometimes called a {\it polyhedral}\rm\  or {\it piecewise linear (PL)}\rm\  approximation. For each segment let $\ell_{ij}$ be the $\zeta$-distance between the endpoints of the segment. Assuming that these lengths are smaller than the injectivity radius of $\zeta$ on $\mathcal D$, the lengths are then the $\zeta$-geodesic distances between the vertices. It means that there is a unique distance realizing  $\zeta$-geodesic in $\mathcal B$ between any two linked vertices. The $\zeta$-geodesics may not agree with the straight triangle sides (of the background Euclidean metric), so in general, the $\zeta$-length of a straight side from $V_i$ to $V_j$ may be greater than $\ell_{ij}$. By Alexandrov's theory, the solutions of the discrete prescribed length equations for a   sequence of such approximating triangulated structures whose triangle diameters tend to zero converges weakly to the solution of the nonlinear problem.

\paragraph{Can the abstract structure be realized as a Euclidean configuration?}
Let us assume that $\mathcal B$ is a PL triangulated disk for which we are given prescribed lengths. 
Assuming strict triangle inequalities hold, we can construct a nondegenerate straight edge triangle $\mathcal T$ in the  Euclidean plane whose side lengths equal the given $\ell_{ij}$  distances between vertices. The affine map $\phi:T_{ijk}\to \mathcal T$ between the two-dimensional triangles determines a Euclidean metric in $T_{ijk}$ by pulling back the Euclidean metric $g_{\mathcal B}=\phi^*(ds^2_{\mathbf E^2})$, which is triangle-wise Euclidean. The induced Euclidean metrics on each triangle glue together to give a metric for the abstract structure $\mathcal B$, which is smooth except possibly at the vertices and continuous everywhere. Note that it may happen that the total Euclidean angle at the interior vertices may not be $2\pi$. The angle deficit called an {\it atom of curvature,} is regarded as a point mass at $V_i$ whose mass is given by
$$
K(V_i)=2\pi-\sum_{\text{$V_i\in T_{ijk}$}}\angle( V_j,V_i,V_k ),
$$
where $\angle (V_j,V_i,V_k)$ is the Euclidean angle between the vectors $V_iV_j$ and $V_iV_k$ at $V_i$. The abstract metric induces an intrinsic distance function between pairs of vertices $V,W\in\mathcal S$, namely
$$
\rho_{\mathcal B}(V,W)=\inf\{\operatorname{length}(\sigma):\text{paths $\sigma$ in $\mathcal B$ from $V$ to $W$}\}
$$
where the infimum is taken over all piecewise smooth curves in $\mathcal B$ from $V$ to $W$ and length is in the piecewise Euclidean metric. The minimizers have to be paths of piecewise linear segments with possible bends only when the paths cross edges. In fact, possible kinks occur only at the vertices because at interior points of edges have full Euclidean neighborhoods where the local length minimizers are straight lines. The metric of $g_{\mathcal B}$ may be recovered from $\rho_{\mathcal B}$. 

The curvature measure $\omega$ is a Borel measure supported on the vertices in $\mathcal B$, and may be given   by
$$
\omega(G)=\sum_{V_i\in G} K(V_i)
$$
for Borel sets $G\subset \mathcal B$.
This notion of curvature of an abstract structure $(\mathcal B,\ell_{ij})$ is due to
A.~D.~Alexandrov \cite[p.~496]{A:1948}, \cite[p.~156]{AZ:1962}. It applies, more generally,  to {\it surfaces with bounded curvature} \cite[p.~6]{AZ:1962} for which, starting from the notion of a  distance function, lengths of arcs, length minimizing (geodesic) arcs, angles between geodesics at a point, angle deficits of geodesic triangles, the curvature measure of a Borel set may be defined. \footnote{Nowadays, geometers are familiar with offspring theory, ``Gromov Convergence'' of length spaces.}

A surface with $\mathcal C^2$  Riemannian metric with bounded Gauss curvature $\mathcal K(x)$ may also be regarded by Alexandrov as a surface of bounded curvature.  The induced distance is the usual one obtained by minimizing the length of paths. The curvature measure is
$$
\omega(G)=\int_G \mathcal K(x)\, dA
$$
taken with respect to induced area form. If we have a sequence of triangulations for a fixed domain $\mathcal B$, then {\it weak convergence in the sense of Alexandrov} means that the induced distance functions converge uniformly of $\mathcal B$. \footnote{Alexandrov's sense of convergence is also used in the context of Monge-Ampere equations, where there is an extensive regularity theory for  Alexandrov solutions (limits of solutions of discrete approximating problems) whose convergence agrees with convergence in the viscosity sense \cite{G:2016}.}

\paragraph{(NP) existence problems for structures.}
Consider the {\it realizability problem:} given an abstract triangulated disk structure $(\mathcal B,\ell_{ij})$ (net for short), is it possible to build a configuration,  a continuous immersion $\phi:\mathcal B\to \mathbf E^2$ such that $\phi$ is linear on each triangle and is a local isometry? In other words, can we develop $\mathcal B$ into the Euclidean plane by gluing together the plane triangles in such a way that at each interior vertex the angles all close up? The answer is clearly ``yes'' assuming that the sum of the angles at every interior vertex is $2\pi$. Globally, the development may wind around and overlap. The overlap may even occur at a single boundary vertex as at a branch point in a Riemann Surface. The metric is obtained by pulling back the Euclidean metric from a single planar region $\phi(\mathcal S)$. Such a configuration then realizes the prescribed lengths of edges.
In analogy to the continuum case, the existence of a configuration of a structure
can be possible if and only if angles at each interior vertex add up to $2\pi$, in other words, the curvature atoms vanish: the structure is {\it flat.} 

\paragraph{Argument using Alexandrov's Polyhedral Realization Theorem.} In simple cases, the existence and uniqueness may be deduced from Alexandrov's theorems about the realizability of polyhedral metrics of the sphere as boundaries of convex polyhedrons.
For a flat disk, if the boundary is also convex, {\it i.e.,} the interior angles of boundary vertices do not exceed $\pi$ then the answer is yes. The sum of the interior angles at the boundary is $2\pi$. Hence the boundary curve may be realized as the boundary of a convex polygon $\Pi$  in the plane. By adding this polygon to the net forming the back side, we get  enough polygons to make a two sided disk which is  topological sphere. All vertices are interior vertices in the two-sided disk. The curvature atoms are nonnegative and add up to $4\pi$ since they are be supported on the vertices of $\Pi$, with total curvature equal to the $2\pi$ contributions from both the back and the front. By Alexandrov's theorem for realizing polyhedral metrics \cite[p.~184]{A:1950}, there is a (degenerate) two-sided flat convex polyhedron in $\mathbf E^3$ whose boundary metric gives the abstract net metric. Omitting the back side corresponding to the  polygon $\Pi$ gives the desired realization. One can generalize to nonconvex boundary by first adding triangles to fill in the convex hull of $\mathcal B$ and then doubling as before. However, this works only for embedded polygons, and not for general nets.

\paragraph{Realizing abstract triangulated structures.} Intuitively, the unique existence of immersion of the abstract structure is evident. The mapping of the first triangle may be moved by rigid motion or reflection to an arbitrary position, but after that, the continuation of the PL immersion of the abstract structure is uniquely determined by the gluing.
The result may only be immersed: namely the image under $\phi_n$ may have self-intersections: the abstract structure may wrap around so that part of the image crosses itself.

We begin by formulating a realization lemma of Alexandrov for a single abstract structure \cite[p.~71]{A:1950}. It will be used to solve the global realizability of a given smooth metric $\zeta$ as well as the existence of a limit for a sequence of abstract structures.

 \medskip
\noindent{\bf Theorem~\ref{th_realize}.\ }{\it Let $(\mathcal B,\ell)$ be an abstract structure which is the $1$-skeleton of a triangulated PL disk $\mathcal B$ with an assignment of edge lengths. Assume that the lengths satisfy a strict triangle inequality and that it has zero curvature atom at each interior vertex. Then there is a  configuration $\phi:\mathcal B\to\mathbf E^2$ that realizes the structure, which says for every $E_{i,j}$, an edge from $V_i$ to $V_j$, then}\rm
$$
| \phi(V_i)-\phi(V_j)|_{\mathbf E^2}=\ell_{i,j}.
$$

\medskip

This relation shows that the pulled back Euclidean metric agrees with the polyhedral metric.
Moreover, the configuration is unique up to rigid motion, which means, if $\tilde\phi$ is another configuration, then there is an isometry (rigid motion) $\mathcal I$ of $\mathbf E^2$ so that $\tilde\phi=\mathcal I \circ \phi$.
The proof is given in the Appendix.

\subsection{Solutions of the Prescribed Green Tensor Equation (NC)}\label{ss_solnNC}

In the first application of the Theorem~\ref{th_realize}, we use the existence of local solutions of (NP) and glue them together to give a global solution. To this end, we assume that $\zeta$ is sufficiently regular and that for each point $X\in\mathcal B$ there is a neighborhood $\mathcal U_X$ such that $(NP)$ may be solved in any geodesic triangle $T_{ijk}\subset \mathcal U_X$ whose vertices are in the neighborhood. It means that there is a 
$\phi_{ijk}\in \mathcal C^{1,\alpha}$ that solves $(NP)$ on $T_{ijk}$. Since the metric s flat, the interior angles are determined by side lengths via cosine law, and thus add up to $2\pi$ going around a vertex because the sum of angles of curves emanating from such vertex do. Note that $\{ U_X\}_{X\in\mathcal B}$ is an open cover of $\mathcal D$.

Suppose we are given a bounded domain $\mathcal B\subset\mathbf E^2$ and a smooth positive definite symmetric matrix function $\zeta_{ij}$ defined on a neighborhood of $\overline{\mathcal B}$ whose curvature is everywhere zero. For simplicity, we may assume that $\mathcal B$ is a geodesic polygon in the $\zeta$ metric. We shall construct a continuous mapping $\phi:\mathcal B \to \mathcal S=\mathbf E^2$ by continuing local solutions. We shall show that it solves $(NP)$ on $\mathcal D$.

We begin by showing that there is an approximating sequence of triangulations to a given $(\mathcal B,\zeta)$ disk with metric.   Start with an initial geodesic triangulation of ${\mathcal B}$ with triangles of diameter less than the injectivity radius of $\zeta$. By barycentric subdivision obtain a sequence of triangulations whose maximum diameter tends to zero.  After finitely many subdivisions, we reach a triangulation whose triangle diameters are smaller than the Lebesgue number of the cover $\{ U_X\}$. We arrange that the edges of the triangles are all minimizing geodesics whose length equals the distance between ending vertices. We also require that triangles be nondegenerate  such that no side length equals the sum of its other two side lengths (has a strict triangle inequality). The last amounts to requiring that the third vertex not be on the geodesic determined by the other two. This can always be arranged vertex-wise by making a small perturbation of the proposed new vertex position.  Call a sequence of nondegenerate triangulations $\mathcal B_{\ell}$, with the property that  the largest diameter of a triangle in $\mathcal B_{\ell}$ tends to zero as $\ell\to\infty$.

In the realization problem for a metric, the barycentric subdivision results in  vertices of $\mathcal B_{\ell}$  that are also vertices of $\mathcal B_{\ell +1}$.
 For this purpose,  the barycentric subdivision of $\mathcal B_{\ell}$ may require a small perturbation of the middle point if needed as explained above.    In this case, in each triangle, there is a Euclidean metric which coincides with $\zeta$ restricted to the edges. For the sufficiently fine subdivision, we have a geodesic triangulation of $\mathcal B_m$. Each triangle comes with an isometry $\phi_T:T\to\mathbf E^2$ determined up to a rigid motion. For example, corresponding points use Fermi coordinates in $T_{ijk}$ and $\mathcal T$. By composing with an appropriate rigid motion $\mathcal R$, the map $\mathcal R \circ\phi_T$ can be pasted to smoothly extend the map built by pasting the maps for triangles one at a time. We have proved the global existence theorem for (NP).

\begin{lemma} \label{lem_existenceNC} Let $\mathcal B$ be a bounded open topological disk in $\mathbf E^2$  with a prescribed $\mathcal C^2$ positive definite matrix function $\zeta$ satisfying the $K\equiv 0$ compatibility condition  defined in a neighborhood of $\overline{\mathcal B}$ with induced metric and  distance $\rho$ such that the boundary $\partial \mathcal B$ is piecewise $\zeta$-geodesic.    There is a sequence of $\zeta$-geodesic triangulations, call them $\mathcal B_k$  such that the largest diameter of the triangles of $D_k$ tends to zero. For $k$ sufficiently large, the induced triangle-wise Euclidean metric and corresponding induced isometry $\mathcal B\to\mathbf E$ solves (NP). 
\end{lemma}

\subsection{The discrete  problem (ND) as an approximation to the continuum problem (NC).}
\label{ss_NDapproxNC}

By pasting together local solutions we obtained a global solution to (NP). However, the construction required knowing geodesic triangles of $\zeta$ so that the flat structure on each $T_{ijk}$  agreed with the flat metric $\zeta$. If we used the PL straight line triangulation of a polygon $\mathcal B\subset  \mathbf E^2$ instead, the edges of the triangles would no longer be geodesics in the $\zeta$ metric. We still define $\ell_{ij}=\rho_{\zeta}(V_i,V_j)$ to be the $\zeta$-distances between vertices, but then the PL-edge segments, not being geodesics will, in general, have a $\zeta$-length greater than $\ell_{ij}$. Thus if we construct the polyhedral metric on the PL disk, it will be flat
because angles at the vertices are determined by triangles in the flat metric $\zeta$. However, the PL metric will not necessarily agree with the $\zeta$-metric in the triangle. This metric still approximates the $\zeta$-metric and its isometric embedding to $\mathbf E^2$ approximates the isometric embedding of the $\zeta$-metric. In this sense, the discrete nonlinear problem approximates the continuum (NP).

\begin{lemma} \label{lem_approx2} Let $\mathcal B$ be a bounded PL triangulated  disk in $\mathbf E^2$  with a prescribed $\mathcal C^2$ positive definite symmetric matrix function $\zeta$  defined in a neighborhood of $\mathcal B$ satisfying the compatibility condition $\mathcal K\equiv 0$  with induced metric and  distance $\rho$ such that the boundary $\partial \mathcal B$ is piecewise $\mathbf E^2$ straight line segments.    There is a sequence of PL triangulations, call them $\mathcal B_k$  such that the largest diameter of the triangles of $\mathcal B_k$ tends to zero. The polyhedral metrics approximate $\zeta$ in the sense of Alexandrov: the induced distance functions $\rho_i\to\rho_{\zeta}$ uniformly in $\mathcal B\times\mathcal B$. After composing by Euclidean isometries if necessary, the PL isometries $\phi_i:\mathcal B_i\to\mathbf E^2$ tend to $\phi_{\zeta}:\mathcal B_i\to\mathbf E^2$ in $\mathcal C^1$ to a solution of (NP). 
\end{lemma}

\begin{proof} Let $\mathcal B$ may be triangulated by PL triangles such that the $\zeta$-diameter of each $T_{ijk}$ is less than the injectivity radius of $\zeta$ and such that  strict triangle inequalities hold. Call this triangulation $B_1$. Call the successive good PL Barycentric subdivisions $B_k$. 
The maximal diameter of the triangles tends to zero as $k\to\infty$. Since $\zeta_{ij}$ is bounded and uniformly positive on $\mathcal B_1$, The $\zeta$-diameters also tend to zero

Fix a point $P_0$ and a direction $e_0$ in the plane. If we designate a basepoint $V_1$ and direction $e$ at $V_1$ common to all triangulations $\mathcal B_{\ell}$, we may use a rigid motion on the configuration $\phi_{\ell}$ constructed to arrange that the point, direction and orientation agree: $\phi_{\ell}(V_1)=P_0$, $d(\phi_{\ell})[V_1](e)=e_0$ and $d(\phi_{\ell})[V_1]$ is orientation preserving.
By assumption or construction, the sum of the angles of triangles adjacent to an interior vertex adds up to $2\pi$.
As before, we can find a development of $(\mathcal B, \zeta_{\ell})$ into the plane and construct the PL mapping $\phi_{\ell}$ which is an isometry to $\mathbf E^2$ on each triangle. 
 We claim that the sequence $\phi_{\ell}$ converges to a limiting map $\phi:\mathcal B\to \mathbf E^2$ which solves the (NC) for the limiting structure.

First, $\phi_{\ell}$ are uniformly Lipschitz. For abstract configurations the $\phi_{\ell}$ are local isometries.  For the realization problem, this follows from the fact that $\zeta_{ij}$ is uniformly bounded by $\Lambda^2\delta_{ij}$, where $\Lambda^2$ is the supremum of all eigenvalues of $\zeta_{ij}(x)$ for all $x\in\overline{\mathcal B}$. It follows that the induced metric $\rho$ is $\sqrt 2\Lambda$-Lipschitz on $\overline{\mathcal B}$ and $\phi_{\ell}$ is $\Lambda$ Lipschitz. 
$$
\rho(x,y)=\int_0^1\sqrt{\zeta_{ij}((1-\tau)x+\tau y)(y^i-x^i)(y^j-x^j)}\, d\tau
\le \Lambda|x-y|
$$
so that in case $\rho(x,y)\ge\rho(x',y')$,
\begin{align*}
|\rho(x,y)-\rho(x',y')|&\le \rho(x,x')+\rho(x',y')+\rho(y',y)-\rho(x',y')\\
&\le  \rho(x,x')+\rho(y',y)\le\Lambda(|x-x'|+|y-y'|)\le \sqrt 2\Lambda|(x-x',y-y')|
\end{align*}
If  $\rho(x,y)\ge\rho(x',y')$, the same inequality follows by swapping roles of $(x,y)$ and $(x',y')$.
Since the maximum stretch in a linear map on a triangle is along one of the edges, $\rho_{\ell}(x,y)\le\Lambda|x-y|$ when restricted to a triangle of $\mathcal B_{\ell}$. If $\sigma_{\ell}$ is a straight line  from $x$ to $y$ in $\overline{\mathcal B}$, let $x=x_1,x_2,\ldots,x_n=y$ be the intersections of $\sigma_{\ell}$ with the edges of the triangles, then by the triangle inequality,
$$
|\phi_{\ell}(x)-\phi_{\ell}(y)|\le\sum_{i=1}^n|\phi_{\ell}(x_i)-\phi_{\ell}(x_{i-1})|
\le \Lambda\sum_{i=1}^n|x_i-x_{i-1}|=\Lambda|x-y|
$$
because $\sigma$ is length realizing in the background Euclidean metric.

It follows that the sequence of maps $\{ \phi_{\ell}\}$ is uniformly bounded and  Lipschitz. Hence a subsequence converges uniformly to $\phi$ which is Lipschitz. In the realization problem, because the functions $\phi_{\ell}$ and $\phi_{\ell'}$, agree on the vertices of $\mathcal B_{\ell}$ (say $\ell\le\ell'$) are uniformly Lipschitz, and diameters of triangles tend to zero, the limit of any two subsequences exists and is equal, hence the whole sequence $\phi_{\ell}$ converges to $\phi$. By the same argument, the whole sequence $\rho_{\ell}\to\rho_{\infty}$ as $\ell\to\infty$.

The curvature atoms are all zero at interior vertices because $\zeta$ is a Euclidean metric on geodesic triangles. Let $V$ be an interior vertex. On the one hand,   the angles between the tangent vectors of triangle sides emanating from an interior vertex of any Riemannian surface add up to $2\pi$. Thus the angles of the $\zeta$ geodesic triangles add to $2\pi$. Because the $\zeta$-geodesic triangles are flat, their angles may be given from side lengths by the cosine law. On the other hand, sides of the PL triangles in the induced PL metric is also determined from the lengths of the sides, hence results in the same angle as for the geodesic triangle.

As these are disks with bounded curvature whose maximum triangle diameters converge to zero, by a theorem of Aleksandrov~\cite[p.79]{AZ:1962}, the metrics converge uniformly $\rho_k\to\rho$. Moreover, the curvature measure converges to the curvature measure of the limiting structure. Thus $\phi_i$ converges weakly in the sense of Alexandrov to a solution of (NP).

Generalizing the approximation sequence for $(\mathcal B_k,\zeta)$, we now consider any convergent sequence of structures whose $\mathcal B$ triangulations become infinitely fine. We assume only that the polyhedral distance functions $\rho_i$ converge uniformly. This limit turns out to be a surface with bounded curvature in the sense of Alexandrov. The configurations of the piecewise Euclidean structures converge to a configuration of the limiting structure. The structures are approximations of this limiting structure. Applied to the approximations from Lemma~\ref{lem_existenceNC}, the limiting configuration solves the global realizability problem for the (NP). Moreover, the compatibility conditions (the flatness of the structures) converges to the compatibility assumed for $\zeta$ (the vanishing of curvature).

The approximate distances $\rho_{\ell}$ converge to the limiting distance $\rho$ which follows from a theorem of \cite[p.~79]{AZ:1962}. In the  $\zeta$-geodesic convex polygon, let $x,y\in\mathcal B$ be any points. Then there holds
$$
-c_1 d \le \rho_{\ell}(x,y)-\rho(x,y)\le c_1d
$$
where $d$ is the largest $\zeta$-diameter of the triangles in the triangulation $\mathcal B_{\ell}$, and the constant $c_1=2+(n-2)\pi$ depends on the number of vertices of $\mathcal B$. (In fact, Alexandrov-Zalgaller's estimate allows arbitrary Euler Characteristic of $\mathcal B$ and arbitrary bound of integral curvature, in Aleksandrov's sense of manifolds with bounded curvature. In our setting, we assume $\mathcal B$ is homeomorphic to a disk and that the PL surfaces are flat.) Thus the sequence of distance functions $\rho_{\ell}$ converges to a limiting function $\rho$ which is a distance function. 

In the realization problem, we claim that $\rho_{\infty}=\rho$ induced from $\zeta$.
The metrics are pullbacks of induced metrics $\zeta_{\ell}=\phi_{\ell}^*ds^2_{\text{Euclidean}}$ which induce distance functions, call them $\rho_{\ell}=\phi_{\ell}^*\rho_{\text{Euclidean}}$. Because $\phi_{\ell}\to\phi$ uniformly on $\mathcal B$, the distance functions converge to the induced metric $\phi^*\rho_{\text{Euclidean}}$. Both limits agree so the limiting map is a solution in Alxandrov's sense of (NC) $\rho=\phi^*\rho_{\text{Euclidean}}$.
Since the limiting map preserves the distance functions, the map itself may be recovered by trilateration. Since a point is uniquely determined by its distance to three general nearby points, it is given by rigid motion in Euclidean coordinates.
This also implies that $\phi$ is $\mathcal C^1$. \end{proof}


\section{Compatibility conditions of (LD) imply those of (LP).}

The linearization of (ND) led to (LD) rather than the discretization of (LC). In this section, we verify that both the interior compatibility condition and the boundary compatibility condition for (LD) implies the compatibility conditions for (LC). We shall consider the limit of a truss as it approximates the continuous material. We shall show that the compatibility condition on the elongations which are induced by a given strain field approximate the compatibility condition for the prescribed strain problem.

\subsection{The wagon wheel condition of (LD) implies the interior compatibility condition of (LC)}

Suppose $\mathcal B\subset\mathbf E^2$ is a PL triangulated disk. Consider the problem determining an infinitesimal  deformation  $u:\mathcal B\to\mathbf E^2$ by prescribing the strains
\begin{equation}\label{eq_prestr}
\text{(LD)}\qquad\qquad\qquad\qquad\qquad\qquad
\frac 12\left(\frac{\partial u^i}{\partial x_j}+\frac{\partial u^j}{\partial x_i}\right)=\epsilon_{ij}\qquad\qquad\qquad\qquad\qquad\qquad
\end{equation}
where $\epsilon_{ij}=\epsilon_{ji}$ is a given symmetric strain field. Were such  $u$ to exist, because it is a map of  Euclidean Spaces  the strain field must necessarily satisfy the {\it continuum compatibility condition} in $\mathcal B$
\begin{equation}\label{eq_2ccc}
\operatorname{Ink}(\epsilon)=\epsilon_{11,22}-2\epsilon_{12,12}+\epsilon_{22,11}=0
\end{equation}
where $\epsilon_{ij,pq}=\frac{\partial^2\epsilon_{ij}}{\partial x_p\,\partial x_q}$,~\eqref{eq_deriveCC}. 
This statement is the linearized equivalent of saying that the pulled back metric of a map between Euclidean Spaces corresponds to a vanishing Riemann curvature.

The infinitesimal deformations equations of a hexagon, $Au=\Lambda$ is a discretization of the continuum equations for prescribed strain
Its compatibility equation approximates continuum compatibility.

\begin{thm} [Expansion of Compatibility Equation for Regular Hexagons]\label{th_pedja} Let $\mathcal B_{3r}\subset \mathbf R^2$ be a disk radius $3r$ about~$0$ and $\mathcal H\subset\overline{\mathcal B_{2r}}$ be a regular hexagon with side length $\delta\le r$ containing~$0$. Let $u\in\mathcal C^4(\mathcal B_{3r},\mathbf R^2)$ be an infinitesimal deformation satisfying the strain equation (\ref{eq_prestr}). 
The wagon wheel condition~\eqref{eq_WW} for the $u$-induced rates of change of distances between vertices of $\mathcal H$ has the Taylor expansion about the origin 
\begin{equation}\label{eq_ccexpansion}
\mathcal W=-\frac 34\left(\epsilon_{11,22}-2\epsilon_{12,12}+\epsilon_{22,11}\right)\delta^2+ 0\cdot \delta^3+\operatorname{o}(\delta^3)
\end{equation}
as $\delta\to 0$  uniformly in $\overline{\mathcal B_{2r}}$ depending on
$\| u \|_{\mathcal C^4(\overline{\mathcal B_{2r}},\mathbf R^2)}$.
So if the discrete compatibility condition $\mathcal W=0$ holds for all $\delta$, then the continuum compatibility conditions~(\ref{eq_2ccc}) hold.
\end{thm}

The conclusion holds for points centered anywhere inside the hexagon. Thus maintaining the wagon wheel condition for refined grids approximates the continuum compatibility condition.
As we remarked after~\eqref{eq_affinehex},
for affine hexagons, the compatibility equation is the wagon wheel condition weighted by the respective side lengths
$$
0=\textstyle\sum_{i=0}^5 \alpha_i\,\dot\alpha_i-\sum_{i=0}^5 \beta_i\dot\beta_i.
$$
The theorem continues to hold for affine hexagons. 
A similar compatibility condition~\eqref{eq_genWW} holds with more complicated weights.
We expect that the expansion of both of these in $\delta$ has the continuum compatibility condition as the lowest order coefficient.

\begin{proof}
Proof of the Theorem depends on expressing the rate of change of distance in terms of strains.
 \begin{lemma}\label{lem_induced}
 Let $\mathcal B_{3r}\subset \mathbf R^2$ be a disk radius $3r$ about the origin and $a_i, a_j\in \mathcal B_{3r}$.  Let $u\in\mathcal C^4(\mathcal B_{3r},\mathbf R^2)$ be an infinitesimal deformation with strains given by (\ref{eq_prestr}). If $\phi(x,t)$ is a deformation such that $\phi(x,0)=x$ and $\dot\phi(x,0)=u(x)$, then
 $$
 \left.\frac d{dt}\right|_{t=0}|\phi(a_i,t)-\phi(a_j,t)|=
 \frac{1}{|a_i-a_j|}\int_0^1(a_i-a_j)^T\epsilon(\gamma(s))(a_i-a_j)\, ds
 $$
 where $\gamma(s)=a_i+s(a_j-a_i)$ for $0\le s\le 1$ is a parameterization of  the line segment from $a_i$ to $a_j$.
 \end{lemma}

\noindent{\it Proof of Lemma.} Let the position of the vertices be denoted $\phi(a_i,t)$ with initial position  $a_i=\phi(a_1,0)$ and initial velocity $U_i=\frac{\partial}{\partial t}\phi(a_1,0)$. Then the elongation

\begin{align*}
L_{ij}&=
\left.\frac d{dt}\right|_{t=0}|\phi(a_i,t)-\phi(a_j,t)|\\
&=
\left.\frac d{dt}\right|_{t=0}\sqrt{(\phi(a_i,t)-\phi(a_j,t))^T (\phi(a_i,t)-\phi(a_j,t))}\\
&=\left.\frac{(\phi(a_i,t)-\phi(a_j,t))^T (\dot\phi(a_i,t)-\dot\phi(a_j,t))}{\sqrt{(\phi(a_i,t)-\phi(a_j,t))^T(\phi(a_i,t)-\phi(a_j,t))}}
\right|_{t=0}\\
&=\frac{(a_i-a_j)^T(u(a_i)- u(a_j))}{|a_i-a_j|}
\end{align*}
By the Fundamental Theorem of Calculus,
\begin{align*}
L_{ij}&=\frac{1}{|a_i-a_j|}\int_0^1(a_i-a_j)^T\frac d{ds} u(\gamma(s))\, ds\\
&=\frac{1}{|a_i-a_j|}\int_0^1(a_i-a_j)^T \nabla u(\gamma(s))\,\dot\gamma(s)\, ds\\
&=\frac{1}{|a_i-a_j|}\int_0^1(a_i-a_j)^T \nabla u(\gamma(s))\,(a_i-a_j)\, ds
\end{align*}
Now the matrix  where $\nabla u$ may be replaced by its symmetrization $\epsilon(\gamma(s))=\frac 12 (\nabla u+(\nabla u)^T)$
in the quadratic form
\begin{align*}
L_{ij}&=\frac{1}{|a_i-a_j|}\int_0^1(a_i-a_j)^T \epsilon(\gamma(s))\,(a_i-a_j)\, ds,
\end{align*}
 proving the lemma.

Finally, the strains are expressed in  Taylor Series about the origin. The elongations of the edges of the hexagon~$\mathcal H$ are computed by integrating the Taylor Series in their expressions. The twelve elongations are put into the wagon wheel condition, and coefficients are collected (using \copyright\textsc{Maple}!) to yield \eqref{eq_ccexpansion}. 
\end{proof}
The theorem was first proved by Krtolica~\cite{K:2016}.

\subsection{Compatibility curve sums  for (LD) implies curve integrals for  (LC)}

Let $\Omega$ be a simply connected subdomain with $\mathcal C^2$ boundary. We can build an approximation  $\Omega_n$ by approximating $\partial\Omega$ by a piecewise linear curve that passes through $n$ equally distant  points $V_{n,1},V_{n,2},\ldots,V_{n,n}\in\partial \Omega$ taken in order around  $\partial \Omega$, attaching inward  facing equilateral triangles to each of the segments, connecting their  interior vertices with edges forming a ring girder $G_n$ along the boundary, and then filling the remainder with an arbitrary triangulation. 
 Then the  compatibility sum for (LD) gives an equation $\mathcal V(G_n,L)=0$ on the prescribed elongations $L$ which is a  weighted sum involving all edges of the double layer, the edges in the girder $G_n$  at most one link from $\partial \Omega_n$. We may partition the girder into $n$ pieces $G_{n,i}$ localized near each of the rim vertices $V_{n,i}$ and split the sum
 $$
 G_n=\bigcup_{i=1}^n G_{n,i};\qquad \mathcal V(G_n,L_n)=\sum_{i=1}^n\mathcal V(G_{n,i}L_n)
 $$
  It turns out, that if we fix a vertex $V_{n,1}=X\in\partial\Omega$ and take an arbitrary strain field $\epsilon$ near $X$, and consider its induced elongations $L_n$,  for the constructed triangulations, then the boundary strain compatibility for (LD) of each localized piece converges to the (LC) boundary  integrand~\eqref{eq_bounintegrand}
 $$
 \mathcal V(G_{n,1},L_n)\, \Delta_n\to \beta(e_1(X))=
 \left[-\frac{\partial\epsilon_{11}}{\partial \bnu}(X)+(\epsilon_{11}(X) -\epsilon_{22}(X))\kappa(X)\right]\, ds
 $$ 
as $n\to\infty$, where $r=\Delta_n=|V_{n,i+1}-V_{n,i}|$ for all $i$ is the common distance between boundary vertices at the $n$-th stage and $\bnu$ is the inward normal.

\begin{figure}[h]
      \begin{center}
          \scalebox{0.5}{\includegraphics{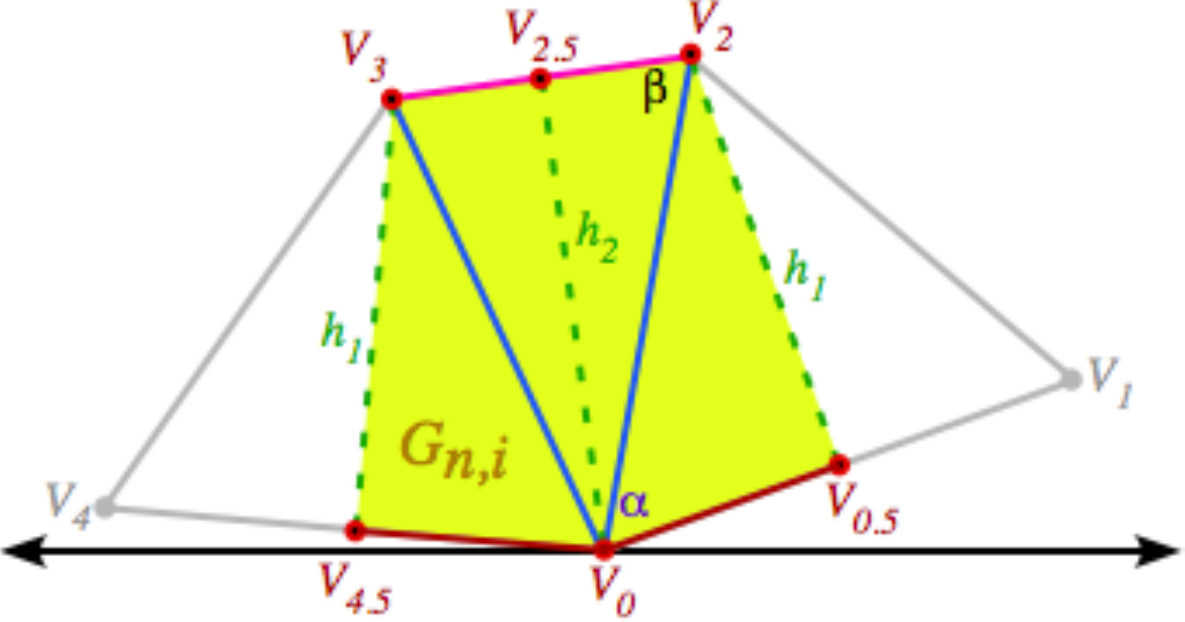}}
 \caption{Piece of a boundary girder $G_{n,i}$. \label{fig_interiorpoint}}
 \end{center}
       \end{figure}

Since we suppose that the boundary is $\mathcal C^3$ we perform the computation for a specific boundary curve that agrees up to the third order to any given boundary curve.

\begin{thm}[Expansion of compatibility condition along a curve]\label{th_Dbouncc}
 Let $ \Omega$ be a subdomain and $\partial\Omega$ a $\mathcal C^3$ curve through the origin $V_0=0$ and tangent to the $x$-axis such that at the origin, its curvature is  $\kappa$, and its derivative of curvature with respect to arclength is $b$.
 Let  $\Omega$ be the region above the curve. Let $B_{\delta}\subset \mathbf R^2$ be a disk radius $\delta$ about the origin. Let $V_1,V_4\in\partial\Omega$  be vertices on both sides of the origin such that $|V_1-V_0|=|V_4-V_0|=r$ and let $V_2$ and $V_3$ be interior vertices above $\partial \Omega$ such that $\bigtriangleup V_0V_1V_2$ and $\bigtriangleup V_0V_3V_4$ are equilateral triangles.
We suppose that $r>0$ is so small that $V_1,\ldots,V_4$ are in $B_{\delta}$. Let $\mathcal T_r$ be a truss such that $V_0$ is adjacent only to vertices $V_1$, $V_2$, $V_3$ and $V_4$.  Let $u\in\mathcal C^4( B_{\delta},\mathbf R^2)$ be an infinitesimal deformation satisfying the strain equation (\ref{eq_prestr}).

If we denote $V_{0.5}$ and $V_{3.5}$ as the midpoints of the sides $V_0V_1$ and $V_4V_0$, resp., then let the localized piece of the boundary girder $G_{n,1}$ near the origin be  the  $V_0V_{0.5}V_2V_3V_{4.5}$ part of the truss $\mathcal T_r=\{ E_{01},E_{02},E_{03},E_{04}, E_{12}, E_{23}, E_{34}\}$. 
The curve compatibility condition for $\partial \Omega$ of Theorem~\eqref{th_bounlay}, where we take half of the contributions from sides $E_{01}$ and $E_{04}$, for the $u$-induced rates of change of distances between vertices of $\mathcal T_r$ has the Taylor expansion about the origin 
\begin{equation}\label{eq_bounlim}
\begin{split}
\mathcal V(G_{n,1},L_n)&=-\epsilon_{11,2}+(\epsilon_{11} -\epsilon_{22})\kappa\\
 &\quad+\left[\frac{\sqrt 3}{12} \epsilon_{11,11} -\frac{\sqrt 3}{4}\epsilon_{22,22}+\left(\frac{3\sqrt 3}4\epsilon_{11,2}+\frac{\sqrt 3}6\epsilon_{12,1} -\frac{\sqrt 3}4\epsilon_{22,2}\right)\kappa\right]r\\
 &\quad +  \left[\left( \frac{1}8\epsilon_{11,11}-\frac{1}6\epsilon_{12,12}-\frac{1}{24}\epsilon_{22,11}-\frac{1}6\epsilon_{22,22}\right)\kappa+\left(\frac{1}2\epsilon_{12,1}-\frac{9}8\epsilon_{22,2}\right)\kappa^2\right.\\
 &\qquad\left.\quad +\frac{b}3\epsilon_{11}+\frac{b\kappa^2}3\epsilon_{12}+\left(\frac{\kappa^3}8-\frac b3\right)\epsilon_{22}\right]r^2+\operatorname{o}(r^2)
\end{split}
\end{equation}
as $r\to 0$.
Hence, in the limit, the discrete curve sum compatibility condition as $r\to 0$ tends to the continuum curve integral compatibility conditions~\eqref{eq_bounintegrand}.

\end{thm}

The distance between $V_2$ and $V_3$ will be smaller or larger than $r$, depending on whether $\kappa>0$ or $\kappa<0$. Note that the third derivative of the boundary influences only the $r^2$ term.

\begin{proof} The proof is similar to Theorem~\ref{th_pedja}. We express the coordinates of the vertices in terms of $r$ and use Lemma~\ref{lem_induced} to compute the induced elongations $L_n$. We apply the boundary compatibility equations from Theorem~\ref{th_bounlay}
expressed as power series in $r$. The result pops out using the computer algebra system \copyright\textsc{Maple}. Here are some details.

For convenience, parameterize the curve in terms of $r$,  the length of the segment from the origin to the point on the curve
$$
(x(r),y(r))=r\bigl(\cos\alpha(r),\sin(\alpha(r)\bigr)
$$
where
$$
\sin\alpha(r)=\frac {\kappa}2 r + \frac b6 r^2.
$$
Approximating $\cos\alpha(r)=\sqrt{1-\sin^2\alpha(r)}$ by the binomial series, and truncating to fourth degree,
$$
\cos\alpha(r)\approx 1-\frac{\kappa^2}8 r^2-\frac{\kappa b}{12}r^3 - \left(\frac{\kappa^4}{128}+\frac{b^2}{72} \right)r^4
$$
The curvature is 
$$
\kappa(r)=\kappa+br+\frac 34\kappa^3r^2 +\frac {37}{24} \kappa^2 b r^3 + {\mathbf O}(r^4)
$$
The points $V_1=(x(r),y(r))$ and $V_4=(x(-r),y(-r))$ are on the curve such that  $|V_1|=|V_4|=r+{\mathbf O}(r^6)$. $V_2$ and $V_3$ are found by rotating $V_1$ and $V_4$ by $\pm 60^{\circ}$. In the curve sum compatibility condition,  $\ell_{23}=|V_3-V_2|$.
$(\ell_{23})^2=\frac 34[1+X(r)]r^2$ may be computed from the Pythagorean formula for $V_3-V_2$ and then the power series of  $\ell_{23}$ and $\ell_{23}^{-1}$ may be computed from the binomial series in $X(r)$.
The support distance is $h_1=\frac{\sqrt 3}2r$ from $V_2$ to $V_0V_1$ and the angle $\alpha=\angle V_1V_0V_2=\frac{\pi}3$. If $V_{3.5}=\frac 12(V_2+V_3)$ is the midpoint, then the support distance $h_2 = |V_{3.3}|$. $h_2^{-1}$ is also found using a binomial expansion. The angle $\beta=\angle V_3V_2V_0=\angle V_2V_3V_0$. A series is deduced from $\cos\beta=\frac{\ell_{23}}{2r}$.

The compatibility condition from Theorem~\ref{th_bounlay}, taking half the contribution of the $V_0V_1$ and $V_0V_4$ sides, is
$$
\mathcal V(G_{n,1})r=\frac 1{2h_1}\left(L_{01}+ L_{40}\right)-\frac 1{h_2}L_{2,3}+
\left(\frac{\cos\beta}{h_2}-\frac{\cos\frac{\pi}3}{h_1}  \right)\left (L_{03}+L_{04}\right).
$$
The lowest order terms of this expression are given by~\eqref{eq_bounlim}.
\end{proof}
Note that this  approximation requires a second order Taylor approximation. The first order terms cancel and that second order terms limit to the compatibility condition.

\section{Appendix}

\subsection{Proof of the genericity of BTP structures, Theorem~\ref{th_BTP}.}\label{ss_aBTP}

\begin{proof}
The rigidity and compatibility count is deduced from the infinitesimal rigidity equations $Aw=0$. If the coordinates of the $i$th vertex are $z_i=(x_i,y_i)$ and the deformation (infinitesimal velocity) $w_i=(u_i,v_i)$ then the single segment satisfies
$$
(x_1-x_2)(u_1-u_2)+(y_1-y_2)(v_1-v_2)=0.
$$
Since $(x_1,y_1)\ne (x_2,y_2)$ the equation is full rank and has three dimensional kernel, thus the segment is infinitesimally rigid. Its compatibility is $c=e-2v+3=1-4+3=0$.

The BTP constructions are illustrated in Figure~\ref{fig_BTP}.
In the bigon construction, the vertices are distinct for trusses $S$ and $T$. Pinning two vertices amounts to adding four equations
\begin{align*}
\begin{aligned}
A(S)W_S &= 0\\
A(T)W_T &= 0
\end{aligned}\qquad\qquad
\begin{aligned}
u_1&=u_3\\
v_1&=v_3
\end{aligned}\qquad\qquad
\begin{aligned}
u_2&=u_4\\
v_2&=v_4.
\end{aligned}
\end{align*}
The number of variables is $2v_S+2v_T$. The number of equations is $e_S+e_T+4$. The kernels of $A(S)$ and $A(T)$ consist of rigid motions. The four extra equations guarantee that the rigid motion is the same for both $A(S)$ and $A(T)$, hence the grand system has 
three dimensional kernel: the bigon is rigid. The Maxwell count of the bigon is
$$
c_{\text{bigon}}=(e_S+e_T+4)-2(v_S+v_T)+3=(e_S-2v_S+3)+(e_T-2v_T+3) +1=c_S+c_T+1.
$$

In the triangle construction, the vertices are distinct for trusses $S$, $T$ and $U$. Connecting the legs of the triangle  amounts to adding six equations
\begin{align*}
\begin{aligned}
A(S)W_S &= 0\\
A(T)W_T &= 0\\
A(U)W_U &= 0
\end{aligned}\qquad\qquad
\begin{aligned}
u_2&=u_3\\
v_2&=v_3\\
u_4&=u_5
\end{aligned}
\qquad\qquad
\begin{aligned}
v_4&=v_5\\
u_6&=u_1\\
v_6&=v_1.
\end{aligned}
\end{align*}
The number of variables is $2v_S+2v_T+2v_U$. The number of equations is $e_S+e_T+e_U+6$. The kernels of $A(S)$, $A(T)$ and $A(U)$ consist of rigid motions. The six extra equations from the three sides and the  non-degeneracy of the triangle  guarantee that the rigid motion is the same for all three $A(S)$, $A(T)$ and $A(U)$, hence the grand system has 
three dimensional kernel: the triangle is rigid. The Maxwell count of the triangle is
\begin{align*}
c_{\text{triangle}}&=(e_S+e_T+e_U+6)-2(v_S+v_T+v_U)+3\\
&=(e_S-2v_S+3)+(e_T-2v_T+3) +(e_U-2v_U+3) =c_S+c_T+c_U.
\end{align*}

In the prism construction, there are three distinct vertices in the  trusses $P$ and $Q$ and two distinct vertices in each of the trusses $R$, $S$ and  $T$. Intuitively, if $P$ and $Q$ connected with just two legs $R$ and $S$ there wii be one  one degree of freedom of motion. The appropriate third leg $T$ will prevent such motion. If, for example, the three connecting legs were parallel, then the prism would admit an infinitesimal shear motion perpendicular to the legs.   Connecting the legs of the prism  amounts to adding twelve equations
\begin{equation}\label{eq_genp}
\begin{aligned}
\begin{aligned}
A(P)W_P &= 0\\
A(Q)W_Q &= 0\\
A(R)W_R &= 0\\
A(S)W_S &= 0\\
A(T)W_T &= 0
\end{aligned}\qquad\qquad
\begin{aligned}
 u_1 &=  u_7\\u_2 &=  u_8\\u_3 &=  u_9\\u_4 &=  u_{10}\\u_5 &=  u_{11}\\u_6 &=  u_{12}
 \end{aligned}\qquad\qquad
 \begin{aligned}
  v_1 &=  v_7\\v_2 &=  v_8\\v_3 &=  v_9\\v_4 &=  v_{10}\\v_5 &=  v_{11}\\ v_6 &=  v_{12}
  \end{aligned}
\end{aligned}
\end{equation}
The number of variables is $2v_P+\cdots +2v_T$. The number of equations is $e_P+\cdots +e_T+12$. The kernels of $A(P)$ through $A(T)$ consist of rigid motions. The twelve extra equations do not necessarily guarantee that the rigid motion is the same for all five $A(S)$ - $A(T)$ unless $R$, $S$ and $T$ are in a non-degenerate position relative to $P$ and $Q$, namely, the system \eqref{eq_genp} has a
three dimensional kernel so the prism is rigid. 
The Maxwell count of the prism is
\begin{align*}
c_{\text{prism}}&=(e_P+\cdots +e_T+12)-2(v_P+\cdots +v_T)+3\\
&=(e_P-2v_P+3)+\cdots + (e_T-2v_T+3) =c_S+\cdots +c_T.
\end{align*}

We may write this nondegeneracy condition as a determinant inequality. View $R$, $S$ and $T$ as segments so, {\it e.g.,} $A(R)$ has the same kernel as 
$$
(x_1-x_4)(u_1-u_4)+(y_1-y_4)(v_1-v_4)=0.
$$
We know that $P$ moves as a rigid body, so its displacement field is given by three parameters $a$, $b$, $c$ corresponding to translation and rotation. In three dimensions, the velocity field of a rotation about the origin at $Z$ is given by cross product with a fixed vector $H$
$$
W(Z)= Z \times H.
$$
Infinitesimal rotation in the $x$-$y$ plane is given by crossing with $H=(0,0,c)$, Adding translation $(a,b)$, the velocity of any rigid motion is thus
\begin{equation}\label{eq_infv}
(u,v)=(a+cy, b-cx).
\end{equation}
The prism is rigid if the legs connecting them make the rigid motions of $P$ and $Q$  coincide. Substituting the unknown motion \eqref{eq_infv} for vertices of $P$ and any fixed motion, say $(u,v)=(0,0)$ for the vertices of $Q$ in the leg equations
\begin{equation}\label{eq_legs}
\begin{aligned}
(x_1-x_4)(u_1-u_4)+(y_1-y_4)(v_1-v_4)&=0\\
(x_2-x_5)(u_2-u_5)+(y_2-y_5)(v_2-v_5)&=0\\
(x_3-x_6)(u_3-u_6)+(y_3-y_6)(v_3-v_6)&=0
\end{aligned}
\end{equation}
gives a homogeneous linear system for $a$, $b$ and $c$. It has a trivial kernel if its determinant is nonvanishing, namely
\begin{equation}\label{eq_prismgen}
\left|\begin{matrix}
x_1-x_4&y_1-y_4&x_1y_4-x_4y_1\\
x_2-x_5&y_2-y_5&x_2y_5-x_5y_2\\
x_3-x_6&y_3-y_6&x_3y_6-x_6y_3
\end{matrix}\right|\ne 0.
\end{equation}
If the legs were parallel, then the first two columns are multiples of one another and the determinant vanishes. But other configurations allow infinitesimal flexes. For example, if the lines determined by the legs meet at the origin, then the areas of the parallelogram determined by the endpoints of the legs all vanish, so the last column is zero.
In this case, a nontrivial flex is given by the velocity field of a rotation about the origin for $P$ and zero for $Q$. The determinant is invariant under translation so that any point may be the meeting point.

In the pinning construction, the vertices are distinct for the truss $T$ and two distinct vertices $z_1$ and $z_2$ share the same coordinates. Pinning two vertices amounts to adding two equations
\begin{align*}
\begin{aligned}
A(T)W_T &= 0
\end{aligned}\qquad\qquad
\begin{aligned}
u_1&=u_2\\
v_1&=v_2
\end{aligned}\qquad\qquad
\end{align*}
The number of variables is $2v_T$. The number of equations is $e_T+2$. The kernels of $A(T)$ consist of rigid motions. The two extra equations restrict the kernel further so the pin is infinitesimally rigid.  The Maxwell count of the pin is
$$
c_{\text{pin}}=(e_T+2)-2v_T+3=(e_T-2v_T+3) +1=c_T+2.
$$
\end{proof}

 \subsection{Proof that wagon wheels form a basis in triangular domains.}\label{ss_hex}
 
 The method of proof is to show that there is a maximal statically determined sub-truss In $X$ that omits exactly $v_i$ edges of $X$, and thus there are $v_i$ compatibility conditions. We remove an edge from each interior hexagon in turn showing that the wagon wheel of that hexagon is independent of the remaining hexagons. 

Let $\mathcal F_Y$ denote the triangles of $X$ which are not in any plate. Define a graph $\mathcal G^Y$ consisting or vertices $\mathcal F_Y$ and edges between any two triangles of $\mathcal F_Y$ which share a common edge. The graph may not be connected. Let $F_i$ denote its connected components. Define $Y_i$ to be the truss made from the union of triangles in $F_i$. It turns out the $Y_i$ are girders connected by the plates.
Thus, after removing the plates, 
$$
\overline{X-\cup_{i=1}^p P_i}=\cup_{j=1}^y Y_j.
$$

\begin{lemma}
 The plates $P_i$ are bounded by a single simple closed curve (are simple trusses).
\end{lemma}
\begin{proof}Any point in $P_i$ is within one unit of a center in $\mathcal G_i$.  Because the straight line path between neighboring centers is also in $P_i$, it is possible to construct a path from any point to a path connecting centers to another point in $P_i$. Thus, $P_i$ is path connected. The boundary of $P_i$ can be at most one closed curve. If not, there are triangles inside the outer boundary of $P_i$ which are not in $X$, contrary to the assumption that $X$ is simply connected.\end{proof}
\begin{lemma}
The $Y_i$ are girders.
\end{lemma}
\begin{proof} $Y_i$ is connected because the graphs $F_i$ are connected. There are no hexagon points in $Y_j$. Thus every vertex is on the boundary. Thus any closed loop in $Y_j$ may be homotoped through $Y_j$ to closed loop $\tilde\gamma$ in the set of boundary paths. As for plates, all lattice points within $\tilde\gamma$ are in $X$. However, any point not already encountered in $\partial Y_j$ would be hexagon points, thus not part of $Y_j$.\end{proof}

\begin{lemma}
Girders are statically determined.
\end{lemma}
\begin{proof}
The girder is infinitesimally rigid. Any single triangle is determined up to a rigid motion. Gluing on another triangle to a rigid structure maintains rigidity because a common edge determines its motions. We shall show that removing any edge from girder results in a flexible structure, hence the girder is statically determined.
A nontrivial flex yields a nontrivial infinitesimal flex.

There are three types of triangles in a girder: those with exactly one, two or three neighboring triangles in the girder.
In the first case, removing a boundary edge leaves another boundary edge which is free to flex. Removing the common edge leaves an empty quadrilateral, which flexes.
In case the triangle has two neighboring triangles,  remembering that the opposite corner is not a hexagon, by removing the boundary edge the opposite corner becomes a hinge that flexes.
In the case that the triangle has three neighboring triangles, by removing an edge leaves an empty quadrilateral. Remembering that none of the vertices of the quadrilateral are hexagons, the quadrilateral flexes.
\end{proof}

$G_i$, being a finite, connected, simply connected subgraph of the triangular lattice has metric geometry. In many ways, it behaves like the Euclidean plane, and arguments from the plane can be applied to $G_i$. Choose a basepoint $c_1$. For each vertex in $P_i$, let $r(c)$ be the $G_i$-distance from the base point. ($r$ is for ``radius.'') Since each edge has unit length, $r(c)$ is the minimal number of edges in an edge-path connecting $c_1$ to $c$ in $G_i$. Any $c\in P_i\backslash G_i$ has $r(c)=r(c_j)+1$ where $c\in H(c_j)$. An edge-path between two vertices which realizes the distance is called  a {\it geodesic.} There may be many geodesics between any two points. However, in these triangular metric graphs, certain convexity properties still hold.  For example,  every point has a neighbor with a radius equal to $r(c)-1$.
The level sets of $r$ are then ``$r$-circles'' about $c_1$. It turns out that the $r$-circles are lines, namely made up of unions of simple paths. We give some proofs.

The triangular truss $H$ has geodesics and distance circles. The lattice is generated by the vectors $e_1=(1,0) $ and $e_2=\frac 12(1,\sqrt 3)$. Put $e_3=e_2-e_1$, $e_4=-e_1$, $e_5=-e_2$,  $e_6=-e_3$ and extend modulo $6$ so $e_7=e_1$ etc. The six edge directions emanating from a vertex are $e_1,\ldots,e_6$. Note that there is a unique geodesic between the points $c$ and $c+re_i$ of length $r$ given by the path $t\mapsto c+te_i$ where $0\le t\le r$. Every other path connecting the endpoints is longer. If a point is in a ``sector'' between the generating rays, say $c=c_1+a e_i+be_{i+1}$ where $a,b\in\mathbb N$ are positive integers, then $r(c)=a+b$ and the set of geodesics connecting $c_1$ to $c$ are zig-zag paths which consist of $a$ steps in the $e_i$ direction and $b$ steps in the $e_{i+1}$ direction, taken in any order. These geodesics sweep out a parallelogram between endpoints. Every other path connecting these endpoints is longer. This means that a geodesic curve either goes straight or turns right or left $60^{\circ}$ at each vertex. 
In the subgraph $G_i$, the paths are restricted to paths connecting points of $G_i$. This could mean that the boundary of $G_i$ may obstruct some of the paths between endpoints, or there may be a critical obstacle, meaning that all geodesics from $c_1$ to $c$ must pass through certain boundary points.  This is the usual situation for variational problems with an obstacle. If the obstacle is effective, then the distance minimizer goes through points of the obstacle. Note that $G_i$ is not very concave. If $G_i$ is to the right then the boundary cannot contain, in order, the points  $c$, $c+e_1$, $c+2e_1$ and $c+e_1+e_6$ because $c+e_1$ and $c+e_1+e_6$  are a unit apart and are  included as an edge  in $\partial G_i$.
Thus the boundary edges here are the points  $c$, $c+e_1$ and $c+e_1+e_6$. In other words, going around the boundary of $g_i$ clockwise the curve may turn right at most $60^{\circ}$ at a vertex. This is the same curvature as a glancing geodesic. Furthermore, three consecutive rights of $60^{\circ}$ don't occur. For example, if $G_i$ is on the left then $c$, $c+e_4$, $c+e_4+e_3$, $c+e_4+e_3+e_2$, $c+e_4+e_3+e_2+e_1$ cannot be consecutive boundary points because $P_i$ then contains the hexagon centered at $c+e_3$.

\begin{lemma} \label{lem_trim} Suppose $c\in G_i$ such that $c\ne c_1$. Then the minimal value of $r$ on $H(c)$ can occur at exactly one or  two neighboring boundary points of $\partial H(c)$.
\end{lemma}
\begin{proof} Thus is clear if $c_1\in H(c)$ so assume $c\notin H(c_1)$.

First, the minimum $c$ cannot be the center point because one of its neighbors must have value $r(c)-1$ on the boundary.
For brevity, let the hexagon be centered at the origin $c=0$, and suppose the minimum occurs at the two non-neighboring points $e_j$ and $e_k$ where $j+2\le k\le j+4\mod 6$. Let $\gamma_j$ and $\gamma_k$ be geodesics from $c_1$ to $e_j$ and $c_1$ to $e_k$, respectively. Then $\gamma_j\gamma_k^{-1}$ is a closed loop in $\mathcal G_i$. Since $\mathcal G_i$ is simply connected, all vertices of $\mathcal H$ interior to the loop also belong to $P_i$. Moreover, hexagons centered on these vertices are also in $P_i$ so the vertices belong to $\mathcal G_i$. Now,  $\gamma_j$ and $\gamma_j$ between the hexagon and the last effective boundary obstacle point (or $c_1$ if no obstacle) makes a closed loop $\sigma$ in $\mathcal H$. Follow the geodesic ray $t\mapsto te_{i+1}$, $t\ge 0$ from $e_j$ to where it meets the shortened loop $\sigma$, say on the $\gamma_i$ side of the obstacle point. Because the ray is strictly minimizing and emanates from the center of the hexagon in a different direction than either $\gamma_i$ (or $\gamma_j$), it is shorter than the corresponding arc on $\gamma_i$. Hence $r(e_{i=1})<r(e_i)$, a contradiction.

It follows that the minima of $r$ on $H(c)$ occur on the boundary of the hexagon and can be taken at most at two neighboring points.
\end{proof}

\begin{lemma} \label{lem_tri}
$r$ may not be equal at all three points of a  unit triangle in $G_i$.
\end{lemma}
\begin{proof} After a rigid motion, for contradiction we may suppose $r=r(0)=r(e_1)=r(e_2)$. Then the three points have neighbors with radius $r-1$. Up to reflection and rotation there are four cases.

Case 1: $r-1=r(e_3)=r(e_6)$. In this case $e_3$ and $e_6$ are points of $\partial H(0)$.  Bu Lemma~\ref{lem_trim}, at least one of $e_4$ or $e_5$ have radius $r-2$, contradicting $r(0)=r$.

Case 2: $r-1=r(e_2+e_3)=r(e_6)$. Let $gamma_6$ and $gamma_7$ be geodesics form $c_1$ to $e_6$ and $e_2+e_3$ resp. Let $\sigma$ be the loop $\gamma_6\gamma_7^{-1}$, at least from the last effective obstacle point. The points inside the loop are in $G_i$, as before. Say that these points are to the left of our five points. The ray $f(t)=t\mapsto te_4$ for $t\ge 0$ meets the loop at either $\gamma_6$ or $\gamma_7$ (or both.) If it is $\gamma_6$ then it is shorter that arcs from $\gamma_6$ joined to segment from $e_6$ to $e_0$ of length $r$. Hence $r(0)<r$ which is a contradiction. If it is $\gamma_7$ then the new  ray $t\mapsto e_2+te_4$ also meets $\gamma_7$ since it is parallel to the old ray and trapped inside the loop $f$ to $\gamma_7$ to $e_2+e_3$ to $e_2$ to $0$.  The new ray is shorter that arcs from $\gamma_7$ joined to segment from $e_2+e_3$ to $e_2$ of length $r$. Hence $r(e_2)<r$ which is a contradiction also.

Case 3: $r-1=r(e_5)=r(2e_1)=r(e_2+e_3)$. Let $\gamma_5$ and $\gamma_7$ be geodesics form $c_1$ to $e_5$ and $e_2+e_3$ resp. Let $\sigma$ be the loop $\gamma_5\gamma_7^{-1}$, at least from the last effective obstacle point. If the loop is to the left, we argue as in Case~2.  If the loop is to the right, the ray $f(t)=t\mapsto te_1$ for $t\ge 0$ meets the loop at either $\gamma_5$ or $\gamma_7$ (or both.) If it is $\gamma 5$ then it is shorter that arcs from $\gamma_5$ joined to segment from $e_5$ to $e_0$ of length $r$. Hence $r(0)<r$ which is a contradiction. If it is $\gamma_7$ then the new  ray $t\mapsto e_2+te_1$ also meets $\gamma_7$ since it is parallel to the old ray and trapped inside the loop $f$ to $\gamma_7$ to $e_2+e_3$ to $e_2$ to $0$.  The new ray is shorter that arcs from $\gamma_7$ joined to segment from $e_2+e_3$ to $e_2$ of length $r$. Hence $r(e_2)<r$ which is a contradiction also.

Case 4: $r-1=r(e_5)=r(2e_1)=r(2e_2)$: is almost the same as case~3.
\end{proof}

This lemma has some immediate consequences. The differential geometric analog in the Euclidean plane is that the gradient of the distance function from a point has unit gradient away from the center and so its level curves are smooth curves. Moreover, the curvature of distance circles is positive.

\begin{lemma} $r$ takes three values on $H(c)$ where $c\in G_i\backslash H(c_1)$. Level sets of $r$ are locally convex line segments.
\end{lemma}
\begin{proof}
Neighbors of the minimum points at radius $r$ have radius $r+1$. The remaining points must have radius $r+2$ by Lemma~\ref{lem_tri}. Thus, locally, if the center of a hexagon has radius $r$ then two neighbors along a line or along a $60^{\circ}$ angle have radius $r$ too. Hence $r$-circles are locally curves which are convex. 
\end{proof}

\begin{lemma}\label{lem_plate}
Plates satisfy Theorem~\eqref{th_simple}.
\end{lemma}
\begin{proof}
 The idea is to construct a maximal  statically determined subtruss $Z$ of $P_i$. The number of edges in $P_i\backslash Z$ is $v_i$, so that it is maximal and the number of compatibility conditions for the truss is $\CCM$.
 The rough idea is to build up the subtruss one hexagon at a time inductively by adding the remainder of each next hexagon minus one edge. Thus this establishes a one-to-one correspondence between the hexagons and the missing edges. The subtruss is infinitesimally rigid. Then one checks that removing any one of the remaining edges results in the truss that admits a nontrivial infinitesimal flex.
 
 Since $\mathcal G_i$ is connected, we begin by ordering the centers $c_j$ of the hexagons, where $j=1,\ldots,v_i$ in such a way that the next center is in the boundary of the union the previous hexagons 
 \begin{equation}\label{eq_armpit}
 c_{j+1}\in \partial Y_j,\qquad\text{were $Y_j= \cup_{k=1}^j\mathcal H(c_k)$}
 \end{equation}
 where $\mathcal H(c_k)$ is the hexagon centered at $c_k$ and $Y_k$ is that portion of $P_i$ consisting of the first $j$ hexagons.  We shall choose centers circlewise,  starting from $c_1$, then points with $r=1$, then $r=2$ until we reach the furthest point. All radii occur because $G_i$ is connected. Recall that the concentric circles are made up of a collection of paths that end the circle exits  $P_i$. Choose $c_2\in\partial H(c_1) \cap G_i$ at one end of a component circle. Then take centers in order along with this component of the bounding circle. Then continue at the start of the next component of the same circle and continue in this fashion until this circle has been exhausted. Then continue in the next higher radius $r$-circle in the same fashion. The result is that the distance function $r(c_i)$ is nondecreasing as a function of $i$. It also follows that the condition \eqref{eq_armpit} is satisfied. Using the fact that $r$-circles are locally convex is important because it implies that each new hexagon extends beyond $Y_i$ and adds three to seven new edges to $Y_{i+1}$, in particular, it adds at least one new interior edge.

 We shall construct a sequence of trusses $Z_j\subset Y_j$ by induction such that $Z_j$ is statically determined. It turns out that in our construction, all vertices of $Y_j$ occur in $Z_j$ and the boundary edges of $Y_j$ remain edges of $Z_j$.
 For the base case, $Y_1$ is a hexagon at the basepoint. Removing any one edge from $Y_1$ will produce a statically determined truss. The wagon wheel condition on this hexagon restricts the allowable elongation for the removed edge. For the sake of definiteness of our construction, let $Z_1$ be $Y_1$ with the edge with the $e_1$ edge from the center removed.
 
 Let us suppose that $Z_j\in Y_j$ be the statically determined subtruss which is $Y_j$ with $j$ edges removed. Let $Z_{j+1}$ be $Z_j$ with all new edges of $H(c_{j+1})$ except for one new interior edge.
 $$
 Z_{j+1}=\Bigl(Z_j \cup\Bigl[ H(c_{j+1})\backslash Y_j\Bigr]\bigr)\backslash (\text{one new interior edge of $H(c_{j+1})$})
 $$
 Note that the new edges of $Z_{j+1}$ are statically determined. If the velocities of the vertices are prescribed on $Z_j$, then they are determined for the new vertices. Note that this uses the fact that the additional edges come from a hexagon and are not two consecutive parallel edges which admit an infinitesimal flex. It also means that an additional wagon wheel condition on the new hexagon is required to restrict the elongation of the removed edge of $H(c_{j+1})$. Hence the new wagon wheel condition is independent of the previous wagon wheel conditions of $Y_j$.

 Let $Z$ denote the structure after adding the $v_i$th hexagon. We claim that the structure is infinitesimally rigid. To see it, notice that by construction, the number of new edge equations in $Z_{j+1}$ exactly equals two times the number of new vertices. Moreover, they are independent of the equations of $Z_j$. Hence the nullity of the matrix for $Z_{j+1}$ is the same as the nullity as the matrix for $Z_1$ which is three, corresponding to its infinitesimal isometries.

 Furthermore, removing one more edge from the new edges of $Z$ makes the structure infinitesimally flexible. Suppose we remove a new edge of $Z_{j}$. This removal introduces a nontrivial flex to $Z_{j}$, hence a nontrivial infinitesimal flex. The vertices of $Z_k$ with $k<j$ have zero velocities, but some new edge vertices are non-vanishing. With each additional hexagon, the velocities of $Z_{j}$ are used as boundary conditions, and the velocities on the new vertices of $Z_{j+1}$ are uniquely determined from the new equations. Similarly, the velocities can be computed at all the new vertices added after $Z_j$. Hence $Z$ has a nontrivial infinitesimal flex.
 \end{proof}
\begin{proof} [Proof of Theorem \ref{th_simple}]
The simple truss $X$ is the union of plates and girders, which are connected along edges.
We claim that the contact region of any two plates or girders consists of a single edge. If two consecutive edges are shared then the hexagon centered at the midpoint is wholly contained in $X$, hence is interior to some plate, and not on the boundary of a plate or a girder.
Similarly, if two non-consecutive edges are shared, then since each piece is simply connected, this forms two disjoint simple closed boundary curves for $X$, contrary to the assumption that there is only one. Note that we assume that the pieces have the same orientation as $\mathcal H$ so that the two pieces in opposite directions at each edge. If one of the gluing reversed the orientation, then a one-sided Moebius strip would have been formed.

If we consider a graph $\mathcal T$  where the vertices are the plates and triangles of girders, and the edges occur between two vertices whenever they share an edge, then $T$ is a tree. If not, $X$ is not simply connected: $X$ is bounded by more than one circle, contrary to assumption.

The union of girders and $Z$'s corresponding to the plates is a maximal statically determined substructure. Ordering the pieces in the tree, so that next piece attaches to the union of previous pieces, as in a tree search, we prove infinitesimal rigidity and statistical determinateness just as in Lemma~\ref{lem_plate}.
Since the hexagons of distinct plates overlap at most on a single boundary edge, the wagon wheel conditions of different plates are independent. Thus all $v_i$ wagon wheel conditions of $X$ are linearly independent.
\end{proof}

\subsection{Proof of Theorem~\ref{th_realize}, existence for (ND).}\label{ss_threal}
\begin{proof} [Proof of Theorem \ref{th_realize}] One proceeds like doing a jigsaw puzzle, constructing the immersion one triangle at a time by extending the immersion on this triangle from what was already built.

We claim that triangles can be ordered in the net so that the next triangle is connected exactly by one edge or exactly by two adjacent edges to the union of the previous triangles. Numbering the triangles in this order then $T_{n+1}$ shares exactly one or two   edges with $U_n=\cup_{j=1}^nT_j$. Such ordering may be chosen in reverse order from the abstract structure by removing first any boundary triangle with two boundary edges from $U_n$ resulting in a smaller disk $U_{n-1}$ with boundary. Continue one at a time. If there aren't any more boundary triangles with two boundary edges, then two cases are possible: either there remain interior vertices in $U_n$ or not. If there are interior vertices, remove a triangle from the boundary of $U_n$ whose boundary edge is opposite an interior vertex. This results in a disk $U_{n-1}$. Then continue removing triangles with two boundary edges as before. Stop when one triangle remains. If there are no more interior vertices in $U_n$, then there must be a boundary triangle with two edges on the boundary, so we proceed as before. To see this, suppose there are $V_i$ and $V_b$ interior and boundary vertices, $E_i$ and $E_b$ interior and boundary edges and $F$ faces. Because each interior edge bounds two triangles and each boundary edge bounds one, the total number of edges is three times the number of faces, less the number of interior edges which have been double-counted
$$
3F-E_i=E_i+E_b.
$$
The Euler characteristic formula of a disk $U_n$ is 
$$
V_i+V_b-E_i-E_b+F=1.
$$
Substituting for $E_i$ and using
 $V_b=E_b$ for disks we get
$$
V_b=F+2-2V_i.
$$
Thus if there are no interior vertices, there must be on average more than one boundary edge per face; in other words, one triangle has at least two boundary edges.

With a right ordering of triangles in hand, we may develop the configuration jig-saw fashion. 
 Say one common edge is  $E_{k,1}\subset T_k$ for $k\le n$. Then the PL map $p_n:U_n\to\mathbf E^2$ is extended to $U_{n+1}$ by the unique  linear map on the opposite side of $p_n(T_k)$ which agrees with $p_n$ on $E_{k,1}$. ({\it i.e.,} we paste the triangles in sequence.) If $T_{n+1}$ shares two edges with $U_n$, then by restriction of the triangulation, these must be two edges that meet at one endpoint, the interior vertex, say $V_k$, $k\le n$. The extension of one side coincides with the extension of the other side because, by the angle condition, the angle from $T_{n+1}$ is the angle deficit of $U_n$ at $V_k$ because the total is $2\pi$. The resulting extension completes $p_{n+1}$ to a PL map on a neighborhood of $V_k$. The image under $p_{n+1}$ is a Euclidean neighborhood of $p_{n+1}(V_k)$ homeomorphic to the neighborhood of $V_k$ in $U_{n+1}$. By choosing the sequence $T_n$ carefully so that each $U_n$ is a disk (which is OK on triangulated disks), this is the only type of common edge situations that are encountered.   Otherwise, one could imagine extending the map on a ring that surrounds some triangles so that filling in the holes will result in triangles with more than two edges in common with $U_n$. Thus we have provided an argument for the existence of an immersion $\phi_{\ell}$ for flat abstract structures.

It remains to argue that for different choices of the ordering of triangles in this construction, the PL map 
$\phi_{\ell}$ from the abstract net is unique up to rigid motion and reflection. Suppose that two configurations are built up starting from the same initial triangle. It suffices to show that the image of any point $X$ in the net is uniquely determined relative to the initial triangle $\phi_{\ell}(T_1)$. Let us consider two sequences of puzzle-pieces. Let $\phi_{\ell}$, the PL map obtained from the order $T_1,\ldots,T_n$ and let $\tilde\phi_{\ell}$ the PL map obtained from a different order  $\ T_{j_1},\ldots,T_{j_n}$, where $j_i$ is a permutation of $\{1,\ldots, n\}$ with $j_1=1$. 
Both $\phi_{\ell}$ and $\tilde\phi_{\ell}$ are continuous. Choose a point $X_0\in T_1=T_{j_1}$. Use an open/closed argument.
Take PL path $\gamma$ from $X_0$ to $X$ in $\mathcal S$. This means that $\gamma$ does not meet a vertex (except possibly $X$) and that restricted to each triangle it crosses, $\gamma$ is linear with nonzero velocity.
Let $t_0=\sup\{ t\in [0,1]: \text{$\phi_{\ell}(\gamma(s))=\tilde\phi(\gamma(s))$ for all $0\le s\le t$.}\}$. If $t_0=1$ then $\phi_{\ell}(X)=\tilde\phi_{\ell}(X)$, so suppose $t_0<1$. Since $\phi_{\ell}=\tilde\phi_{\ell}$ on $T_1$, the two functions agree while $\gamma$ is in $T_1$ so $t_0>0$. Since $\gamma$ has positive velocity in each $T_i$ it crosses, there is an $\epsilon>0$ so that $\gamma(t)$ is in the same triangle, say $T_i$ for $t_0-\epsilon<t<t_0$.
If $\gamma(t_0)$ is an interior point, there is a $\delta>0$ so that $\gamma(t)$ is in the same triangle for $t_0-\epsilon<t<t_0+\delta$. But if the linear functions $\phi_{\ell}$ and $\tilde\phi_{\ell}$ agree on the first part of the triangle, they must agree on the second part.  If $\gamma(t_0)$ is a boundary point of a triangle then by construction of $\gamma$, it is an interior point of some edge $E_{ij}$. Since both $\phi_{\ell}$ and $\tilde\phi_{\ell}$ are constructed by gluing the same Euclidean triangle along the edge $\phi_{\ell}(E_{ij})$ so that $\phi_{\ell}$ and $\tilde\phi_{\ell}$ continue to agree on $T_j$. Hence there is a $\delta>0$ so that $\gamma(t)$ is in $T_i\cup T_j$ for $t_0-\epsilon<t<t_0+\delta$. But if the linear functions $\phi_{\ell}$ and $\tilde\phi_{\ell}$ agree on both of these triangles. The upshot in both cases is that $\phi_{\ell}(t)$ and $\tilde\phi_{\ell}(t)$ agree for $0\le t<t_0+\delta$, contradicting the statement that $t_0$ is the largest interval of agreement.
\end{proof}

\end{document}